\documentclass[12pt,english,american]{article}
\usepackage[T1]{fontenc}
\usepackage[latin9]{inputenc}
\usepackage{geometry}
\usepackage{color}
\usepackage{babel}
\usepackage{amsthm}
\usepackage{amstext}
\usepackage{amssymb}
\usepackage{mathrsfs}
\usepackage{amsmath}
\usepackage{cancel}   
\usepackage[unicode=true,pdfusetitle,
 bookmarks=true,bookmarksnumbered=false,bookmarksopen=false,
 breaklinks=true,pdfborder={0 0 0},backref=false,colorlinks=true]
 {hyperref}
\hypersetup{
 linkcolor=blue,citecolor=blue, urlcolor=blue}

\makeatletter

\def\qed{$\Box$\medskip}

\newcommand{\beq}{\begin{equation}}
\newcommand{\eeq}{\end{equation}}
\newcommand{\beqa}{\begin{eqnarray}}
\newcommand{\eeqa}{\end{eqnarray}}
\newcommand{\ben}{\begin{arabicenumerate}}
\newcommand{\een}{\end{arabicenumerate}}

\def\bel{\begin{lem} } 
\def\eel{\end{lem} }
\def\bet{\begin{thm}}
\def\eet{\end{thm}}
\def\bed{\begin{defn}}
\def\eed{\end{defn} }

\def\bec{\begin{cor}}
\def\eec{\end{cor}}
\def\ber{\begin{rem}}
\def\eer{\end{rem}}

\usepackage{enumitem}		
\theoremstyle{plain}
\newtheorem{thm}{\protect\theoremname}[section]
\theoremstyle{definition}
\newtheorem{defn}[thm]{\protect\definitionname}
\theoremstyle{plain}

\theoremstyle{plain}
\theoremstyle{remark}
\newtheorem{rem}[thm]{\protect\remarkname}
\theoremstyle{plain}
\newtheorem{lem}[thm]{\protect\lemmaname}
\theoremstyle{plain}
\newtheorem{cor}[thm]{\protect\corollaryname}

\usepackage{txfonts,refstyle,xcolor}

\newref{con}{name = Conjecture\ }
\newref{prop}{name = Proposition\ }
\newref{def}{name = Definition\ }
\newref{sec}{name = Section\ }
\newref{sub}{name = Section\ }
\newref{thm}{name = Theorem\ }
\newref{lem}{name = Lemma\ }
\newref{cor}{name = Corollary\ }
\newref{fig}{name = Figure\ }
\newref{rem}{name =  Remark\ }

\usepackage{bbm}
\newcommand{\charf}{\mathbbm{1}}

\usepackage[all]{xy}

\newcommand{\xyR}[1]{%
     \makeatletter
     \xydef@\xymatrixrowsep@{#1}
     \makeatother
}

\newcommand{\xyC}[1]{%
     \makeatletter
     \xydef@\xymatrixcolsep@{#1}
     \makeatother
}

\newcommand{\ncol}[1]{\color{normalcolor}}

\makeatother

\addto\captionsamerican{\renewcommand{\corollaryname}{Corollary}}
\addto\captionsamerican{\renewcommand{\definitionname}{Definition}}
\addto\captionsamerican{\renewcommand{\lemmaname}{Lemma}}
\addto\captionsamerican{\renewcommand{\propositionname}{Proposition}}
\addto\captionsamerican{\renewcommand{\remarkname}{Remark}}
\addto\captionsamerican{\renewcommand{\theoremname}{Theorem}}
\addto\captionsenglish{\renewcommand{\corollaryname}{Corollary}}
\addto\captionsenglish{\renewcommand{\definitionname}{Definition}}
\addto\captionsenglish{\renewcommand{\lemmaname}{Lemma}}
\addto\captionsenglish{\renewcommand{\propositionname}{Proposition}}
\addto\captionsenglish{\renewcommand{\remarkname}{Remark}}
\addto\captionsenglish{\renewcommand{\theoremname}{Theorem}}
\providecommand{\corollaryname}{Corollary}
\providecommand{\definitionname}{Definition}
\providecommand{\lemmaname}{Lemma}
\providecommand{\propositionname}{Proposition}
\providecommand{\remarkname}{Remark}
\providecommand{\theoremname}{Theorem}

\addto\captionsamerican{\renewcommand{\corollaryname}{Corollary}}
\addto\captionsamerican{\renewcommand{\definitionname}{Definition}}
\addto\captionsamerican{\renewcommand{\lemmaname}{Lemma}}
\addto\captionsamerican{\renewcommand{\propositionname}{Proposition}}
\addto\captionsamerican{\renewcommand{\remarkname}{Remark}}
\addto\captionsamerican{\renewcommand{\theoremname}{Theorem}}
\addto\captionsenglish{\renewcommand{\corollaryname}{Corollary}}
\addto\captionsenglish{\renewcommand{\definitionname}{Definition}}
\addto\captionsenglish{\renewcommand{\lemmaname}{Lemma}}
\addto\captionsenglish{\renewcommand{\propositionname}{Proposition}}
\addto\captionsenglish{\renewcommand{\remarkname}{Remark}}
\addto\captionsenglish{\renewcommand{\theoremname}{Theorem}}
\providecommand{\corollaryname}{Corollary}
\providecommand{\definitionname}{Definition}
\providecommand{\lemmaname}{Lemma}
\providecommand{\propositionname}{Proposition}
\providecommand{\remarkname}{Remark}
\providecommand{\theoremname}{Theorem}

\begin{document}
\title{Block-diagonalization of infinite-volume lattice Hamiltonians with unbounded interactions} 
  \author{S. Del Vecchio\footnote{Dipartimento di Matematica, Universit\`a di Roma ``Tor Vergata", Italy
/ email: delvecchio@mat.uniroma2.it}\,,  J. Fr\"ohlich\footnote{Institut f\"ur Theoretiche Physik, ETH-Z\"urich , Switzerland / email: juerg@phys.ethz.ch}\,, A. Pizzo \footnote{Dipartimento di Matematica, Universit\`a di Roma ``Tor Vergata", Italy
/ email: pizzo@mat.uniroma2.it}}

\date{30/08/2021}

\maketitle

\abstract{In this paper we extend the \emph{local} iterative Lie-Schwinger block-diagonalization \mbox{method --} 
introduced in  \cite{DFPR3} for quantum lattice systems with bounded interactions in \textit{arbitrary} dimension--  to 
systems with unbounded interactions, i.e., systems of bosons. We study Hamiltonians that can be written as the
sum of a gapped operator consisting of a sum of on-site terms and a perturbation given by 
\emph{relatively bounded} (but unbounded) interaction potentials of short range multiplied by a real coupling constant $t$. 
For sufficiently small values of $\vert t \vert$ \textit{independent} of the size of the lattice, we prove that the
spectral gap above the ground-state energy of such Hamiltonians remains strictly positive.

\noindent
As in \cite{DFPR3}, we iteratively construct a sequence of 
\emph{local} block-diagonalization steps based on unitary conjugations of the original Hamiltonian and inspired by 
the Lie-Schwinger procedure. To control the ranges and supports of the effective potentials generated in the course 
of our block-diagonalization steps, we use methods introduced in \cite{DFPR3} for Hamiltonians with bounded 
interactions potentials. However, due to the unboundedness of the interaction potentails, weighted operator norms 
must be introduced, and some of the steps of the inductive proof by which we control the weighted norms of the effective 
potentials require special care to cope with matrix elements of unbounded operators.

\noindent
We stress that no \emph{``large-field problems''} appear in our construction. In this respect our operator methods turn 
out to be an efficient tool to separate the low-energy spectral region of the Hamiltonian from other spectral regions, 
where the unbounded nature of the interaction potentials would become manifest.  }\\
\vspace{5mm}
\newpage

\section{Introduction}

In \cite{DFPR3} we have studied families of quantum lattice systems of  fermions describing insulating materials 
in two or more dimensions. In the present paper  we extend the results obtained in \cite{DFPR3} to systems of bosons.
We consider Hamiltonians describing tight-binding models of 
particles hopping on a lattice $\mathbb{Z}^{d}, d\geq 2$. These  Hamiltonians read as the sum of an \textit{unperturbed operator}, $K_0$, and a 
\textit{perturbation}, $K_{I}$, consisting of a collection of short range \textit{interaction potentials}.  The operator $K_0$ 
is written as a sum of on-site terms, $H_{\mathbf{i}}$, i.e., the operators $H_{\mathbf{i}}$ depend on the degrees of freedom associated with
single sites $\mathbf{i} \in \mathbb{Z}^{d}$. Whereas in \cite{DFPR3} the operators constituting  $K_{I}$ are bounded, here, in presence of bosons, we consider and control unbounded interactions, more precisely interactions that are locally relatively-bounded w.r.t. $K_0$ in a sense specified in Sect. \ref{models}. 
Our study concerns the low-energy spectrum of the Hamiltonians 
of these systems. The main result is to  show that the ground-state energies of these Hamiltonians are 
separated from the rest of their energy spectrum by a strictly positive gap. 

Our analysis relies on a method introduced in \cite{FP} to iteratively  block-diagonalize the Hamiltonians with respect to the 
ground-state subspace of $K_0$. The block-diagonalization is accomplished by a sequence of unitary 
conjugations of the Hamiltonians. To this end we iteratively construct an 
anti-self-adjoint operator $S\equiv S(t)= -S(t)^{*}, t\in \mathbb{R},$ such that the ground-state of $K_0$, denoted by $\Omega$, is still the ground-state of the operator 
$e^{S}\big(K_0 + t\cdot K_{I} \big)e^{-S}$, and if we restrict this operator to the subspace orthogonal to $\Omega$ its spectrum is strictly 
above the ground-state energy, for values of $|t|$ ($t$ is the coupling constant) sufficiently small but independent of the size of the lattice.
The construction of $S=S(t)$ is inspired by a novel technique introduced in \cite{FP} for quantum chains and in \cite{DFPR3} for systems in arbitrary spatial dimensions larger than $1$. In \cite{DFPR1} the scheme of \cite{FP} was extended to one dimensional bosons systems with relatively bounded interactions analogous to those discussed in the present paper.

Our technique yields a unified (i.e., both for fermions and bosons, both for self-adjoint and complex Hamiltonians \cite{DFPR2}) multi-scale, iterative perturbation scheme enabling us to successively block-diagonalize the 
Hamiltonians associated with sequences of bounded, connected subsets of the lattice. In one dimension, 
such subsets are intervals. But, for $d>1$, the number of connected subsets of a given cardinality, $R$, containing a fixed point 
of the lattice grows exponentially in $R$, and this calls for a refinement of 
the methods in \cite{FP}. Indeed there is a qualitative difference between dimension one and dimensions larger than one: In dimension one, starting from a family of intervals corresponding to supports of interaction potentials, the connected sets corresponding to the supports of the interaction potentials created in the next step in our block-diagonalization procedure are again intervals.  In higher dimensions, however, starting from interaction potentials whose supports are rectangles, the growth process of the supports of interaction potentials created in our block-diagonalization would produce connected subsets of arbitrary shapes. The number of growth processes leading to shapes containing $n$ lattice bonds scales like $n!$, which leads to combinatorial divergences.
In \cite{DFPR3} we succeeded in overcoming this difficulty when extending our methods to models in arbitrary dimensions  $d > 1$. Our block-diagonalization method involves subtle growth processes of the supports of local interaction potentials in configuration space. The core of our method is geometric, i.e., it amounts to control the growth processes precisely enough to keep control of the new interaction potentials associated with the new shapes. The key idea is to associate what we call \emph{minimal rectangles} with arbitrary finite regions in the lattice (the smallest rectangles containing the given region) and then lump together all interaction potentials whose supports correspond to the same minimal rectangle. In this way we are able to dramatically reduce the number of shapes created in the course of our block-diagonalization procedure, at the price of a modest, but manageable overestimation of the (weighted) norms of the interaction potentials that are being created in the process. 

Similarly to the extension of the method of \cite{FP} for quantum chains to one-dimensional boson systems, the scheme in \cite{DFPR3}  also requires some modifications since weighted operator norms are introduced for the unbounded interaction potentials. Though these  modifications do not affect the geometric features of the strategy,  they are technically demanding at various steps of the proof by induction.

We focus our attention on unperturbed operators 
$K_0$ with a unique ground-state, $\Omega$, and a positive energy gap above their ground-state energy. But our method can be extended to families of operators with degenerate ground-state energies. Indeed, in \cite{FP}, our scheme has been employed to deal with small perturbations of the Hamiltonian of the Kitaev chain, which has a degenerate groundstate. The extension to unbounded operators for quantum chains (e.g. the massive $\phi^4$ model on a one-dimensional lattice) has been discussed in \cite{DFPR1}, where we specify the general structure of the degenerate ground-state subspace that allows us to implement our block-diagonalization scheme. Under the same assumption on the ground-state subspace, our method works in any dimensions.

The motivation of our analysis comes from recent studies of Hamiltonians of \textit{``topological insulators''} 
appearing in the characterization of
\textit{``topological phases''}, see e.g.   \cite{BN, BH, BHM, NSY2, NSY3}. However, the scope of our techniques is actually more general as shown in the present paper focused on boson systems. Indeed our methods provide a unified treatment of all these systems (fermion, boson, and spin systems), based on the same algorithms. Notably, a key advantage of our methods is that they completely avoid \emph{large-field problems}, which often render the analysis of models involving bosons very cumbersome. 
Several results can be found in the literature for models with bounded interactions -- see \cite{DS}, where fermionic path integral methods have been used for the same purpose; \cite{DRS}, inspired by KAM theory; \cite{NSY1}, \cite{H}, \cite{MZ}, where quasi-adiabatic flows have been constructed to establish results related to ours.  Only in \cite{Y} unbounded interactions are considered and similar results have been obtained by using cluster expansions based on operator methods (see also \cite{KT} for the same technique but applied to bounded interactions). 

\noindent
Methods similar to ours have become quite popular in work on many-body localisation; see \cite{I1, I2}.



{\bf{Acknowledgements.}}
S.D.V. and A.P.  acknowledge the MIUR Excellence Department Project awarded to the Department of Mathematics, University of Rome Tor Vergata, CUP E83C18000100006.
S.D.V. is supported by MUR CUPE83C18000100006.

\subsection{Ultralocal Lattice Hamiltonians in $d\geq 2$ dimensions}\label{models}
\setcounter{equation}{0}

Let $\Lambda_{N}^{d}\subset\mathbb{Z}^d$ be a finite,  $d$-dimensional square lattice, where each side consists of $N$ vertices, with $d\geq 2$. Each vertex is labeled by a multi-index $\bold{i}:=(i_1, \dots, i_d)$ where $i_j\in (1, \dots, N)$, $j=1,\dots,d$.  The Hilbert space of pure state vectors of the lattice quantum systems we consider is $$\mathcal{H}^{(N)}:=\bigotimes_{\bold{i}\in \Lambda_{N}^{d}}  \,\mathcal{H}_{\bold{i}}$$ where $\mathcal{H}_{\bold{i}}\simeq \mathcal{H}, \, \forall \bold{i}\in \Lambda_N^d$, and where $\mathcal{H}$ is a separable Hilbert space.   Let $H$ be a (possibly unbounded) non-negative operator such that $0$ is an eigenvalue of $H$ with corresponding eigenvector $\Omega \in \mathcal{H}$, and 
$$H \upharpoonright_{\lbrace \mathbb{C} \Omega \rbrace^{\perp}} \geq \charf \,,$$
where $\charf $ is the identity operator.

\noindent
We define, for $\bold{i}\in \Lambda_{N}^{d}$,
\begin{equation}\label{H_i}
H_{\bold{i}}:= ( \bigotimes_{\bold{j} \in \Lambda_{N}^{d}\setminus \{\bold{i}\}} \charf_{\bold{j}})\otimes  \underset{\underset{\bold{i}^{th} \text{slot}}{\uparrow}}{H} \,
\end{equation}
where $\charf_{\bold{j}}$ is the identity in the Hilbert space $\mathcal{H}_{\bold{j}}$.
Denote by $P_{\Omega_{\bold{i}}}$ the orthogonal projection onto the subspace
\begin{equation}\label{vacuum_i}
 ( \bigotimes_{\bold{j}\in \Lambda_{N}^{d}\setminus \{\bold{i}\}}\mathcal{H}_{\bold{j}})\otimes  \underset{\underset{\bold{i}^{th} \text{slot}}{\uparrow}}{\{\mathbb{C}\Omega\}} 
\subset \mathcal{H}^{(N)}\,, \quad \text{  and}\quad   P_{\Omega_{\bold{i}}}^{\perp} := \charf - P_{\Omega_{\bold{i}}}\,.
\end{equation}
Thus we have
\begin{equation*} 
H_{\bold{i}}=P^{\perp}_{\Omega_{\bold{i}}}\,H_\bold{i}\, P^{\perp}_{\Omega_{\bold{i}}}+P_{\Omega_{\bold{i}}}\, H_\bold{i} \, P_{\Omega_{\bold{i}}}
\end{equation*}
and
\begin{equation}\label{gaps}
P_{\Omega_{i}}\,H_\bold{i}\, P_{\Omega_{\bold{i}}}=0\,,\quad P^{\perp}_{\Omega_{\bold{i}}}\, H_{\bold{i}}\, P^{\perp}_{\Omega_{\bold{i}}}\geq P^{\perp}_{\Omega_{\bold{i}}}\,.
\end{equation}

We study lattice quantum systems with Hamiltonians of the form
\begin{equation}\label{Hamiltonian}
K_{\Lambda_N^d}= K_{\Lambda_N^d}(t):= \sum_{\bold{i}\in \Lambda^{d}_N}H_{\bold{i}}+ t \sum_{\underset{k:=|\bold{k}| \leq \bar{k}}{J_{\bold{k},\bold{i}}\subset J_{\bold{N}-\bold{1},\bold{1}}}} V_{J_{\bold{k},\bold{i}}}\,, 
\end{equation}
where:

\begin{enumerate}
\item[i)] $t \in \mathbb{R}$ is the coupling constant; 
\item[ii)] 
$J_{\bold{k},\bold{q}}\equiv J_{k_1, \dots, k_d\,;\,q_1,\dots,q_d}\subseteq \Lambda_N^{d}$ denotes the rectangle contained in $\Lambda_N^{d}$ whose sides have lengths $k_1, k_2, \dots, k_d$, and such that the coordinates of the $2^d$  vertices are sets of $d$ numbers with either $q_j$ or $ q_j+k_j$ at the j-th position, for all $1\leq j \leq d$; thus we have $\Lambda_{N}^{d} \equiv J_{\bold{N}-\bold{1},\bold{1}}$ where $\bold{N}-\bold{1}=(N-1, \dots, N-1)$ and $\bold{1}=(1, \dots, 1)$;
\item[iii)]
\begin{equation}
 |\bold{k}|:=\sum_{i=1}^{d}k_i\,
\end{equation}
i.e., $|\bold{k}|$ denotes the sum of the sides;
\item[iv)] $\bar{k} < \infty$ is an arbitrary, but fixed integer;
\item[v)]   $V_{J_{\bold{k},\bold{i}}}$ is a symmetric (possibly unbounded) operator acting on $\mathcal{H}^{(N)}$, localized in $J_{\bold{k},\bold{i}}$,  in the sense that 
\begin{equation}\label{potential}
V_{J_{\bold{k},\bold{i}}} \,\, \text{  acts as the identity on  }\,\, \bigotimes_{\bold{j}\in \Lambda_{N}^{d}\,;\, \bold{j}\notin J_{\bold{k},\bold{i}}}  \,\mathcal{H}_{\bold{j}}\,.
\end{equation}

\end{enumerate}
$J_{\bold{k},\bold{i}}$ will be referred to as the ``support'' of $V_{J_{\bold{k},\bold{i}}}$.\\
 Regarding the domains of these operators, we assume that 
\begin{equation}
D((H_{J_{\bold{k},\bold{i}}}^0)^{\frac{1}{2}})\subseteq D(V_{J_{\bold{k},\bold{i}}}),\quad \quad \end{equation}
  where $H_{J_{\bold{k},\bold{i}}}^0:=\sum_{\bold{l}\in J_{\bold{k},\bold{i}}}H_{\bold{l}}$. 
  Furthermore, we assume that the interaction potentials are uniformly relatively bounded with respect
  to the unperturbed Hamiltonian in the sense of quadratic forms, namely
  \begin{equation}\label{klmn-cond}
  |\langle \phi\,,\, V_{J_{\bold{k},\bold{i}}} \phi \rangle|\leq a_{\bar{k}}\langle \phi\,,\,(H_{J_{\bold{k},\bold{i}}}^0+1)\phi\rangle \,,
\end{equation}
for any $\phi \in D((H_{J_{\bold{k},\bold{i}}}^0)^{\frac{1}{2}})$, for some $N$-independent constant $a_{\bar{k}}>0$.
  
   Under these assumptions, 
the symmetric operator in (\ref{Hamiltonian})  is defined and bounded from below on the domain $D(H^0_{J_{\bold{N}-\bold{1},\bold{1}}})$. $K_{\Lambda_N^d}(t)$ can thus be extended via Friedrichs extension to a densely defined self-adjoint operator whose domain $D(K_{\Lambda_N^d})$ is such that $D((H_{J_{\bold{N}-\bold{1},\bold{1}}}^0)) \subseteq D(K_{\Lambda_N^d})\subseteq D((H_{J_{\bold{N}-\bold{1},\bold{1}}}^0)^{\frac{1}{2}})$. 
Under our hypotheses on the interaction potentials, and using the inequality
\begin{equation}
\sum_{J_{\bold{k},\bold{i}} \subset J_{\bold{N}-\bold{1},\bold{1}}}H_{J_{\bold{k},\bold{i}}}^0\leq \Big\{\prod_{j=1}^{d}(k_j+1)\Big\}\, \sum_{\bold{i}\in \Lambda^{d}_N} H_{\bold{i}}\,,
\end{equation}
for fixed $\bold{k}$, it is easily verified that this self-adjoint extension coincides with the self-adjoint operator defined through the KLMN theorem,  induced {\color{blue}by} the symmetric quadratic form associated with  (\ref{Hamiltonian}).

\begin{rem}
The results of this paper actually can be proven in the same way also for the slightly weaker assumption that the $V_{J_{\bold{k},\bold{i}}}$ are only symmetric quadratic forms on the form domain $Q(V_{J_{\bold{k},\bold{i}}})\supseteq D((H_{J_{\bold{N}-\bold{1},\bold{1}}}^0)^{\frac{1}{2}})$ (as opposed to symmetric operators) and that (\ref{klmn-cond}) is satisfied for any $\phi\in Q(V_{J_{\bold{k},\bold{i}}})$. The only difference is that one must use the KLMN theorem to define the selfadjoint operator $K_{\Lambda_N^d}$ and consequently there is no control on its domain other than $D(K_{\Lambda_N^d})\subseteq D((H_{J_{\bold{N}-\bold{1},\bold{1}}}^0)^{\frac{1}{2}})$.
\end{rem}
The constraint in (\ref{klmn-cond}) readily implies
that 
\begin{equation}\label{weighted}
\|(H_{J_{\bold{k},\bold{i}}}^0+1)^{-\frac{1}{2}}V_{J_{\bold{k},\bold{i}}}(H_{J_{\bold{k},\bold{i}}}^0+1)^{-\frac{1}{2}}\|\leq a_{\bar{k}}\, ,
\end{equation}
thus we introduce the weighted norm
\begin{equation}
\|V_{J_{\bold{k},\bold{i}}}\|_{H^0}:=\|(H_{J_{\bold{k},\bold{i}}}^0+1)^{-\frac{1}{2}}V_{J_{\bold{k},\bold{i}}}(H_{J_{\bold{k},\bold{i}}}^0+1)^{-\frac{1}{2}}\|\,,
\end{equation}
where the weight $(H_{J_{\bold{k},\bold{i}}}^0+1)^{-\frac{1}{2}}$ actually depends on the rectangle $J_{\bold{k},\bold{i}}$ even though it is not made explicit in the notation $\|\cdot\|_{H^0}$.

\subsection{Main result}
The main result proven in this paper is the following (see Theorem \ref{main-res}).

{\bf{Theorem.}}
\textit{If the coupling constant $t\in \mathbb{R}$ is small enough, more precisely $\vert t \vert < t_d$, for a sufficiently small, 
but positive and $N-$ independent constant $t_d$, and assuming that conditions (\ref{gaps}), (\ref{potential}) and (\ref{klmn-cond}) hold, the Hamiltonian $K_{\Lambda_N^d}$  defined in (\ref{Hamiltonian}) has the following properties 
\begin{enumerate}
\item[(i)]{ $K_{\Lambda_N^d}$ has a unique ground-state; and}
\item[(ii)]{ the energy spectrum of $K_{\Lambda_N^d}$ has a strictly positive gap,  $\Delta_{N}(t) $,   bounded below by $\frac{1}{2}$, above the 
ground-state} energy\,.
\end{enumerate}
These properties hold for arbitrary values of $N< \infty$.
}
\vspace{3mm}

The families of models to which our results apply include the anharmonic quantum crystal models described by Hamiltonians of the form
\begin{equation}
K^{crystal}_{\Lambda_N^d}:=\sum_{\bold{i}\in \Lambda^{d}_N}\Big(-\frac{d^2}{dx^2_{\bold{i}}}+V(x_{\bold{i}}))\Big)+t\sum_{j=1}^{d}\hat{\sum_{\bold{i}\,,\,\bold{i}+\bold{h}_j \in \Lambda^{d}_N}}\,W(x_{\bold{i}},x_{\bold{i}+\bold{h}_j})=:\sum_{\bold{i}\in \Lambda^{d}_N}H_{\bold{i}}+t\sum_{j=1}^{d}\hat{\sum_{\bold{i}\,,\,\bold{i}+\bold{h}_j \in \Lambda^{d}_N}}\,W(x_{\bold{i}},x_{\bold{i}+\bold{h}_j})
\end{equation}
where $\bold{h}_j:=(0, \dots, i_j = 1, \dots, 0)$ and the hat in the sum $\hat{\sum}_{\bold{i}\,,\,\bold{i}+\bold{h}_j \in \Lambda^{d}_N}$ means that, concerning the sites on the boundary, it is restricted  to pairs $(\bold{i}\,,\,\bold{i}+\bold{h}_j )$ of nearest sites, i.e., no periodic condition is imposed. The operator $K^{crystal}_{\Lambda_N^d}$
acts on the Hilbert space $\mathcal{H}^{N}:=\bigotimes_{\bold{i}\in \Lambda_{N}^{d}}  \,L^2(\mathbb{R}\,,\,dx_{\bold{i}})$, with the assumptions $V(x_{\bold{i}})\geq 0$,  $V(x_{\bold{i}})\to \infty$, for $|x_{\bold{i}}|\to \infty$,  $D((H_{\bold{i}}+H_{\bold{i}+\bold{h}_j})^{\frac{1}{2}})\subseteq D(W(x_{\bold{i}},x_{\bold{i}+\bold{h}_j}))$, and $W(x_{\bold{i}},x_{\bold{i}+\bold{h}_j})$ form-bounded by $H_{\bold{i}}+H_{\bold{i}+\bold{h}_j}$.  The class described above includes the $\phi^4-$model on the $d$-dimensional lattice, corresponding to $V(x_{\bold{i}})=x_{\bold{i}}^2+x_{\bold{i}}^4$ and $W(x_{\bold{i}},x_{\bold{i}+\bold{h}_j})=x_{\bold{i}}\cdot x_{\bold{i}+\bold{h}_j}$.
\\

{\bf{Organization of the paper.}} 
In Sect.  \ref{proofs}, we describe in detail the Local Lie-Schwinger procedure applied to the present context. In Sect. \ref{graphs}, we recall the notion of \textit{``minimal rectangles''} that plays a crucial role in our analysis. In Sect. \ref{globalstrat}, we explain the global strategy to obtain our main result, while Sect. \ref{defialg} contains the complete definition of the Lie-Schwinger block-diagonalization algorithm.
\noindent
\noindent
In Sect. \ref{sec:gap} we show how to provide a lower bound for the spectral gap $\Delta_{N}(t)$.
\noindent
Sect. \ref{mainsec} contains the main technical result of the paper (Theorem \ref{th-norms}), namely the proof of convergence of our procedure, with some technical Lemmas being deferred to Section \ref{teclem} and to Appendix \ref{appA} and \ref{appB}.
The final result of this paper, Theorem \ref{main-res}, follows from Theorem \ref{th-norms}.
\\

{\bf{Notation}}\\


\noindent
1) Throughout the paper, the same symbol is used for the operator $O_{\bold{j}}$ acting on $\mathcal{H}_{\bold{j}}$ and the corresponding operator $$ O_{\bold{j}} \otimes \charf_{J_{\bold{k},\bold{q}}\setminus \{\bold{j}\}}$$ which acts on $ \bigotimes_{\bold{i}\in J_{\bold{k},\bold{q}} }\mathcal{H}_{\bold{i}}$, for $ \bold{j}\in J_{\bold{k},\bold{q}}$. Similarly, with a slight abuse of notation, we use the same notation for an operator
$O_{J_{\bold{l}, \bold{i}}}$ acting on $\mathcal{H}_{J_{\bold{l}, \bold{i}}}:=\bigotimes_{\bold{j}\in J_{\bold{l}, \bold{i}}} \mathcal{H}_{\bold{j}}$ and the corresponding operator acting on the whole Hilbert space $\mathcal{H}^{(N)}$ which is obtained out of $O_{J_{\bold{l}, \bold{i}}}$ by tensoring by the identity operator on all the remaining sites.
\\

\noindent
2) We use the notation ``$\subset$" to denote strict inclusion, otherwise the notation  ``$\subseteq$" is used. 
\\

\noindent
3) The multiplicative constant which is implicit in the symbol $\mathcal{O}(\cdot)$ can possibly be dependent on the spatial dimension $d$.
\\

\section{The Local Lie-Schwinger Block-Diagonalization Algorithm}\label{proofs}

For expository purposes, the Hamiltonian that we shall study in the following sections is of the type
\begin{equation}\label{initial-ham}
K_{\Lambda^{d}_N}:=\sum_{\bold{i}\in \Lambda^{(d)}_N}H_{\bold{i}}+t\sum_{j=1}^{d}\sum_{q_1=1}^{N}\dots \sum_{q_j=1}^{N-1} \dots \sum_{q_d=1}^{N}\,V_{J_{\bold{1}_j,\bold{q}}}
\end{equation}
where 
\begin{equation}
(\bold{1}_j,\bold{q}):=(0,\dots, k_j=1, \dots,0\,;\,q_1,\dots,q_d)\,, \label{def-1_j}
\end{equation}
namely we restrict our study to nearest neighbours interaction terms. Any finite range interaction can be also treated in the same way. 
Furthermore without loss of generality we assume that $a_{\bar{k}}=\frac{1}{2}$, where $a_{\bar{k}}$ is the constant appearing in (\ref{klmn-cond}).

\subsection{Minimal rectangles $J_{\bold{k},\bold{q}}$}\label{graphs}

Recall that by $J_{\bold{k},\bold{q}}\equiv J_{k_1, \dots, k_d\,;\,q_1,\dots,q_d}$ we denote the rectangle in $\Lambda_N^{d}$ whose sides have lengths $k_1, k_2, \dots, k_d$ and such that the coordinates of its $2^d$  vertices are the $d$-tuples of integers with either $q_j$ or $ q_j+k_j$ at the j-th position, for all $1\leq j \leq d$. Recall further that by $|\bold{k}|$ we denote the sum of the sides, i.e.,
\begin{equation}
|\bold{k}|:=\sum_{i=1}^{d}k_i\,.
\end{equation}
As in \cite{DFPR3}, we consider the pairs $(\bold{k},\bold{q})$ labeling rectangles in $\Lambda_{N}^d$ to be ordered with a total ordering, $\succ$, defined as follows:
\begin{equation}\label{ordering}
(\bold{k}',\bold{q}') \succ (\bold{k},\bold{q})
\end{equation} 
 if
\begin{itemize}
\item $\sum_{j=1}^{d} k'_{j}> \sum_{j=1}^{d}k_j$ 
\item or if $\sum_{j=1}^{d} k'_{j}=\sum_{j=1}^{d}k_j$ and for some $0\leq j\leq d$ there holds $k'_l=k_l$ for $ l < j$ and $k_j>k'_j$
\item or if, for all $l$, $k'_l=k_l$ and for some $0\leq j\leq d$  there holds $q'_l=q_l$ for $ l > j$ and $q'_j>q_j$\,.
\end{itemize}
%
%
%


\begin{defn}
Let $J_{\bold{k},\bold{q}}$ and $J_{\bold{k}',\bold{q}'}$ be two rectangles in $\Lambda_N^d$ with nonempty intersection. The  \emph{minimal rectangle} associated with $J_{\bold{k},\bold{q}}\cup J_{\bold{k}',\bold{q}'}$ is defined to be the \textit{smallest} rectangle
containing  $J_{\bold{k},\bold{q}}$ and $J_{\bold{k}',\bold{q}'}$. Note that its corners  are the $2^d$ numbers with either
\begin{equation}
 \min\{q_j,q'_j\}\,, \quad\text{or}\quad \max\{q_j+k_j,q'_j+k'_j\}\,
\end{equation}
at the $j$-th position. The minimal rectangle associated with $J_{\bold{k},\bold{q}}$ and $J_{\bold{k}',\bold{q}'}$ is denoted 
by 
\begin{equation}
[J_{\bold{k},\bold{q}}\cup J_{\bold{k}',\bold{q}'}]\,.
\end{equation}
\end{defn}
\begin{defn}
Let $J_{\bold{k},\bold{q}}\subset J_{\bold{l},\bold{i}}$. We define 
\begin{equation}
\mathcal{G}^{(\bold{k},\bold{q})}_{J_{\bold{l},\bold{i}}}:=\{ \,\, J_{\bold{k}',\bold{q}'}\subseteq J_{\bold{l},\bold{i}}\quad \text{such that} \quad [J_{\bold{k},\bold{q}}\cup J_{\bold{k}',\bold{q}'}]=J_{\bold{l},\bold{i}}\,\,\}\,.
\end{equation}
Note that compared to the definition of $\mathcal{G}^{(\bold{k},\bold{q})}_{J_{\bold{l},\bold{i}}}$ in \cite{DFPR3} here there is no constraint $J_{\bold{k}',\bold{q}'}\neq J_{\bold{l},\bold{i}}$.
\end{defn}
\begin{rem}\label{shapes}
The number of shapes\footnote{The term "shape" here means an equivalence class of rectangles that can be obtained from one another by translation on the lattice.} of rectangles $J_{\bold{l},\bold{i}}$ at fixed $|\bold{l}|=l$ can be bounded from above by  $(l+1)^{d-1}= O(l^{d-1})$. As a consequence:
\begin{itemize}
\item[a)] the number of rectangles
$J_{\bold{k}, \bold{q}}\subset J_{\bold{r}, \bold{i}}$ with fixed circumference $k$ is bounded by\\
$(r+1)^d (k+1)^{d-1}= O(r^d\, k^{d-1})$;
\item[b)]  the number of rectangles
$J_{\bold{k}', \bold{q}'}\subset J_{\bold{r}, \bold{i}}$ is bounded by $(r+1)^d\sum_{k=1}^r (k+1)^{d-1}= O (r^{2d})$;
 \item[c)] the number of rectangles in $\mathcal{G}^{(\bold{k},\bold{q})}_{J_{\bold{r},\bold{i}}}$ is bounded by
$2d(r+1)^{d-1}\sum_{k=1}^r (k+1)^{d-1}= O(r^{2d-1})$.
\end{itemize}
\end{rem}

\subsection{Transformed Hamiltonians}\label{globalstrat}
To prove our main result, Theorem \ref{main-res}, we employ the scheme of the local Lie-Schwinger block-diagonalization algorithm, developed in \cite{FP}, \cite{DFPR1}, \cite{DFPR2}, \cite{DFPR3}, which is here adapted to suit the present situation. We here briefly recap the strategy of the algorithm.

Recall that we consider the pairs $(\bold{k},\bold{q})$ to be totally ordered with the relation $\succ$ defined in Section \ref{graphs}.
For each $(\bold{k},\bold{q})$ with $(\bold{N}-\bold{1},\bold{1}) \succeq (\bold{k},\bold{q})\succeq  (\bold{0},\bold{N})$ - which we call the \emph{step} of the algorithm- we want to associate a potential term, $V^{(\bold{k},\bold{q})}_{J_{\bold{l},\bold{i}}}$,  to each rectangle $J_{\bold{l},\bold{i}}$, so that in the step $(\bold{k},\bold{q})$ we have the effective Hamiltonian 
\begin{eqnarray}\label{effectivehamil}
K_{\Lambda_N^{d}}^{(\bold{k},\bold{q})}
& =&
\sum_{\bold{i}\in \Lambda^{(d)}_N}H_{\bold{i}}+t\sum_{\bold{k}_{(1)}'\,,\,\bold{q}'}V^{(\bold{k},\bold{q})}_{J_{\bold{k}_{(1)}',\bold{q}'}}+t\sum_{\bold{k}_{(2)}'\,,\,\bold{q}'}V^{(\bold{k},\bold{q})}_{J_{\bold{k}_{(2)}',\bold{q}'}}+\dots+t\sum_{\bold{k}_{(|\bold{k}|)}'\,,\,\bold{q}'\,;\,(\bold{k}_{(|\bold{k}|)}'\,,\,\bold{q}')\prec (\bold{k},\bold{q})}V^{(\bold{k},\bold{q})}_{J_{\bold{k}'_{(|\bold{k}|)},\bold{q}'}}\quad\quad\quad \label{K-tranf-1}\\
& &+tV^{(\bold{k},\bold{q})}_{J_{\bold{k},\bold{q}}}\\
& &
+t\sum_{\bold{k}_{(|\bold{k}|)}'\,,\,\bold{q}'\,;\,(\bold{k}_{(|\bold{k}|)}'\,,\,\bold{q}')\succ (\bold{k},\bold{q})}V^{(\bold{k},\bold{q})}_{J_{\bold{k}'_{(|\bold{k}|)},\bold{q}'}}+t\sum_{\bold{k}_{(|\bold{k}|+1)}'\,,\,\bold{q}'}
V^{(\bold{k},\bold{q})}_{J_{\bold{k}'_{(|\bold{k}|+1)}},\bold{q}'}+
\dots+tV^{(\bold{k},\bold{q})}_{J_{\bold{N}-\bold{1},\bold{1}}},\label{K-tranf-2}
\end{eqnarray}
where 
\begin{itemize}
\item The index $\bold{k}_{(j)}'$ labels all the shapes of rectangles $J_{\bold{k}',\bold{q}'}$ such that $|\bold{k}'|=j$;
\item
The operator $V^{(\bold{k},\bold{q})}_{J_{\bold{l},\bold{i}}}$ acts as the identity on the spaces $\mathcal{H}_{\bold{j}}$ for $\bold{j} \notin J_{\bold{l},\bold{i}}$.  In general  $V^{(\bold{k},\bold{q})}_{J_{\bold{l},\bold{i}}}$ is $t$-dependent though this is not explicit in our notation. 
\end{itemize}
 Obviously we start with $V^{(\bold{0},\bold{N})}_{J_{\bold{l},\bold{i}}}:=V_{J_{\bold{l},\bold{i}}}$ and thus $K_{\Lambda_N^d}^{(\bold{0},\bold{N})}:=K_N$.

We will show that it is possible to define the potentials $V^{(\bold{k},\bold{q})}_{J_{\bold{l},\bold{i}}}$ so that the following properties are satisfied. There is $t_d>0$ (independent of $N$ and of $(\bold{k},\bold{q})$) such that for every $|t|<t_d$:
\begin{enumerate}
\item The effective Hamiltonian $K_{\Lambda_N^{d}}^{(\bold{k},\bold{q})}$ at step $(\bold{k},\bold{q})$ is obtained by a unitary conjugation of the effective Hamiltonian at the previous step $(\bold{k},\bold{q})_{-1}$, i.e. $K_{\Lambda_N^{d}}^{(\bold{k},\bold{q})_{-1}}$;
\item
For all rectangles $J_{\bold{l},\bold{i}}$ with $(\bold{k},\bold{q})\succ (\bold{l},\bold{i})$ and for the rectangle $J_{\bold{l},\bold{i}} = J_{\bold{k},\bold{q}} $ the associated $V^{(\bold{k},\bold{q})}_{J_{\bold{l},\bold{i}}}$ is block-diagonal w.r.t. the decomposition of the identity into the sum of the operators
\begin{equation}\label{pro-minus-multi}
P^{(-)}_{J_{\bold{l},\bold{i}}}:=\charf_{\mathcal{H}^{(N^{d})}\ominus \mathcal{H}_{J_{\bold{l},\bold{i}}}}\otimes  \Big(\bigotimes_{\bold{j}\in J_{\bold{l},\bold{i}}}  P_{\Omega_{\bold{j}}}\Big)\,,
\end{equation}
\begin{equation}\label{pro-plus-multi}
P^{(+)}_{J_{\bold{l},\bold{i}}}:=\charf_{\mathcal{H}^{(N^{d})}\ominus \mathcal{H}_{J_{\bold{l},\bold{i}}}}\otimes  \Big(\bigotimes_{\bold{j}\in J_{\bold{l},\bold{i}}} P_{\Omega_{\bold{j}}}\Big)^{\perp}\,.
\end{equation}
Furthermore each potential is not changed anymore by the algorithm once it is block-diagonalized, i.e., $V^{(\bold{k},\bold{q})}_{J_{\bold{l},\bold{i}}}=V^{(\bold{l},\bold{i})}_{J_{\bold{l},\bold{i}}}$ for every $(\bold{k},\bold{q})\succeq (\bold{l},\bold{i})$.
\item The effective potentials $V^{(\bold{k},\bold{q})}_{J_{\bold{l},\bold{i}}}$ are small in the sense that
\begin{equation}\label{ass-2}
\|(H_{J_{\bold{l},\bold{i}}}^0+1)^{-\frac{1}{2}}V^{{(\bold{k},\bold{q})}}_{J_{\bold{l},\bold{i}}}(H_{J_{\bold{l},\bold{i}}}^0+1)^{-\frac{1}{2}}\|=:\|V^{{(\bold{k},\bold{q})}}_{J_{\bold{l},\bold{i}}}\|_{H^0} \leq t^{\frac{l-1}{4}}\,.
\end{equation} 
Note that the larger is the support of the potential, the smaller is the associated weighted norm.
\end{enumerate}

\begin{rem}\label{remark-decomp}
Note that if $V^{(\bold{k},\bold{q})}_{J_{\bold{l},\bold{i}}}$ is block-diagonal w.r.t. the decomposition of the identity into $$P^{(+)}_{J_{\bold{l},\bold{i}}}+P^{(-)}_{J_{\bold{l},\bold{i}}}\,,$$
i.e., $$V^{(\bold{k},\bold{q})}_{J_{\bold{l},\bold{i}}}=P^{(+)}_{J_{\bold{l},\bold{i}}}V^{(\bold{k},\bold{q})}_{J_{\bold{l},\bold{i}}}P^{(+)}_{J_{\bold{l},\bold{i}}}+P^{(-)}_{J_{\bold{l},\bold{i}}}V^{(\bold{k},\bold{q})}_{J_{\bold{l},\bold{i}}}P^{(-)}_{J_{\bold{l},\bold{i}}}\,\,,$$  then for  $J_{\bold{l}',\bold{i}'}$ with $J_{\bold{l},\bold{i}}\subset J_{\bold{l}',\bold{i}'}$, it is also block-diagonal with respect to the decomposition of the identity 
$$P^{(+)}_{J_{\bold{l}',\bold{i}'}}+P^{(-)}_{J_{\bold{l}',\bold{i}'}}.$$
 Indeed we have
$$P^{(+)}_{J_{\bold{l}',\bold{i}'}}\Big[P^{(+)}_{J_{\bold{l},\bold{i}}}V^{(\bold{k},\bold{q})}_{J_{\bold{l},\bold{i}}}P^{(+)}_{J_{\bold{l},\bold{i}}}+P^{(-)}_{J_{\bold{l},\bold{i}}}V^{(\bold{k},\bold{q})}_{J_{\bold{l},\bold{i}}}P^{(-)}_{J_{\bold{l},\bold{i}}}\Big]P^{(-)}_{J_{\bold{l}',\bold{i}'}}=0$$
since for the first term we can use 
\begin{equation}
P^{(+)}_{J_{\bold{l},\bold{i}}}\,P^{(-)}_{J_{\bold{l}',\bold{i}'}}=0\,
\end{equation}
while for the second one we can use
\begin{equation}
P^{(-)}_{J_{\bold{l},\bold{i}}}V^{(\bold{k},\bold{q})}_{J_{\bold{l},\bold{i}}}P^{(-)}_{J_{\bold{l},\bold{i}}}\,P^{(-)}_{J_{{\bold{l}',\bold{i}'}}}=P^{(-)}_{J_{\bold{l}',\bold{i}'}}P^{(-)}_{J_{\bold{l},\bold{i}}}V^{(\bold{k},\bold{q})}_{J_{\bold{l},\bold{i}}}P^{(-)}_{J_{\bold{l},\bold{i}}}P^{(-)}_{J_{\bold{l}',\bold{i}'}}
\end{equation}
and
\begin{equation}
P^{(+)}_{J_{\bold{l}',\bold{i}'}}P^{(-)}_{J_{\bold{l}',\bold{i}'}}=0\,.
\end{equation}
\end{rem}

For fixed $(\bold{k},\bold{q})$ we define the Hamiltonian
\begin{equation}\label{def-G}
G_{J_{\bold{k},\bold{q}}}:=\sum_{\bold{i}\subset J_{\bold{k},\bold{q}} }H_{\bold{i}}+t\sum_{J_{\bold{k}_{(1)}',\bold{q}'}\subset J_{\bold{k},\bold{q}}} V^{(\bold{k},\bold{q})_{-1}}_{J_{\bold{k}_{(1)}',\bold{q}'}}+\dots+t\sum_{J_{\bold{k}'_{(|\bold{k}|-1)}},\bold{q}' \subset J_{\bold{k},\bold{q}}}V^{(\bold{k},\bold{q})_{-1}}_{J_{\bold{k}'_{(|\bold{k}|-1)},\bold{q}'}}.
\end{equation}
Note that due to Remark \ref{remark-decomp} and the above Property 2, $G_{J_{\bold{k},\bold{q}}}$ is block-diagonal w.r.t. the decomposition of  the identity 
\begin{equation}
P^{(+)}_{J_{\bold{k},\bold{q}}}+P^{(-)}_{J_{\bold{k},\bold{q}}}\,,
\end{equation} i.e., 
\begin{equation}
G_{J_{\bold{k},\bold{q}}}=P^{(+)}_{J_{\bold{k},\bold{q}}}G_{J_{\bold{k},\bold{q}}}P^{(+)}_{J_{\bold{k},\bold{q}}}+P^{(-)}_{J_{\bold{k},\bold{q}}}G_{J_{\bold{k},\bold{q}}}P^{(-)}_{J_{\bold{k},\bold{q}}}\,.
\end{equation}
We also define
\begin{equation} \label{def-E-bis}
E_{J_{\bold{k},\bold{q}}}:=\langle \bigotimes_{\bold{j}\in J_{\bold{k},\bold{q}}} \Omega_{\bold{j}}\,,\, G_{J_{\bold{k},\bold{q}}}  \bigotimes_{\bold{j}\in J_{\bold{k},\bold{q}}} \Omega_{\bold{j}} \rangle \,,
\end{equation}
hence $$G_{J_{\bold{k},\bold{q}}}P^{(-)}_{J_{\bold{k},\bold{q}}}=E_{J_{\bold{k},\bold{q}}}P^{(-)}_{J_{\bold{k},\bold{q}}}\,.$$

Assuming the above properties 2. and 3. (at step $(\bold{k},\bold{q})_{-1}$), it is not difficult to show (see Section \ref{sec:gap}) that $G_{J_{\bold{k},\bold{q}}}$ has a spectral gap $\Delta_{J_{\bold{k},\bold{q}}}$ bounded from below by $\frac{1}{2}$. Clearly, due to Property 1., the spectral properties of our original Hamiltonian $K_{\Lambda_N^d}=:K_{\Lambda_N^d}^{(\bold{0},\bold{N})}$ are the same as the spectral properties of the final Hamiltonian $K_{\Lambda_N^d}^{(\bold{N}-\bold{1},\bold{1})}$. Furthermore
$$K_{\Lambda_{N}^d}^{(\bold{N}-\bold{1},\bold{1})}=G_{J_{\bold{N}-\bold{1},\bold{1}}}+t\,V_{J_{\bold{N}-\bold{1},\bold{1}}}^{(\bold{N}-\bold{1},\bold{1})}$$
i.e., $K_{\Lambda_{N}^d}^{(\bold{N}-\bold{1},\bold{1})}$ agrees with $G_{J_{\bold{N}-\bold{1},\bold{1}}}$ up to a single "small" (in the sense of Property 3.) effective interaction potential which is block-diagonal with respect to $P^{(+)}_{J_{\bold{N}-\bold{1},\bold{1}}}$ and $P^{(-)}_{J_{\bold{N}-\bold{1},\bold{1}}}$. The problem of estimating the spectral gap of the original Hamiltonian $K_{\Lambda_N^d}$ is thus easily redirected to the estimate of the spectral gap for $G_{J_{\bold{N}-\bold{1},\bold{1}}}$, which is obtained by induction on the steps of the algorithm $(\bold{k},\bold{q})$.

\subsection{Definition and Consistency of the Algorithm}\label{defialg}

Here  we define the effective interaction potentials $$V^{(\bold{k},\bold{q})}_{J_{\bold{l},\bold{i}}}$$ iteratively in terms of the analogous terms at the previous step $(\bold{k},\bold{q})_{-1}$. The recursive definition of the effective interaction potentials is inspired by the so-called Lie-Schwinger block-diagonalization method (see \cite{DFFR}), applied at step $(\bold{k},\bold{q})$ to the Hamiltonian $G_{J_{\bold{k},\bold{q}}}+t\,V^{(\bold{k},\bold{q})_{-1}}_{J_{\bold{k},\bold{q}}}$, where $G_{J_{\bold{k},\bold{q}}}$ is treated as the unperturbed Hamiltonian and $t\,V^{(\bold{k},\bold{q})_{-1}}_{J_{\bold{k},\bold{q}}}$ as the small perturbation. 
For $B$ a symmetric operator with domain $D(B)$ and $A$ bounded operator such that $AD(B)\subseteq D(B)$, we use the following notation
\begin{equation}
ad\, A\,(B):=[A\,,\,B]\,
\end{equation}
and, for $n\geq 2$,
\begin{equation}
ad^n A\,(B):=[A\,,\,ad^{n-1} A\,(B)]\,.
\end{equation}

The reader is warned that the following definition may at first seem formal, but can be indeed rigorously shown to be well-posed - see Remark \ref{well-defi-alg}.

\begin{defn}\label{def-interactions-multi}
Setting
\begin{equation}
V_{J_{\bold{0},\bold{i}}}^{(\bold{0},\bold{N})}= H_{\bold{i}}\quad ,\quad
V_{J_{\bold{1}_j,\bold{q}}}^{(\bold{0},\bold{N})}= V_{J_{\bold{1}_j,\bold{q}}}\quad,\quad
V_{J_{\bold{k},\bold{i}}}^{(\bold{0},\bold{N})} =0\,\,\text{for}\,\, |\bold{k}|\geq 2\,,
\end{equation}

we define $V^{(\bold{k},\bold{q})}_{J_{\bold{l},\bold{i}}}$ as the symmetric operator with domain $D((H^0_{J_{\bold{l},\bold{i}}})^{1/2})$:
\begin{itemize}
\item[a)]
if  $J_{\bold{k},\bold{q}} \nsubset J_{\bold{l},\bold{i}}$, 
\begin{equation}
V^{(\bold{k},\bold{q})}_{J_{\bold{l},\bold{i}}}:=V^{(\bold{k},\bold{q})_{-1}}_{J_{\bold{l},\bold{i}}}\;
\end{equation}
\item[b)]
if $J_{\bold{l},\bold{i}}= J_{\bold{k},\bold{q}}$,
\begin{equation}
V^{(\bold{k},\bold{q})}_{J_{\bold{l},\bold{i}}}:= \sum_{j=1}^{\infty}t^{j-1}(V^{(\bold{k},\bold{q})_{-1}}_{J_{\bold{l},\bold{i}}})^{diag}_j \, ,
\end{equation}
where $(V^{(\bold{k},\bold{q})_{-1}}_{J_{\bold{l},\bold{i}}})_j$ is defined in (\ref{formula-v_j}) and \emph{diag} means the diagonal part w.r.t. the decomposition of the identity into 
\begin{equation}\label{decomp-id}
P^{(+)}_{J_{\bold{k},\bold{q}}}+P^{(-)}_{J_{\bold{k},\bold{q}}}\,;
\end{equation}
\item[c)]
if $J_{\bold{k},\bold{q}}\subset J_{\bold{l},\bold{i}}$,
\begin{eqnarray}\label{construction-conn}
V^{(\bold{k},\bold{q})}_{J_{\bold{l},\bold{i}}} &:= &V^{(\bold{k},\bold{q})_{-1}}_{J_{\bold{l},\bold{i}}}\,+\sum_{J_{\bold{k}',\bold{q}'}\in \mathcal{G}^{(\bold{k},\bold{q})}_{J_{\bold{l},\bold{i}}}}\,\sum_{n=1}^{\infty}\frac{1}{n!}\,ad^{n}S_{J_{\bold{k},\bold{q}}}(V^{(\bold{k},\bold{q})_{-1}}_{J_{\bold{k}',\bold{q}'}})\, \label{main-def-V-multi} 
\end{eqnarray}
(Note that $\mathcal{G}^{(\bold{k},\bold{q})}_{J_{\bold{l},\bold{i}}}$ is not empty only if the rectangle $J_{\bold{k},\bold{q}}$ has at least one vertex contained in $J_{\bold{l},\bold{i}}$);
\end{itemize}
where $S_{J_{\bold{k},\bold{q}}}$ is the bounded operator defined recursively as
\begin{equation}\label{formula-S}
S_{J_{\bold{k},\bold{q}}}:=\sum_{j=1}^{\infty}t^j(S_{J_{\bold{k},\bold{q}}})_j\,
\end{equation}
with
\begin{itemize}
\item
\begin{equation}\label{S-def}
(S_{J_{\bold{k},\bold{q}}})_j:=\frac{1}{G_{J_{\bold{k},\bold{q}}}-E_{J_{\bold{k},\bold{q}}}}P^{(+)}_{J_{\bold{k},\bold{q}}}\,(V^{(\bold{k},\bold{q})_{-1}}_{J_{\bold{k},\bold{q}}})_j\,P^{(-)}_{J_{\bold{k},\bold{q}}}-h.c.\,,
\end{equation}
where $G_{J_{\bold{k},\bold{q}}}$ and $E_{J_{\bold{k},\bold{q}}}$ are defined in (\ref{def-G}) and (\ref{def-E-bis}) respectively,
\item 
$(V^{(\bold{k},\bold{q})_{-1}}_{J_{\bold{k},\bold{q}}})_1=V^{(\bold{k},\bold{q})_{-1}}_{J_{\bold{k},\bold{q}}}$ and, for $j\geq 2$,
\begin{eqnarray}\label{formula-v_j}
& &(V^{(\bold{k},\bold{q})_{-1}}_{J_{\bold{k},\bold{q}}})_j\\
&:= &\sum_{p\geq 2, r_1\geq 1 \dots, r_p\geq 1\, ; \, r_1+\dots+r_p=j}\frac{1}{p!}\text{ad}\,(S_{J_{\bold{k},\bold{q}}})_{r_1}\Big(\text{ad}\,(S_{J_{\bold{k},\bold{q}}})_{r_2}\dots (\text{ad}\,(S_{J_{\bold{k},\bold{q}}})_{r_p}(G_{J_{\bold{k},\bold{q}}}))\dots \Big)\\
& &+\sum_{p\geq 1, r_1\geq 1 \dots, r_p\geq 1\, ; \, r_1+\dots+r_p=j-1}\frac{1}{p!}\text{ad}\,(S_{J_{\bold{k},\bold{q}}})_{r_1}\Big(\text{ad}\,(S_{J_{\bold{k},\bold{q}}})_{r_2}\dots (\text{ad}\,(S_{J_{\bold{k},\bold{q}}})_{r_p}(V^{(\bold{k},\bold{q})_{-1}}_{J_{\bold{k},\bold{q}}}))\dots \Big)\,.\quad\quad\quad\quad
\end{eqnarray}
\end{itemize}

\end{defn}
\begin{rem}\label{decompdefi}
Note that, for any  $(\bold{k},\bold{q})$ such that $(\bold{k},\bold{q})\succeq (\bold{l},\bold{i})$ and for any $(\bold{l}^\prime,\bold{i}^\prime)$ with $J_{\bold{l}^\prime,\bold{i}^\prime}\supseteq J_{\bold{l},\bold{i}}$, the interaction potential $V^{(\bold{k},\bold{q})}_{J_{\bold{l},\bold{i}}}$ is block-diagonal with respect to the projections $P^{(+)}_{J_{\bold{l}^\prime,\bold{i}^\prime}}$ and $P^{(-)}_{J_{\bold{l}^\prime,\bold{i}^\prime}}$ by points $(b)$ and $(a)$ of Definition \ref{def-interactions-multi} and by Remark \ref{remark-decomp}.
\end{rem}
\begin{rem}\label{well-defi-alg}
Showing that the effective interaction potentials $V^{(\bold{k},\bold{q})}_{J_{\bold{l},\bold{i}}}$ are well-defined symmetric operators on $D((H^0_{J_{\bold{l},\bold{i}}})^{1/2})$ and that the $S_{J_{\bold{l},\bold{i}}}$ as in Definition \ref{def-interactions-multi} are well defined bounded operators requires a great deal of effort and is proven inductively combining the main technical results of the present paper. The scheme that is used to verify this is analogous to \cite{DFPR1}.  Precisely, it follows from a recursive use of Theorem \ref{th-norms} and Lemma \ref{control-LS}, starting with our hypothesis on the initial Hamiltonian $K_{\Lambda_N^d}$, i.e., (\ref{klmn-cond}), observing also that  $S_{J_{\bold{k},\bold{q}}}D((H^0_{J_{\bold{l},\bold{i}}})^{1/2})\subseteq D((H^0_{J_{\bold{l},\bold{i}}})^{1/2})$ as {\color{blue}it} is shown in the following Lemma.
\end{rem}

\begin{lem}
Under the same hypothesis of Lemma \ref{control-LS}, the domain  $D((H^0_{J_{\bold{l},\bold{i}}})^{\frac{1}{2}})$ is invariant under  $S_{J_{\bold{k},\bold{q}}}$.
\end{lem}
\emph{Proof}
For any $\varphi \in D((H^0_{J_{\bold{l},\bold{i}}})^{\frac{1}{2}})$ we claim that
\begin{eqnarray}
& &\|(H^0_{J_{\bold{l},\bold{i}}})^{\frac{1}{2}}\,S_{J_{\bold{k},\bold{q}}}\,\varphi\| \label{domain-in}\\
&=&\|(H^0_{J_{\bold{l},\bold{i}}})^{\frac{1}{2}}\,\frac{1}{(H^0_{J_{\bold{l},\bold{i}}\setminus J_{\bold{k},\bold{q}}}+1)^{\frac{1}{2}}}(H^0_{J_{\bold{l},\bold{i}}\setminus J_{\bold{k},\bold{q}}}+1)^{\frac{1}{2}}S_{J_{\bold{k},\bold{q}}}\, \varphi\| \\
&=&\|(H^0_{J_{\bold{l},\bold{i}}})^{\frac{1}{2}}\,\frac{1}{(H^0_{J_{\bold{l},\bold{i}}\setminus J_{\bold{k},\bold{q}}}+1)^{\frac{1}{2}}}  S_{J_{\bold{k},\bold{q}}} \,(H^0_{J_{\bold{l},\bold{i}}\setminus J_{\bold{k},\bold{q}}}+1)^{\frac{1}{2}}\varphi\|\\
&\leq &\|\frac{(H^0_{J_{\bold{l},\bold{i}}})^{\frac{1}{2}}}{(H^0_{J_{\bold{l},\bold{i}}\setminus J_{\bold{k},\bold{q}}}+1)^{\frac{1}{2}}(H^0_{ J_{\bold{k},\bold{q}}}+1)^{\frac{1}{2}}}\|\,\|(H^0_{J_{\bold{k},\bold{q}}}+1)^{\frac{1}{2}}S_{J_{\bold{k},\bold{q}}}\|\,\|(H^0_{J_{\bold{l},\bold{i}}\setminus J_{\bold{k},\bold{q}}}+1)^{\frac{1}{2}}\varphi \| \\
&\leq &C_{\varphi}\,, \label{domain-fin}
\end{eqnarray}
for some constant $C_{\varphi}$ depending on $\varphi$, where we have exploited  estimate (\ref{S-Hest}) in Lemma \ref{control-LS}, the spectral theorem for commuting self-adjoint operators, and the assumption that $\varphi \in D((H^0_{J_{\bold{l},\bold{i}}})^{\frac{1}{2}})$.\quad\quad \qed

Theorem \ref{th-potentials} stated below shows how the effective Hamiltonian $K_{\Lambda_N^d}^{(\bold{k},\bold{q})}$ (\ref{effectivehamil}) on step $(\bold{k},\bold{q})$ arising from the effective potentials in Definition \ref{def-interactions-multi} is related to the effective Hamiltonian at the previous step $K_{\Lambda^d_N}^{(\bold{k},\bold{q})_{-1}}$. Its proof is done following the rationale of Theorem 4.3 in \cite{DFPR1}, and thus we have omitted it.

\begin{rem}\label{self-adjointness}
For $t\geq0$ small enough (and not dependent on $(\bold{k},\bold{q})$ or $N$), from Theorem \ref{th-norms}
and using the bound
\begin{equation}
\sum_{J_{\bold{k},\bold{i}} \subset J_{\bold{N}-\bold{1},\bold{1}}}H_{J_{\bold{k},\bold{i}}}^0\leq \Big\{\prod_{j=1}^{d}(k_j+1)\Big\}\, \sum_{\bold{i}\in \Lambda^{d}_N} H_{\bold{i}}\,,
\end{equation}
the Hamiltonian $K_{\Lambda_N^d}^{(\bold{k},\bold{q})}$  is seen to be defined as a symmetric operator that is bounded from below on $D(H^0_{J_{\bold{k},\bold{q}}})$. Thus it has
 a self-adjoint extension  (again denoted by $K_{\Lambda_N^d}^{(\bold{k},\bold{q})}$ ) with domain $D(K_{\Lambda_N^d}^{(\bold{k},\bold{q})})$ with $D((H^0_{J_{\bold{N}-\bold{1},\bold{1}}}))\subseteq D(K_{\Lambda_N^d}^{(\bold{k},\bold{q})}) \subseteq D((H^0_{J_{\bold{N}-\bold{1},\bold{1}}})^{\frac{1}{2}})$. The same procedure can be applied to the symmetric quadratic form $G_{J_{\bold{k},\bold{q}}}$ in (\ref{def-G}) to obtain a self-adjoint operator (also denoted $G_{J_{\bold{k},\bold{q}}}$) with domain $D(G_{J_{\bold{k},\bold{q}}})$ such that  $D((H^0_{J_{\bold{k},\bold{q}}})) \subseteq D(G_{J_{\bold{k},\bold{q}}})\subseteq D((H^0_{J_{\bold{k},\bold{q}}})^{\frac{1}{2}})$.
\end{rem}

\begin{thm}\label{th-potentials}
There exists $t_d>0$ such that for every $t\in\mathbb{R}$ with $|t|<t_d$,
the operator $K_{\Lambda_N^d}^{(\bold{k},\bold{q})}= K_{\Lambda_N^d}^{(\bold{k},\bold{q})}(t)$
is self-adjoint on the domain $e^{S_{J_{\bold{k},\bold{q}}}}D(K_N^{(\bold{k},\bold{q})_{-1}})$ and coincides with $e^{S_{J_{\bold{k},\bold{q}}}}\,K_N^{(\bold{k},\bold{q})_{-1}}\,e^{-S_{J_{\bold{k},\bold{q}}}}$. 
\end{thm}

Thus Theorem \ref{th-potentials} tells us that the spectral properties of the fully block-diagonalized (with respect to $P^{(+)}_{J_{\bold{N}-\bold{1},\bold{1}}}$ and $P^{(-)}_{J_{\bold{N}-\bold{1},\bold{1}}}$) final Hamiltonian resulting from the local Lie-Schwinger block-diagonalization algorithm (i.e., the self-adjoint operator $K_{\Lambda_N^d}^{(\bold{N}-\bold{1},\bold{1})}$) are equivalent to  those of our original Hamiltonian $K_{\Lambda_N^d}$.

\subsection{Gap of the local Hamiltonians $G_{J_{\bold{k},\bold{q}}}$: Main argument}\label{sec:gap}

In this section we show explicitly how the local Lie-Schwinger block-diagonalization algorithm is tailored to provide spectral gap estimates, in particular how to obtain an estimate of the spectral gap of the Hamiltonians $G_{J_{\bold{k},\bold{q}}}$. 


%
Similarly to the one dimensional case, see \cite{DFPR1}, and to the bounded $d$-dimensional case \cite{DFPR3}, it is not difficult to prove that, under the assumption
\begin{equation}\label{ass-2-multi}
\|(H_{J_{\bold{l},\bold{i}}}^0+1)^{-\frac{1}{2}}V^{{(\bold{k},\bold{q})}_{-1}}_{J_{\bold{l},\bold{i}}}(H_{J_{\bold{l},\bold{i}}}^0+1)^{-\frac{1}{2}}\|=:\|V^{(\bold{k},\bold{q})_{-1}}_{J_{\bold{l},\bold{i}}}\|_{H^0} \leq t^{\frac{l-1}{4}}\,\quad,\quad l=|\bold{l}|:=l_1+l_2+\cdots+l_d\,,
\end{equation} 
for $(0\leq)t< t_d$ where $t_d$ depends on the lattice dimension but is independent of  $(\bold{k},\bold{q})$, and $N$ (see Theorem \ref{th-norms}), the Hamiltonian $G_{J_{\bold{k}, \bold{q}}}$ (\ref{def-G}) has a spectral gap $\Delta_{J_{\bold{k}, \bold{q}}} \geq \frac{1}{2}$ in the interval $[0, t_d)$.

%

\noindent
Thus, for this section, we set our inductive hypothesis to be
\begin{equation}\label{ass-2}
\|V^{{(\bold{k},\bold{q})}_{-1}}_{J_{\bold{l},\bold{i}}}\|_{H^0} \leq t^{\frac{l-1}{4}}\,.
\end{equation} 
(In Theorem \ref{th-norms},  starting from the potential terms $V_{J_{\bold{1}_j,\bold{q}}}^{(\bold{0},\bold{N})}= V_{J_{\bold{1}_j,\bold{q}}}$, (\ref{ass-2}) is established by induction.) According to the scheme described in Definition \ref{def-interactions-multi} (see Remark \ref{decompdefi}), for any  $k>l$,  $V^{{(\bold{k},\bold{q})}_{-1}}_{J_{\bold{l},\bold{i}}}$ is block-diagonalized, i.e., 
\begin{equation}\label{informal-in}
V^{{(\bold{k},\bold{q})}_{-1}}_{J_{\bold{l},\bold{i}}}=P^{(+)}_{J_{\bold{l},\bold{i}}}V^{{{(\bold{k},\bold{q})}_{-1}}}_{J_{\bold{l},\bold{i}}}P^{(+)}_{J_{\bold{l},\bold{i}}}+P^{(-)}_{J_{\bold{l},\bold{i}}}V^{{(\bold{k},\bold{q})}_{-1}}_{J_{\bold{l},\bold{i}}}P^{(-)}_{J_{\bold{l},\bold{i}}}\,.
\end{equation}
Hence  we can  write
\begin{eqnarray}
& &P^{(+)}_{J_{\bold{k},\bold{q}}}\,\Big[\sum_{i\in J_{\bold{k},\bold{q}} }H_i+t\sum_{J_{{\bold{l}_{(l)}},\bold{i}} \subset J_{\bold{k},\bold{q}}} V^{{(\bold{k},\bold{q})}_{-1}}_{J_{\bold{l},\bold{i}}}\Big]P^{(+)}_{J_{\bold{k},\bold{q}}}\\
&=&P^{(+)}_{J_{\bold{k},\bold{q}}}\,\Big[\sum_{i\in J_{\bold{k},\bold{q}} }H_i+t\sum_{J_{{\bold{l}_{(l)}},\bold{i}} \subset J_{\bold{k},\bold{q}}} P^{(+)}_{J_{\bold{l},\bold{i}}}V^{{(\bold{k},\bold{q})}_{-1}}_{J_{\bold{l},\bold{i}}}P^{(+)}_{J_{\bold{l},\bold{i}}}+t\sum_{J_{{\bold{l}_{(l)}},\bold{i}} \subset J_{\bold{k},\bold{q}}} P^{(-)}_{J_{\bold{l},\bold{i}}}V^{{(\bold{k},\bold{q})}_{-1}}_{J_{\bold{l},\bold{i}}}P^{(-)}_{J_{\bold{l},\bold{i}}}\Big]P^{(+)}_{J_{\bold{k},\bold{q}}}\,.\quad\quad\quad \label{1.57}
\end{eqnarray}
For $\psi \in D((H_{J_{\bold{r},\bold{i}}}^0)^{\frac{1}{2}})$ we estimate
\begin{eqnarray}
& &|\langle \psi\,,\,P^{(+)}_{J_{\bold{r},\bold{i}}}V^{{(\bold{k},\bold{q})}_{-1}}_{J_{\bold{r},\bold{i}}}P^{(+)}_{J_{\bold{r},\bold{i}}} \psi \rangle| \label{in-est-V}\\
&= &|\langle \psi\,,\,P^{(+)}_{J_{\bold{r},\bold{i}}}\,(H_{J_{\bold{r},\bold{i}}}^0)^{\frac{1}{2}}(\frac{H_{J_{\bold{r},\bold{i}}}^0+1}{H_{J_{\bold{r},\bold{i}}}^0})^{\frac{1}{2}}(H_{J_{\bold{r},\bold{i}}}^0+1)^{-\frac{1}{2}}V^{{(\bold{k},\bold{q})}_{-1}}_{J_{\bold{r},\bold{i}}}(H_{J_{\bold{r},\bold{i}}}^0+1)^{-\frac{1}{2}}(\frac{H_{J_{\bold{r},\bold{i}}}^0+1}{H_{J_{\bold{r},\bold{i}}}^0})^{\frac{1}{2}}(H_{J_{\bold{r},\bold{i}}}^0)^{\frac{1}{2}}\,P^{(+)}_{J_{\bold{r};\bold{i}}}\psi \rangle| \quad\quad \quad \\
&\leq  &2\cdot t^{\frac{r-1}{4}}\,\langle \psi\,,\,P^{(+)}_{J_{\bold{r},\bold{i}}}\,H_{J_{\bold{r},\bold{i}}}^0\,P^{(+)}_{J_{\bold{r},\bold{i}}}\psi \rangle\\
&\leq &2\cdot t^{\frac{r-1}{4}}\,\langle \psi\,,\,H_{J_{\bold{r},\bold{i}}}^0\psi \rangle \label{fin-est-V}
\end{eqnarray}
where we have used the assumption in (\ref{ass-2}) and
\begin{equation}
\|P^{(+)}_{J_{\bold{r},\bold{i}}}\,(\frac{H_{J_{\bold{r},\bold{i}}}^0+1}{H_{J_{\bold{r},\bold{i}}}^0})^{\frac{1}{2}}\|\leq \sqrt{2},
\end{equation}
which follows from (\ref{gaps}).

\noindent
Next we observe that, by Remark \ref{shapes},
\begin{equation}\label{ineq-inter-00}
\sum_{J_{{\bold{l}_{(l)}},\bold{i}}\subset J_{\bold{k},\bold{q}}}H^0_{J_{\bold{l},\bold{i}}}\leq (l+1)^{2d-1} H^0_{J_{\bold{k},\bold{q}}} \,.
\end{equation}
Thus  we find that
\begin{equation}\label{ineq-inter-0}
\sum_{J_{{\bold{l}_{(l)}},\bold{i}}\subset J_{\bold{k},\bold{q}}}P^{(+)}_{J_{\bold{r},\bold{i}}}\leq  (l+1)^{2d-1}\sum_{\bold{j}\in  J_{\bold{k},\bold{q}}}\charf_{ \Lambda^{(d)}_N \setminus \bold{j}}\otimes P^{\perp}_{\Omega_{\bold{j}}}\leq (l+1)^{2d-1} H^0_{J_{\bold{k},\bold{q}}}\,
\end{equation}
using the inequality proven  in Corollary \ref{op-ineq-2} combined with (\ref{gaps}).
Due to the estimate in (\ref{in-est-V})-(\ref{fin-est-V}),  and using inequality (\ref{ineq-inter-00}),  we have that
\begin{equation}\label{ineq-cor}
\pm \sum_{J_{{\bold{l}_{(l)}},\bold{i}} \subset J_{\bold{k},\bold{q}}} P^{(+)}_{J_{\bold{l},\bold{i}}}V^{{(\bold{k},\bold{q})}_{-1}}_{J_{\bold{l},\bold{i}}}P^{(+)}_{J_{\bold{l},\bold{i}}}\leq 2\cdot t^{\frac{l-1}{4}}\, (l+1)^{2d-1} H^0_{J_{\bold{k},\bold{q}}}\,.
\end{equation}
Hence, recalling that $t>0$ and combining (\ref{ass-2}) with (\ref{ineq-cor}), we conclude that
\begin{eqnarray}
(\ref{1.57})
&\geq &  P^{(+)}_{J_{\bold{k},\bold{q}}}\,\Big[(1- 2 t\cdot t^{\frac{l-1}{4}}\, (l+1)^{2d-1})\,H^0_{J_{\bold{k},\bold{q}}}\Big]P^{(+)}_{J_{\bold{k},\bold{q}}} \\
& &+P^{(+)}_{J_{\bold{k},\bold{q}}}\,\Big[t\sum_{J_{{\bold{l}_{(l)}},\bold{i}} \subset J_{\bold{k},\bold{q}}}P^{(-)}_{J_{\bold{l},\bold{i}}} V^{{(\bold{k},\bold{q})}_{-1}}_{J_{\bold{l},\bold{i}}}P^{(-)}_{J_{\bold{l},\bold{i}}}\Big]P^{(+)}_{J_{\bold{k},\bold{q}}}\label{second-line}\\
&=& P^{(+)}_{J_{\bold{k},\bold{q}}}\,\Big[(1-2 t\cdot t^{\frac{l-1}{4}}\, (l+1)^{2d-1})\,H^0_{J_{\bold{k},\bold{q}}}\Big]P^{(+)}_{J_{\bold{k},\bold{q}}}\\
& &+P^{(+)}_{J_{\bold{k},\bold{q}}}\,\Big[t\sum_{J_{{\bold{l}_{(l)}},\bold{i}} \subset J_{\bold{k},\bold{q}}}\langle  V^{{(\bold{k},\bold{q})}_{-1}}_{J_{\bold{l},\bold{i}}} \rangle P^{(-)}_{J_{\bold{l},\bold{i}}}\Big]P^{(+)}_{J_{\bold{k},\bold{q}}}\,,\label{third-line}\label{informal-fin}
\end{eqnarray}
where 
\begin{equation} 
 \langle  V^{{(\bold{k},\bold{q})}_{-1}}_{J_{\bold{l},\bold{i}}} \rangle:=\langle \bigotimes_{\bold{j}\in J_{\bold{l},\bold{i}}} \Omega_{\bold{j}}\,,\, V^{{(\bold{k},\bold{q})}_{-1}}_{J_{\bold{l},\bold{i}}} \bigotimes_{\bold{j}\in J_{\bold{l},\bold{i}}} \Omega_{\bold{j}} \rangle \,.
\end{equation}

We are ready for  the main  lemma of this section.
\begin{lem}\label{gap}
Assuming the bound in (\ref{ass-2}), and choosing $t>0$ such that
\begin{equation}
1-3 t \sum_{l=1}^{\infty}  t^{\frac{l-1}{4}}\, (l+1)^{2d-1}>0\,,
\end{equation}
the inequality
\begin{equation}
P^{(+)}_{J_{\bold{k},\bold{q}}}(G_{J_{\bold{k},\bold{q}}}-E_{J_{\bold{k},\bold{q}}})P^{(+)}_{J_{\bold{k},\bold{q}}}
\geq\Big(1-3 t \sum_{l=1}^{\infty}  t^{\frac{l-1}{4}}\, (l+1)^{2d-1}\Big)H^{0}_{J_{\bold{k},\bold{q}}}\,\,P^{(+)}_{J_{\bold{k},\bold{q}}} \label{final-eq-1}
\end{equation}
holds in the sense of quadratic forms on the domain $D((H_{J_{\bold{k},\bold{q}}}^0)^{\frac{1}{2}})$, where $E_{J_{\bold{k},\bold{q}}}$ is defined in (\ref{def-E-bis}).
\end{lem}

\noindent
\emph{Proof}
Proceeding as in (\ref{informal-in})-(\ref{informal-fin}), 
we have
\begin{eqnarray}
& &P^{(+)}_{J_{\bold{k};\bold{q}}}G_{J_{\bold{k},\bold{q}}}P^{(+)}_{J_{\bold{k},\bold{q}}} \\
&\geq &P^{(+)}_{J_{\bold{k},\bold{q}}}\,\Big\{\Big[1-2 t \sum_{l=1}^{k-1}  t^{\frac{l-1}{4}}\, (l+1)^{2d-1})\Big]\,H^0_{J_{\bold{k},\bold{q}}}\Big\}\,P^{(+)}_{J_{\bold{k},\bold{q}}} \quad\quad\quad\quad \nonumber \\
& &+ P^{(+)}_{J_{\bold{k},\bold{q}}}\,\Big[t\sum_{J_{{\bold{l}_{(1)}},\bold{i}} \subset J_{\bold{k},\bold{q}}}\langle  V^{{{(\bold{k},\bold{q})}_{-1}}}_{J_{\bold{l},\bold{i}}} \rangle  P^{(-)}_{J_{\bold{l},\bold{i}}}+\dots+t\sum_{J_{{\bold{l}_{(k-1)}},\bold{i}}\subset  J_{\bold{k},\bold{q}}}\langle V^{{(\bold{k},\bold{q})}_{-1}}_{J_{\bold{l},\bold{i}}}\rangle P^{(-)}_{J_{\bold{l},\bold{i}}} \Big]P^{(+)}_{J_{\bold{k},\bold{q}}} \,.\quad\quad\quad 
\end{eqnarray}
Next, using $P^{(-)}_{J_{\bold{l},\bold{i}}}+P^{(+)}_{J_{\bold{l},\bold{i}}}=\charf $, 
\begin{eqnarray}
& &P^{(+)}_{J_{\bold{k};\bold{q}}}G_{J_{\bold{k},\bold{q}}}P^{(+)}_{J_{\bold{k},\bold{q}}} \label{gap-1} \\
&\geq &P^{(+)}_{J_{\bold{k},\bold{q}}}\,\Big[\Big[1-2 t \sum_{l=1}^{k-1}  t^{\frac{l-1}{4}}\, (l+1)^{2d-1})\Big]\,H^0_{J_{\bold{k},\bold{q}}}\Big]\,P^{(+)}_{J_{\bold{k},\bold{q}}} \quad\quad\quad\quad \nonumber \\
& &+ P^{(+)}_{J_{\bold{k},\bold{q}}}\,\Big[-t\sum_{J_{{\bold{l}_{(1)}},\bold{i}} \subset J_{\bold{k},\bold{q}}}\langle  V^{{{(\bold{k},\bold{q})}_{-1}}}_{J_{\bold{l},\bold{i}}} \rangle  P^{(+)}_{J_{\bold{l},\bold{i}}}+\dots-t\sum_{J_{{\bold{l}_{(k-1)}},\bold{i}}\subset  J_{\bold{k},\bold{q}}}\langle V^{{(\bold{k},\bold{q})}_{-1}}_{J_{\bold{l},\bold{i}}}\rangle P^{(+)}_{J_{\bold{l},\bold{i}}} \Big]P^{(+)}_{J_{\bold{k},\bold{q}}}\\
&&+ P^{(+)}_{J_{\bold{k},\bold{q}}}\,\Big[t\sum_{J_{{\bold{l}_{(1)}},\bold{i}} \subset J_{\bold{k},\bold{q}}}\langle  V^{{{(\bold{k},\bold{q})}_{-1}}}_{J_{\bold{l},\bold{i}}} \rangle  +\dots+t\sum_{J_{{\bold{l}_{(k-1)}},\bold{i}}\subset  J_{\bold{k},\bold{q}}}\langle V^{{(\bold{k},\bold{q})}_{-1}}_{J_{\bold{l},\bold{i}}}\rangle \Big]P^{(+)}_{J_{\bold{k},\bold{q}}}  \,.\quad\quad\quad 
\end{eqnarray}
Finally, using (\ref{ineq-inter-0}) and $|\langle V^{{(\bold{k},\bold{q})}_{-1}}_{J_{\bold{l},\bold{i}}}\rangle|\leq \|V^{{(\bold{k},\bold{q})}_{-1}}_{J_{\bold{l},\bold{i}}}\|_{H^0}$,\begin{eqnarray}
& &P^{(+)}_{J_{\bold{k};\bold{q}}}G_{J_{\bold{k},\bold{q}}}P^{(+)}_{J_{\bold{k},\bold{q}}} \\
&\geq &P^{(+)}_{J_{\bold{k},\bold{q}}}\,\Big\{\Big[1-3 t \sum_{l=1}^{k-1}  t^{\frac{l-1}{4}}\, (l+1)^{2d-1})\Big]\,H^0_{J_{\bold{k},\bold{q}}}\Big\}\,P^{(+)}_{J_{\bold{k},\bold{q}}} \quad\quad\quad\quad \nonumber \\
&&+ P^{(+)}_{J_{\bold{k},\bold{q}}}\,\Big[t\sum_{J_{{\bold{l}_{(1)}},\bold{i}} \subset J_{\bold{k},\bold{q}}}\langle  V^{{{(\bold{k},\bold{q})}_{-1}}}_{J_{\bold{l},\bold{i}}} \rangle  +\dots+t\sum_{J_{{\bold{l}_{(k-1)}},\bold{i}}\subset  J_{\bold{k},\bold{q}}}\langle V^{{(\bold{k},\bold{q})}_{-1}}_{J_{\bold{l},\bold{i}}}\rangle \Big]P^{(+)}_{J_{\bold{k},\bold{q}}}\\
&=&P^{(+)}_{J_{\bold{k},\bold{q}}}\,\Big\{\Big[1-3 t \sum_{l=1}^{k-1}  t^{\frac{l-1}{4}}\, (l+1)^{2d-1})\Big]\,H^0_{J_{\bold{k},\bold{q}}}\Big\}\,P^{(+)}_{J_{\bold{k},\bold{q}}}+E_{J_{\bold{k},\bold{q}}}P^{(+)}_{J_{\bold{k},\bold{q}}}\, ,\quad\quad\quad\quad \label{gap-2}
\end{eqnarray}
where we have used the definition in (\ref{def-E-bis}) in the last step.
\qed

\noindent

Lemma \ref{gap} implies that under the assumption in (\ref{ass-2-multi}) the Hamiltonian $G_{J_{\bold{k},\bold{q}}}$ has a gap that can be estimated from below by $\frac{1}{2}$ for $t>0$ sufficiently small but independent of $N$ and $(\bold{k},\bold{q})$. This is stated in the Corollary below.
\begin{cor}\label{cor-gap}
Assuming Lemma \ref{gap},  for $t>0$ sufficiently small, dependent on $d$ but independent of $N$ and $(\bold{k},\bold{q})$, the Hamiltonian $G_{J_{\bold{k},\bold{q}}}$ has a gap $\Delta_{J_{\bold{k},\bold{q}}}\geq \frac{1}{2}$ above the ground state energy $$E_{J_{\bold{k},\bold{q}}}=t\sum_{J_{\bold{k}_{(1)}',\bold{q}'}\subset J_{\bold{k},\bold{q}}} \langle V^{(\bold{k},\bold{q})_{-1}}_{J_{\bold{k}_{(1)}',\bold{q}'}}\rangle +\dots+t\sum_{J_{\bold{k}'_{(j_{\bold{k}-1})},\bold{q}'} \subset J_{\bold{k},\bold{q}}} \langle V^{(\bold{k},\bold{q})_{-1}}_{J_{\bold{k}'_{(j_{\bold{k}-1})},\bold{q}'}} \rangle 
$$ corresponding to the the ground state vector $\bigotimes_{\bold{i}\in J_{\bold{k},\bold{q}}}\Omega_{\bold{i}}\,$, due to the identity
\begin{eqnarray}
& &P^{(-)}_{J_{\bold{k},\bold{q}}}G_{J_{\bold{k},\bold{q}}}P^{(-)}_{J_{\bold{k},\bold{q}}}\\
&= &P^{(-)}_{J_{\bold{k},\bold{q}}}\,\Big[t\sum_{J_{\bold{k}_{(1)}',\bold{q}'}\subset J_{\bold{k},\bold{q}}} \langle V^{(\bold{k},\bold{q})_{-1}}_{J_{\bold{k}_{(1)}',\bold{q}'}}\rangle P^{(-)}_{J_{\bold{k}_{(1)}',\bold{q}'}}  +\dots+t\sum_{J_{\bold{k}'_{(|\bold{k}|-1)},\bold{q}'} \subset J_{\bold{k},\bold{q}}}\langle V^{(\bold{k},\bold{q})_{-1}}_{J_{\bold{k}'_{(|\bold{k}|-1)}},\bold{q}'}}\rangle P^{(-)}_{J_{\bold{k}_{(|\bold{k}|-1)}',\bold{q}'}} \Big] P^{(-)}_{J_{\bold{k},\bold{q}}\nonumber \\
&= &P^{(-)}_{J_{\bold{k},\bold{q}}}\,\Big[t\sum_{J_{\bold{k}_{(1)}',\bold{q}'}\subset J_{\bold{k},\bold{q}}} \langle V^{(\bold{k},\bold{q})_{-1}}_{J_{\bold{k}_{(1)}',\bold{q}'}}\rangle  +\dots+t\sum_{J_{\bold{k}'_{(|\bold{k}|-1)},\bold{q}'} \subset J_{\bold{k},\bold{q}}}\langle V^{(\bold{k},\bold{q})_{-1}}_{J_{\bold{k}'_{(|\bold{k}|-1)}},\bold{q}'}}\rangle  \Big] P^{(-)}_{J_{\bold{k},\bold{q}}\,.\nonumber 
\end{eqnarray}
\end{cor}

\section{Control of the weighted norms $\|V^{(\bold{k},\bold{q})}_{J_{\bold{r},\bold{i}}}\|_{H^0}$}\label{mainsec}

Following the scheme of \cite{DFPR3}, in Theorem \ref{th-norms} we shall prove by induction that, for every pair $(\mathbf{r},\mathbf{i})$, an upper bound 
of the form
\begin{equation}
 \|V^{(\bold{k},\bold{q})}_{J_{\bold{r},\bold{i}}}\|_{H^0}\leq C_{j} \frac{t^{\frac{r-1}{3}}}{r^{\,\rho_{j}}}\quad,\quad j=1,2,3\,, \label{strat}
\end{equation} 
holds true, at all steps $(\bold{k},\bold{q})$ up to step $(\bold{r},\bold{i})$, where $C_j$ and the exponent $\rho_{j}\equiv \rho_j(d)>0$ depend on the regime $\mathfrak{R}j$ introduced below; 
the regimes, $\mathfrak{R}1, \mathfrak{R}2$, and $\mathfrak{R}3$,  depend on 
the relative magnitude of the circumferences $k=|\bold{k}|$ and $r=|\bold{r}|$ as follows. 


\begin{itemize}
\item[$\mathfrak{R}1$)] The first regime deals with the case of rectangles labelled by $(\bold{k},\bold{q})$ that are ``small'' as compared
 to the rectangle labelled by $(\bold{r},\bold{i})$, more precisely with pairs $(\bold{k}, \bold{q})$ such that $k \leq \lfloor r^{\frac{1}{4}} \rfloor $. 

\item[$\mathfrak{R}2$)] The second regime is associated with rectangles labelled by pairs $(\bold{k},\bold{q})$ such that
$\lfloor  r^{\frac{1}{4}} \rfloor \leq k\leq r-\lfloor r^{\frac{1}{4}}\rfloor$. 
\item[$\mathfrak{R}3$)] The third regime deals with ``large'' rectangles $(\bold{k},\bold{q})$, more precisely
$ r-\lfloor r^{\frac{1}{4}}\rfloor  \leq k \leq r$. \end{itemize}

For the analysis of $\mathfrak{R}1$ and $\mathfrak{R}2$ in Theorem \ref{th-norms} we follow the same strategy of the corresponding theorem for the bounded case in \cite{DFPR3}, but some nontrivial modifications are required due to the involvement of the weighted norms. $\mathfrak{R}3$ is the case where the most effort is required compared to the bounded case in \cite{DFPR3}. We recall that in this regime, the mechanism that is used is based on \emph{large denominators}. 
This means that the contributions in (\ref{main-def-V-multi}) corresponding to potentials 
$V^{(\bold{k},\bold{q})_{-1}}_{J_{\bold{k}',\bold{q}'}}$  that are already block-diagonal are collected and then estimated in terms of a sum of projections $P^{(+)}_{J_{\bold{k}',\bold{q}'}}$ controlled,   through an induction, by 
the denominator appearing  in the expression of $(S_{J_{\bold{r}, \bold{i}}})_{1}$ (see formula (\ref{S-def})). In the proof by induction for this last regime, new auxiliary quantities displayed in (\ref{R3-1}) are used; due to the unboundedness of the interaction, the inductive control of this mechanism is more challenging compared to the bounded case. The estimates that are required for the inductive control of these auxiliary quantities are derived in Lemma \ref{inductive-aux}. \\


\begin{thm}\label{th-norms}
There exists $t_d>0$ such that for every $0\leq t < t_{d}$ and $(\bold{N}-\bold{1},\bold{1})\succeq (\bold{k},\bold{q})\succeq (\bold{0,\bold{N}})$, the Hamiltonians $G_{J_{\bold{k},\bold{q}}}$ and $K_{N^d}^{(\bold{k},\bold{q})}$ are well defined self-adjoint operators, and for any rectangle $J_{\bold{r},\bold{i}}$, with $r=|\bold{r}|\geq 1$, and for $x_d:=20d$, we have: 

\noindent
S1) 

\noindent
Let $(\bold{k},\bold{q})_* := (\bold{k}_*,\bold{q}_*)$ be defined for some $(\bold{k}_*,\bold{q}_*)$ such that $ |\bold{k}_*|=\lfloor r^{\frac{1}{4}} \rfloor$, where $\lfloor \cdot \rfloor$ is the integer part. If  $(\bold{k},\bold{q})\prec (\bold{k},\bold{q})_*$, then
 \begin{equation}\label{inductive-reg-1}
 \|V^{(\bold{k},\bold{q})}_{J_{\bold{r},\bold{i}}}\|_{H^0}\leq \frac{t^{\frac{r-1}{3}}}{r^{x_d+2d}}\,;
 \end{equation}

\noindent
 Let $ (\bold{k},\bold{q})_{**}:= (\bold{k}_{**},\bold{q}_{**})$ be defined for some $(\bold{k}_{**},\bold{q}_{**})$  such that $|\bold{k}_{**}| = r-\lfloor r^{\frac{1}{4}} \rfloor $. If $ (\bold{k},\bold{q})_{**}\succ (\bold{k},\bold{q})\succeq (\bold{k},\bold{q})_{*} $, then 
\begin{equation}
\|V^{(\bold{k},\bold{q})}_{J_{\bold{r},\bold{i}}}\|_{H^0}\leq 2 \cdot \frac{t^{\frac{r-1}{3}}}{r^{x_d+2d}}\,;\label{inductive-reg-2}
\end{equation}
 
\noindent
  If $ (\bold{r},\bold{i})\succ (\bold{k},\bold{q}) \succeq  (\bold{k},\bold{q})_{**}$, then
\begin{equation}
\| (\frac{1}{\sum_{\bold{j}\in J_{\bold{r},\bold{i}}} H_{\bold{j}}+1})^{\frac{1}{2}}(\frac{1}{\sum_{\bold{j}\in J_{\bold{r},\bold{i}}} P^{\perp}_{\Omega_{\bold{j}}}+1})^{\frac{1}{2}}\,P^{(\#)}_{J_{\bold{r},\bold{i}}}V^{(\bold{k},\bold{q})}_{J_{\bold{r},\bold{i}}}P^{(\hat{\#})}_{J_{\bold{r},\bold{i}}}\,(\frac{1}{\sum_{\bold{j}\in J_{\bold{r},\bold{i}}} P^{\perp}_{\Omega_{\bold{j}}}+1})^{\frac{1}{2}}(\frac{1}{\sum_{\bold{j}\in J_{\bold{r},\bold{i}}} H_{\bold{j}}+1})^{\frac{1}{2}}\|\leq  3\cdot\frac{t^{\frac{r-1}{3}}}{r^{x_d+2d}}\,, \quad \#, \hat{\#}=\pm \,,   \label{R3-1}
\end{equation}
and
\begin{equation}
\| V^{(\bold{k},\bold{q})}_{J_{\bold{r},\bold{i}}}\|_{H^0}\leq  48  \cdot \frac{t^{\frac{r-1}{3}}}{r^{x_d}}\,; \label{R3-2}
\end{equation}

\noindent
 If $(\bold{k},\bold{q})\succeq  (\bold{r},\bold{i})$, then
\begin{equation}
\| V^{(\bold{k},\bold{q})}_{J_{\bold{r},\bold{i}}}\|_{H^0}\leq  96  \cdot \frac{t^{\frac{r-1}{3}}}{r^{x_d}}\,.  \label{R3-3}
\end{equation}

\noindent
S2)  

\noindent
$G_{J_{(\bold{k},\bold{q})_{+1}}}$ has spectral gap  $\Delta_{J_{(\bold{k},\bold{q})_{+1}}}\geq \frac{1}{2}$ above its ground state energy,  where $G_{J_{\bold{k},\bold{q}}}$ is defined in (\ref{def-G})  for $|\bold{k}|\geq 2$,  and $$G_{J_{(\bold{1}_j,\bold{q})_{+1}}}:=H^{(0)}_{J_{(\bold{1}_j,\bold{q})_{+1}}}:=\sum_{\bold{i}\in J_{(\bold{1}_j,\bold{q})_{+1}}}H_{\bold{i}}$$  provided $(\bold{1}_j,\bold{q})_{+1}$ is of the form $(\bold{1}_{j'},\bold{q}')$ for some $j'$ and $\bold{q}'$; $ (\bold{1}_j,\bold{q})$ is defined in (\ref{def-1_j}).
\end{thm}

\noindent
\emph{Proof.}

The proof is by induction on the diagonalization step $(\bold{k},\bold{q})$. For each $(\bold{r},\bold{i})$ we will prove S1) and S2) from $(\bold{k},\bold{q})=(\bold{0},\bold{N})$ up to $(\bold{k},\bold{q})=(\bold{N-1},\bold{1})$; (note that in step $(\bold{k}, \bold{q})$  S2) concerns the Hamiltonian $G_{J_{(\bold{k},\bold{q})_{+1}}}$, and that in step $(\bold{k},\bold{q})=(\bold{N-1},\bold{1})$ it  is not defined). Namely, we assume that S1) holds for all $V^{(\bold{k}',\bold{q}')}_{J_{\bold{r},\bold{i}}}$ with $(\bold{k}',\bold{q}') \prec (\bold{k},\bold{q})$ and  S2) for all $(\bold{k}',\bold{q}') \prec (\bold{k},\bold{q})$. Then we prove that they hold true for all $V^{(\bold{k},\bold{q})}_{J_{\bold{r},\bold{i}}}$ and for $G_{J_{(\bold{k},\bold{q})_{+1}}}$.  By Lemma \ref{control-LS}, this implies that $S_{J_{(\bold{k},\bold{q})_{+1}}}$ is a well-defined bounded operator, and that $K_{N^d}^{(\bold{k},\bold{q})}$ is {\color{blue}a} well defined self-adjoint operator (see Remark \ref{self-adjointness}).

\noindent
For $(\bold{k},\bold{q})= (\bold{0},\bold{N})$,  S1) follows by direct computation,  since 
$$\|V_{J_{\bold{1}_j,\bold{q}}}^{(\bold{0},\bold{N})}\|_{H^0}=\| V_{J_{\bold{1}_j,\bold{q}}}\|_{H^0}\leq1\,,$$
and
$V_{J_{\bold{r},\bold{i}}}^{(\bold{0},\bold{N})}=0$ otherwise; S2) is trivial since, by definition, $(\bold{0},\bold{N})_{+1}=(\bold{1}_1, \bold{1})$ and $G_{J_{\bold{1}_1,\bold{1}}}=H^{(0)}_{J_{\bold{1}_1,\bold{1}}}$ (where $\bold{1}_j$ is defined in (\ref{def-1_j})).


\emph{Warning}: Many positive constants are introduced throughout the proof. We shall denote universal constants by $c,C$ and $d$-dependent constants by $c_d, C_d$. Their value may change from line to line.
\\

\noindent
\emph{Induction step in the proof of S1)}
 
 The cases where $J_{\bold{r},\bold{i}}$ is such that $r=1$ or $r=2$ can be treated exactly as the analogous cases of Theorem 5.1 in \cite{DFPR3}. Thus in the following we assume $r>2$.

\noindent
As explained at the beginning of the present section, in order to control the norm $\|V^{(\bold{k},\bold{q})}_{J_{\bold{r},\bold{i}}} \|_{H^0}$   we distinguish three regimes, $\mathfrak{R}1$, $\mathfrak{R}2$ and $\mathfrak{R}3$,  which depend on the relative magnitude between  $k=|\bold{k}|$ and $r=|\bold{r}|$. These are respectively associated with (\ref{inductive-reg-1}), (\ref{inductive-reg-2}), and (\ref{R3-1})-(\ref{R3-2})-(\ref{R3-3}). 
\noindent

We recall how the induction is used in the following analysis. Assuming that (\ref{inductive-reg-1}), (\ref{inductive-reg-2}), (\ref{R3-1}), (\ref{R3-2}), and (\ref{R3-3}) hold true for the potentials associated with rectangles $J_{\bold{l}',\bold{i}'}$ such that  $(\bold{l}', \bold{i}')\prec (\bold{r}, \bold{i})$,  in steps $(\bold{k}', \bold{q}')\prec (\bold{k}, \bold{q})$, we prove  that, depending on the considered regime, (\ref{inductive-reg-1}), (\ref{inductive-reg-2}), and  (\ref{R3-1}) hold, respectively,   in step $(\bold{k}, \bold{q})$ for the potential associated with $J_{\bold{r},\bold{i}}$; as a consequence, if (\ref{R3-1}) is verified then also (\ref{R3-2}) and (\ref{R3-3}) hold (in step $(\bold{k}, \bold{q})$)\,.
\\

\noindent

\noindent
\underline{\emph{Regime $\mathfrak{R}1)$} }
\\
\noindent
The analysis of $\mathfrak{R}1$ performed for the bounded case in \cite{DFPR3} is quite robust and applies to the present situation, modulo some modifications in order to replace the norms $\|\cdot\|$ with the weighted norms $\|\cdot\|_{H^0}$.
We recall that the strategy can be outlined as follows.
\begin{itemize}
\item The potential $V^{(\bold{k},\bold{q})}_{J_{\bold{r},\bold{i}}}$ is re-expanded according to the set of prescriptions given by the tree diagram in Definition \ref{def-tree} below. The weighted norm of each single summand  $\mathfrak{b}$ -- where $\mathfrak{b}$ stands for branch-operator (see (\ref{opbranch-0})) -- of the re-expansion is bounded from above in Lemma \ref{weightcontrol}.
\item  A path visiting rectangles (Definition \ref{pathsdef}), $\Gamma_\mathfrak{b}$, is assigned injectively to each summand $\mathfrak{b}$, with the properties listed in Lemma \ref{conn-rect-2}. A weight $w_\Gamma$ is assigned to each path visiting rectangles, in such a way that 
\begin{equation}\label{first-point}
\|\mathfrak{b}\|_{H^0}\leq w_{\Gamma_\mathfrak{b}},
\end{equation}
so that
\begin{equation}
\|V^{(\bold{k},\bold{q})}_{J_{\bold{r},\bold{i}}}\|_{H^0}\leq \sum_{\mathfrak{b}}\|\mathfrak{b}\|_{H^0}\leq \sum_{\Gamma_{\mathfrak{b}}} w_{\Gamma_\mathfrak{b}}.
\end{equation}
This allows us to bound the weighted norm of $V^{(\bold{k},\bold{q})}_{J_{\bold{r},\bold{i}}}$ by estimating the number of involved paths $\Gamma$, each contributing with weight $w_\Gamma$.
\end{itemize}

The procedure to state (\ref{first-point}) differs from the bounded case of \cite{DFPR3} due to the involvement of weighted norms. The definition of the weights along with the rest of the proof has a geometric content that is independent of  the type (bounded or unbounded) of interaction. Hence this part of the proof is done in the same way as \cite{DFPR3} Section 3 and Theorem 5.1, but, for the convenience of the reader, in the following we shall quickly review it, deferring some parts to the appendix.

In order to streamline our formulae, we set the notation
\begin{equation}\label{Acal}
\sum_{n=1}^{\infty}\frac{1}{n!}\,ad^{n}S_{J_{\bold{k},\bold{q}}}(\dots)=:\mathcal{A}_{J_{\bold{k},\bold{q}}}(\dots)\,.
\end{equation}
In order to establish notation we here recall how the re-expansion of the potential is performed by means of the tree diagram mentioned above.

\begin{defn}\label{def-tree}
We associate with $V^{(\bold{k},\bold{q})}_{J_{\bold{r},\bold{i}}}$ a tree diagram in the following way.
\noindent
\begin{enumerate}
\item The levels of the tree used to identify the contributions to the re-expansion of a potential $V^{(\bold{k},\bold{q})}_{J_{\bold{r},\bold{i}}}$ are labeled by  $(\bold{k^{\prime},\bold{q}^{\prime}})$, with  $(\bold{k^{\prime},\bold{q}^{\prime}})$ such that $ (\bold{k,\bold{q}}) \succeq (\bold{k^{\prime},\bold{q}^{\prime}})\succeq (\bold{0},\bold{N})$. We say that such a tree is \textit{rooted} at level $(\mathbf{k}, \mathbf{q})$.
\item There is a single vertex at the top of a tree rooted at level $(\bold{k},\bold{q})$; it is labeled by the symbol 
$V^{(\bold{k},\bold{q})}_{J_{\bold{r},\bold{i}}}$ of the potential.
\item The vertices at level $(\bold{k}^{\prime},\bold{q^{\prime}})_{-1}$ of a tree rooted at level $(\mathbf{k}, \mathbf{q})$ are determined by the vertices of the tree at level $(\bold{k}^{\prime},\bold{q^{\prime}})$ in the following way:
Each vertex $\mathfrak{v}\equiv \mathfrak{v}_{V_{J_{\bold{s},{u}}}^{(\bold{k}^{\prime},\bold{q}^{\prime})}}$  at level 
$(\bold{k^{\prime},\bold{q}^{\prime}})$, labeled by $V_{J_{\bold{s},{u}}}^{(\bold{k}^{\prime},\bold{q}^{\prime})}$, is linked 
to two sets of descendants (vertices) at level $(\bold{k^{\prime},\bold{q}^{\prime}})_{-1}$ with the following properties: The two sets of vertices 
are \textit{empty} if $(\bold{s},\bold{u})=(\bold{k}^{\prime},\bold{q}^{\prime})$; otherwise
\begin{itemize}
\item the leftmost set of vertices actually consists of a single vertex, which is labeled by the potential 
$V^{(\bold{k}^{\prime},\bold{q}^{\prime})_{-1}}_{J_{\bold{s},\bold{u}}}$;
\item the rightmost set of vertices is empty if $J_{\bold{k}^{\prime},\bold{q}^{\prime}}\nsubset J_{\bold{s},\bold{u}}$; otherwise it contains a vertex for each element $J_{\mathbf{s}', \mathbf{u}'}$ belonging to
$\mathcal{G}^{(\bold{k}^{\prime},\bold{q}^{\prime})}_{J_{\bold{s},\bold{u}}}$, and this vertex is labeled by $V^{(\bold{k}^{\prime},\bold{q}^{\prime})_{-1}}_{J_{\bold{s}^{\prime},\bold{u}^{\prime}}}$.
\end{itemize}

\item Each vertex $\mathfrak{v}$ at level $(\bold{k}^{\prime},\bold{q}^{\prime})$ is connected by an edge to its descendants at level $(\bold{k}^{\prime},\bold{q}^{\prime})_{-1}$. Edges are labelled by rectangles, or carry no label, in the following way:
\begin{itemize}
\item [e-i)] the edge connecting a vertex $\mathfrak{v}$ at level $(\bold{k}^{\prime},\bold{q}^{\prime})$ to its leftmost descendant
 at level $(\bold{k}^{\prime},\bold{q}^{\prime})_{-1}$ has no label.
It stands for the map
$$V^{(\bold{k}^{\prime},\bold{q}^{\prime})}_{J_{\bold{s},\bold{u}}} \rightarrow V^{(\bold{k}^{\prime},\bold{q}^{\prime})_{-1}}_{J_{\bold{s},\bold{u}}}\,, $$ 
where $V^{(\bold{k}^{\prime},\bold{q}^{\prime})}_{J_{\bold{s},\bold{u}}}$ is the potential labelling $\mathfrak{v}$ and $V^{(\bold{k}^{\prime},\bold{q}^{\prime})_{-1}}_{J_{\bold{s},\bold{u}}} $ labels its leftmost descendant at level $(\bold{k}^{\prime},\bold{q}^{\prime})_{-1}$;
\item [e-ii)] each edge $\mathfrak{e}$ connecting the vertex $\mathfrak{v}$ at level $(\bold{k}^{\prime},\bold{q}^{\prime})$ to other descendants at level $(\bold{k}^{\prime},\bold{q}^{\prime})_{-1}$ is labeled by  a rectangle $J_{\bold{k}^{\prime},\bold{q}^{\prime}}$. It stands for the map
$$V^{(\bold{k}^{\prime},\bold{q}^{\prime})}_{J_{\bold{s},\bold{u}}} \rightarrow \mathcal{A}_{J_{\bold{k}^{\prime},\bold{q}^{\prime}}}(V^{(\bold{k}^{\prime},\bold{q}^{\prime})_{-1}}_{J_{\bold{s}^{\prime},\bold{u}^{\prime}}})\,,$$
where $V^{(\bold{k}^{\prime},\bold{q}^{\prime})}_{J_{\bold{s},\bold{u}}}$ labels the vertex $\mathfrak{v}$ and $V^{(\bold{k}^{\prime},\bold{q}^{\prime})_{-1}}_{J_{\bold{s}^{\prime},\bold{u}^{\prime}}}$ is the potential labelling the vertex connected to $\mathfrak{v}$ by the edge $\mathfrak{e}$.
\end{itemize}

\item A leaf of the tree is a vertex at some level $(\bold{k}^{\prime},\bold{q}^{\prime})$ that has no descendants, i.e., that is not connected to any vertex at level $(\bold{k}^{\prime},\bold{q}^{\prime})_{-1}$ by any edge. Note that a leaf of the tree 
is labeled by a potential of the type $V^{(\bold{k}^{\prime\prime},\bold{q}^{\prime\prime})}_{J_{\bold{k}^{\prime\prime},\bold{q}^{\prime\prime}}}$ for some $(\bold{k}^{\prime\prime},\bold{q}^{\prime\prime})\succeq (\bold{0},\bold{N})$.
%
\item A branch of a tree rooted at $(\mathbf{k}, \mathbf{q})$ is an ordered connected set of edges with the following properties:
\begin{itemize}
\item the first edge of a branch has the vertex at level $(\bold{k},\bold{q})$ as an endpoint;
\item the last edge of a branch has a leaf at some level $(\bold{k}^{\prime\prime},\bold{q}^{\prime\prime})$ as an endpoint (referred to as the leaf of the branch);
\item there is a single edge connecting vertices at levels $(\bold{k}^{\prime},\bold{q}^{\prime})$ and $(\bold{k}^{\prime},\bold{q}^{\prime})_{-1}$ for every $(\bold{k}^{\prime},\bold{q}^{\prime})$ with $(\bold{k},\bold{q})\succeq (\bold{k}^{\prime},\bold{q}^{\prime})\succ (\bold{k}^{\prime\prime},\bold{q}^{\prime\prime})$.
\end{itemize}




\item With each branch  $\mathfrak{b}$ of a tree 
we associate a set,   $\mathcal{R}_\mathfrak{b}$, of rectangles
consisting of i) those rectangles labelling the edges of $\mathfrak{b}$, and ii) the rectangle $J_{\bold{k}^{\prime\prime},\bold{q}^{\prime\prime}}$ indicating the support of the potential labelling the leaf of $\mathfrak{b}$. 

The set $\mathcal{R}_\mathfrak{b}$ inherits the ordering relation (\ref{ordering}), hence its elements can be enumerated  by a map 
$$i\in \big\{1,\cdots ,\vert \mathcal{R}_{\mathfrak{b}}\vert \big\}\rightarrow J_{\bold{k}^{(i)},\bold{q}^{(i)}}\in \mathcal{R}_\mathfrak{b}$$ 
with $(\bold{k}^{(i)},\bold{q}^{(i)})\succ (\bold{k}^{(i+1)},\bold{q}^{(i+1)})$ and where  $\vert \mathcal{R}_\mathfrak{b}\vert$ is the cardinality of the set $\mathcal{R}_\mathfrak{b}$. Note that $J_{\bold{k}^{(| \mathcal{R}_{\mathfrak{b}}|)},\bold{q}^{(|\mathcal{R}_{\mathfrak{b}}|)}}$ is the rectangle associated with the potential labelling the leaf of $\mathfrak{b}$. 

\item To every branch $\mathfrak{b}$ we can associate the \textit{``branch operator''}, also denoted by $\mathfrak{b}$,
\begin{equation}\label{opbranch-0}
\mathfrak{b}:=\mathcal{A}_{J_{\bold{k}^{(1)},\bold{q}^{(1)}}}(\,\mathcal{A}_{J_{\bold{k}^{(2)},\bold{q}^{(2)}}} (\cdots  \mathcal{A}_{J_{\bold{k}^{(|\mathcal{R}_b | -1)},\bold{q}^{(| \mathcal{R}_b| -1)}}}(V_{\mathcal{L}_\mathfrak{b}})\cdots )\, )\,,
\end{equation}
where $V_{\mathcal{L}_\mathfrak{b}}:= V^{(\bold{k}^{(| \mathcal{R}_{\mathfrak{b}}|)},\bold{q}^{(|\mathcal{R}_{\mathfrak{b}}|)})}_{J_{\bold{k}^{(| \mathcal{R}_{\mathfrak{b}}|)},\bold{q}^{(|\mathcal{R}_{\mathfrak{b}}|)}}}$ is the potential labelling the leaf of $\mathfrak{b}$.

The set of branches whose corresponding branch operators are non-zero is denoted by 
$\mathcal{B}_{V^{(\bold{k},\bold{q})}_{J_{\bold{r},\bold{i}}}}$. We use the notation $\mathfrak{b}$ both for a branch and its corresponding branch operator.
\end{enumerate}
\qed
\end{defn}

The weighted norm of a single branch operator $\mathfrak{b}\in \mathcal{B}_{V^{(\bold{k},\bold{q})}_{J_{\bold{r},\bold{i}}}}$, contributing to the expansion of $V^{(\bold{k},\bold{q})}_{J_{\bold{r},\bold{i}}}$, is estimated in Lemma \ref{weightcontrol} as follows.
\noindent

For $\mathfrak{b}\in\mathcal{B}_{V^{(\bold{k},\bold{q})}_{J_{\bold{r},\bold{i}}}}$,
\begin{equation}\label{bnorm}
\|\mathfrak{b}\|_{H^0}:=\|(H^0_{J_{\bold{r},\bold{i}}}+1)^{-\frac{1}{2}}\,\mathfrak{b}\, (H^0_{J_{\bold{r},\bold{i}}}+1)^{-\frac{1}{2}}\|\leq t^{\frac{r-1}{3}}\, \prod_{R\in\mathcal{R}_{\mathfrak{b}}} ((c+1)\frac{t^{\frac{1}{3}}}{(\rho(R))^{x_d}})
\end{equation}
where $c$ is a universal constant and $\rho(R)$ is the size of $R\in\mathcal{R}_{\mathfrak{b}}$, i.e.,  $R=J_{\bold{s},\bold{u}}$ for some ${\bold{s},\bold{u}}$ and $\rho(R)=s$.

 Now, in order to estimate the number of summands $\mathfrak{b}\in  \mathcal{B}_{V^{(\bold{k},\bold{q})}_{J_{\bold{r},\bold{i}}}}$ appearing in the re-expansion provided in Definition \ref{def-tree}, note that by Property P-iv) in Lemma \ref{treeprops}, for any two distinguished $\mathfrak{b}_1,\mathfrak{b}_2\in  \mathcal{B}_{V^{(\bold{k},\bold{q})}_{J_{\bold{r},\bold{i}}}}$, the corresponding sets of rectangles $\{\mathcal{R}_{\mathfrak{b}_{1}}\}$, $\{\mathcal{R}_{\mathfrak{b}_{2}}\}$, defined in point 7. of Definition \ref{def-tree}, do not agree. This allows us by Lemma \ref{conn-rect-2} to assign injectively to each $\mathfrak{b}\in  \mathcal{B}_{V^{(\bold{k},\bold{q})}_{J_{\bold{r},\bold{i}}}}$ a path of rectangles $\Gamma_{\mathfrak{b}}$ (see Definition \ref{pathsdef}).

Given any path of rectangles $\Gamma$, to each  step $\mathcal{S}_{\Gamma}\ni\sigma=(J_{\bold{s}^{(i)},\bold{u}^{(i)}},J_{\bold{s}^{(i+1)},\bold{u}^{(i+1)}})$ (see Definition \ref{pathsdef}) we  assign the weight  
\begin{equation}\label{stepweight}
w_\sigma:=\Big((c+1)\frac{t^{1/3}}{s_\sigma^{x_d}}\Big)^{1/2}
\end{equation}
where $s_\sigma:=\max\{s^{(i)},s^{(i+1)}\}$, with $w_\sigma<1$ for $t>0$ sufficiently small. 

Let $\mathfrak{b}\in\mathcal{B}_{V^{(\bold{k},\bold{q})}_{J_{\bold{r},\bold{i}}}}$, then from Lemma \ref{pathweightest}
\begin{equation}
\|\mathfrak{b}\|_{H^0} \leq t^{\frac{r-1}{3}}\cdot \prod_{\sigma\in\mathcal{S}_{\Gamma_{\mathfrak{b}}}}w_\sigma ,
\end{equation}
where $\Gamma_{\mathfrak{b}}$ is the path associated with $\mathfrak{b}$ constructed in Lemma \ref{conn-rect-2}, $\mathcal{S}_{\Gamma_{\mathfrak{b}}}$ is the set of steps of $\Gamma_{\mathfrak{b}}$.
%
\noindent
Hence, summing over all branches, we get
\begin{equation}
\|V^{(\bold{k},\bold{q})}_{J_{\bold{r},\bold{i}}}\|_{H^0}\leq \sum_{\mathfrak{b}\in \mathcal{B}_{V^{(\bold{k},\bold{q})}_{J_{\bold{r},\bold{i}}}}}\|\mathfrak{b}\|_{H^0}
\leq \sum_{ \mathfrak{b}\in \mathcal{B}_{V^{(\bold{k},\bold{q})}_{J_{\bold{r},\bold{i}}}}}t^{\frac{r-1}{3}}\cdot \prod_{\sigma\in\mathcal{S}_{\Gamma_{\mathfrak{b}}}}w_\sigma
\end{equation}
which can be bounded from above by estimating the number of weighted paths $\Gamma_{\mathfrak{b}}$ as follows
\begin{eqnarray}
& & \sum_{\mathfrak{b}\in \mathcal{B}_{V^{(\bold{k},\bold{q})}_{J_{\bold{r},\bold{i}}}}}t^{\frac{r-1}{3}}\cdot \prod_{\sigma\in\mathcal{S}_{\Gamma_{\mathfrak{b}}}}w_\sigma\\
&\leq&C_d\cdot r^{2d-1}\cdot t^{\frac{r-1}{3}}\cdot \sum_{j= \lfloor c_d \cdot r/k \rfloor}^{\infty} \Big(\sum_{\rho, \rho'=1}^{k} \Big((c+1)\frac{t^{1/3}}{(\max\{\rho,\rho^\prime\})^{x_d}}\Big)^{1/2} D_{\rho,\rho'}\Big)^j\,\,\label{est-r1-in}
\end{eqnarray}

where:
\begin{itemize}
\item \begin{equation}\label{directions}
D_{\rho,\rho'}:=\mathfrak{C}_d \cdot \rho^d \cdot  \rho^{\prime d-1}\,,
\end{equation}
where $\mathfrak{C}_d$ is a $d$-dependent constant, is an upper bound on the number of possible directions 
of a path $\Gamma=\{J_{\bold{s}^{(i)},\bold{u}^{(i)}}\}_{i=1}^n$, extended by one more step as specified here: given the 
path $\Gamma=\{J_{\bold{s}^{(i)},\bold{u}^{(i)}}\}_{i=1}^n$, the number of paths 
$\Gamma^+=\{J_{\bold{s}^{'(i)},\bold{u}^{'(i)}}\}_{i=1}^{n+1}$ of length $l_{\Gamma^+}=n$, whose first $n$ elements
agree with $\Gamma$ (i.e., 
$ \{J_{\bold{s}^{(i)},\bold{u}^{(i)}}\}_{i=1}^{n}=\{J_{\bold{s}^{\prime (i)},\bold{u}^{\prime(i)}}\}_{i=1}^{n}$) 
and for which $s^{\prime (n+1)}:=s'$ and $s^{(n)}:=s$, is bounded from above by $D_{s,s'}$. \\
\item
$$\sum_{\rho'=1}^{k} \Big((c+1)\frac{t^{1/3}}{(\max\{\rho,\rho^\prime\})^{x_d}} \Big)^{1/2} D_{\rho,\rho'} $$
accounts for all the weighted directions for a step from a rectangle of size $\rho$.
\item
the term $C_d\cdot r^{2d-1}$ is a bound\footnote{ It is enough to consider the volume of the rectangle $J_{\bold{r},\bold{i}}$ and Remark \ref{shapes}.} on the number of possible initial rectangles of a fixed path $\Gamma_{\mathfrak{b}}$;
\item
the sum over $j$ is the sum over the number of steps of  $\Gamma_{\mathfrak{b}}$ which by construction is bounded from below by $\lfloor c_d \cdot r/k \rfloor$, due to P-ii) in Lemma \ref{treeprops};
\item we have also used the fact that the correspondence $\mathfrak{b}\rightarrow \Gamma_{\mathfrak{b}}$ is injective, as the expression in (\ref{est-r1-in}) counts each single path only once.
 
\end{itemize}

\noindent
Finally, we can bound
\begin{eqnarray}
(\ref{est-r1-in}) \label{est-r1-in-bis}&\leq &C_d\cdot r^{2d-1}\cdot t^{\frac{r-1}{3}}\cdot \sum_{j= \lfloor c_d \cdot r/k \rfloor}^{\infty}\Big((c+1)^{1/2} \cdot t^{\frac{1}{6}}\cdot   2 \sum_{\rho=1}^{k} 
 \frac{\rho \cdot D_{\rho,\rho}}{\rho^{x_d/ 2} }  \Big)^j\,\\\
&\leq &C_d\cdot r^{2d-1}\cdot t^{\frac{r-1}{3}}\cdot t^{\frac{1}{12}\cdot \lfloor c_d \cdot r/k \rfloor}\cdot \sum_{j= \lfloor c_d \cdot r/k \rfloor}^{\infty}\Big((c+1) \cdot t^{\frac{1}{12}}\cdot 2 \sum_{\rho=1}^{k}  \frac{\rho \cdot D_{\rho,\rho}}{\rho^{x_d/2}}\Big)^j\,\quad\nonumber \\
&\leq &\frac{t^{\frac{r-1}{3}}}{r^{x_d+2d}}\,, \label{est-r1-fin}
\end{eqnarray}
where $t\geq 0$ is chosen small enough such that (recall $k\leq \lfloor r^{\frac{1}{4}} \rfloor$)
\begin{equation}
C_d\cdot r^{4d-1+x_d} \cdot  t^{\frac{1}{12}\cdot \lfloor c_d \cdot r/k \rfloor}\cdot \sum_{j=  \lfloor c_d \cdot r/k \rfloor}^{\infty}\,\Big((c+1)^{1/2} \cdot t^{\frac{1}{12}}\cdot 2 \sum_{\rho=1}^{k}  \frac{\rho \cdot D_{\rho,\rho}}{\rho^{x_d/2}}\Big)^j<1.\,
\end{equation}

\noindent
\underline{\emph{Regime $\mathfrak{R}2)$} }
\\

\noindent
The proof of $\mathfrak{R}2)$ is done analogously to the bounded case in \cite{DFPR3} but we also need some estimates introduced in \cite{DFPR1} where we treated the unbounded interactions in one space dimension; for completeness, here we review some of the steps and make the modifications explicit.

For $(\bold{k},\bold{q})$ in this regime and $(\bold{k},\bold{q})\succeq (\bold{s},\bold{u})\succ (\bold{k}_*,\bold{q}_*) $, where $(\bold{k}_*,\bold{q}_*)$ is the greatest rectangle  of regime $\frak{R}1$ with respect to the ordering $\succ$, we iteratively use
\begin{eqnarray}\label{R2exp}
\|V^{(\bold{s},\bold{u})}_{J_{\bold{r},\bold{i}}}\|_{H^0}
&\leq &\|V^{(\bold{s},\bold{u})_{-1}}_{J_{\bold{r},\bold{i}}}\|_{H^0}+\|\sum_{J_{\bold{k}',\bold{q}'}\in \mathcal{G}^{(\bold{s},\bold{u})}_{J_{\bold{r},\bold{i}}}}\,\sum_{n=1}^{\infty}\frac{1}{n!}\,ad^{n}S_{J_{\bold{s},\bold{u}}}(V^{(\bold{s},\bold{u})_{-1}}_{J_{\bold{k}',\bold{q}'}})\|_{H^0}\,,
\end{eqnarray}
to get
\begin{eqnarray}\label{R2exp2}
\|V^{(\bold{k},\bold{q})}_{J_{\bold{r},\bold{i}}}\|_{H^0}
&\leq&\|V^{(\bold{k}_*,\bold{q}_*)}_{J_{\bold{r},\bold{i}}}\|_{H^0}+\sum_{(\bold{k},\bold{q})\succeq (\bold{s},\bold{u})\succ (\bold{k}_*,\bold{q}_*) } \|\sum_{J_{\bold{k}',\bold{q}'}\in \mathcal{G}^{(\bold{s},\bold{u})}_{J_{\bold{r},\bold{i}}}}\,\sum_{n=1}^{\infty}\frac{1}{n!}\,ad^{n}S_{J_{\bold{s},\bold{u}}}(V^{(\bold{s},\bold{u})_{-1}}_{J_{\bold{k}',\bold{q}'}})\|_{H^0}\,.\nonumber
\end{eqnarray}
We first show 
\begin{equation}
\|\sum_{J_{\bold{k}',\bold{q}'}\in \mathcal{G}^{(\bold{s},\bold{u})}_{J_{\bold{r},\bold{i}}}}\,\sum_{n=1}^{\infty}\frac{1}{n!}\,ad^{n}S_{J_{\bold{s},\bold{u}}}(V^{(\bold{s},\bold{u})_{-1}}_{J_{\bold{k}',\bold{q}'}})\|_{H^0} \leq (c\, t)\, \|V^{(\bold{s},\bold{u})}_{J_{\bold{s},\bold{u}}} \|_{H^0}\, \|V^{(\bold{s},\bold{u})_{-1}}_{J_{\bold{k}^\prime,\bold{q}^\prime}} \|_{H^0}
\end{equation}
for a universal constant $c$.
To do this, we estimate the norms of terms of the type
\begin{equation}
(H_{J_{\bold{r},\bold{i}}}^0+1)^{-\frac{1}{2}}\,S_{J_{\bold{s},\bold{u}}}\dots S_{J_{\bold{s},\bold{u}}} V^{(\bold{s},\bold{u})_{-1}}_{J_{\bold{k}',\bold{q}'}}\, S_{J_{\bold{s},\bold{u}}}\dots S_{J_{\bold{s},\bold{u}}}\,(H_{J_{\bold{r},\bold{i}}}^0+1)^{-\frac{1}{2}}
\end{equation}
that we re-write as
\begin{equation}
(H_{J_{\bold{r},\bold{i}}}^0+1)^{-\frac{1}{2}}\,S_{J_{\bold{s},\bold{u}}}\dots S_{J_{\bold{s},\bold{u}}}(H_{J_{\bold{k}^\prime,\bold{q}^\prime}}^0+1)^{\frac{1}{2}}(H_{J_{\bold{k}^\prime,\bold{q}^\prime}}^0+1)^{-\frac{1}{2}} V^{(\bold{s},\bold{u})_{-1}}_{J_{\bold{k}',\bold{q}'}}\, S_{J_{\bold{s},\bold{u}}}\dots S_{J_{\bold{s},\bold{u}}}\,(H_{J_{\bold{k}^\prime,\bold{q}^\prime}}^0+1)^{-\frac{1}{2}}\,.
\end{equation}
In particular, let us show how to bound
\begin{equation}
(H_{J_{\bold{r},\bold{i}}}^0+1)^{-\frac{1}{2}}\,S_{J_{\bold{s},\bold{u}}}\dots S_{J_{\bold{s},\bold{u}}}(H_{J_{\bold{k}^\prime,\bold{q}^\prime}}^0+1)^{\frac{1}{2}}\,.
\end{equation}
We insert $\charf=(H_{J_{\bold{k}^\prime,\bold{q}^\prime}\setminus J_{\bold{s},\bold{u}}}^0+1)^{\frac{1}{2}}(H_{J_{\bold{k}^\prime,\bold{q}^\prime}\setminus J_{\bold{s},\bold{u}}}^0+1)^{-\frac{1}{2}}$ and exploit $[H_{J_{\bold{k}^\prime,\bold{q}^\prime}\setminus J_{\bold{s},\bold{u}}}^0\,,\,S_{J_{\bold{s},\bold{u}}}]=0$ that holds since the two supports, $J_{\bold{k}^\prime,\bold{q}^\prime}\setminus J_{\bold{s},\bold{u}}$ and $J_{\bold{s},\bold{u}}$, are nonoverlapping by construction. Here $H_{J_{\bold{k}^\prime,\bold{q}^\prime}\setminus J_{\bold{s},\bold{u}}}$ is of course naturally defined even if $J_{\bold{k}^\prime,\bold{q}^\prime}\setminus J_{\bold{s},\bold{u}}$ is not necessarily a rectangle. Thus
\begin{eqnarray}
& &(H_{J_{\bold{r},\bold{i}}}^0+1)^{-\frac{1}{2}}(H_{J_{\bold{k}^\prime,\bold{q}^\prime}\setminus J_{\bold{s},\bold{u}}}^0+1)^{\frac{1}{2}}(H_{J_{\bold{k}^\prime,\bold{q}^\prime}\setminus J_{\bold{s},\bold{u}}}^0+1)^{-\frac{1}{2}}\,S_{J_{\bold{s},\bold{u}}}\dots S_{J_{\bold{s},\bold{u}}}(H_{J_{\bold{r},\bold{i}}}^0+1)^{\frac{1}{2}}\nonumber \\
&=&(H_{J_{\bold{r},\bold{i}}}^0+1)^{-\frac{1}{2}}(H_{J_{\bold{k}^\prime,\bold{q}^\prime}\setminus J_{\bold{s},\bold{u}}}^0+1)^{\frac{1}{2}}\,S_{J_{\bold{s},\bold{u}}}\dots S_{J_{\bold{s},\bold{u}}}(H_{J_{\bold{k}^\prime,\bold{q}^\prime}\setminus J_{\bold{s},\bold{u}}}^0+1)^{-\frac{1}{2}}(H_{J_{\bold{r},\bold{i}}}^0+1)^{\frac{1}{2}}\,.\nonumber
\end{eqnarray}
Thus the result in (\ref{R2exp2}) is obtained by making use of:
\begin{itemize}
\item
the results from Lemma \ref{control-LS} that we can exploit due to the inductive hypothesis for S2)
\begin{equation}\label{bound-S}
\|S_{J_{\bold{s},\bold{u}}}\|\leq C\,t\,
 \| V^{(\bold{s},\bold{u})_{-1}}_{J_{\bold{s},\bold{u}}}\|_{H_0}
\end{equation}
\begin{equation}
\|S_{J_{\bold{s},\bold{u}}}(H^0_{J_{\bold{s},\bold{u}}}+1)^{\frac{1}{2}}\| \leq C\,t\,
 \| V^{(\bold{s},\bold{u})_{-1}}_{J_{\bold{s},\bold{u}}}\|_{H_0}\,;
\end{equation}
\item
the operator norm bound
\begin{equation}
\|(H_{J_{\bold{r},\bold{i}}}^0+1)^{-\frac{1}{2}}(H_{J_{\bold{k}^\prime,\bold{q}^\prime}\setminus J_{\bold{s},\bold{u}}}^0+1)^{\frac{1}{2}}\|\leq 1
\end{equation}
that follows  from the spectral theorem for commuting operators and from the inclusion $J_{\bold{k}^\prime,\bold{q}^\prime}\setminus J_{\bold{s},\bold{u}}\subset J_{\bold{r},\bold{i}}$;
\item the operator norm bound
\begin{equation}
\|(H^0_{J_{\bold{s},\bold{u}}}+1)^{-\frac{1}{2}}(H_{J_{\bold{k}^\prime,\bold{q}^\prime}\setminus J_{\bold{s},\bold{u}}}^0+1)^{-\frac{1}{2}}(H_{J_{\bold{k}^\prime,\bold{q}^\prime}}^0+1)^{\frac{1}{2}}\|\leq 1
\end{equation}
that follows  from the spectral theorem for commuting operators.
\end{itemize}

By the inductive hypotheses (\ref{inductive-reg-1}), (\ref{inductive-reg-2}),  (\ref{R3-2}),  and (\ref{R3-3}), together with (\ref{R2exp2}) for $t\geq 0$ sufficiently small, we get from (\ref{R2exp})
\begin{eqnarray}
\|V^{(\bold{k},\bold{q})}_{J_{\bold{r},\bold{i}}}\|_{H^0}
&\leq&\|V^{(\bold{k}_*,\bold{q}_*)}_{J_{\bold{r},\bold{i}}}\|_{H^0}\\
& &+\sum_{s=\lfloor r^\frac{1}{4} \rfloor}^{ r-\lfloor r^\frac{1}{4} \rfloor }\sum_{s_{1}=0}^{s}\sum_{s_{2}=0}^{s-s_{1}}\dots \sum_{s_{d}=0}^{s-s_{1}-\dots -s_{d-1}}\delta_{s_1+s_2+\dots +s_d-s}\, \cdot c_d \cdot r^{2d-1} \cdot t\cdot \frac{t^{\frac{s-1}{3}} }{s^{x_d}}\cdot \frac{t^{\frac{r-s-1}{3}} }{(r-s)^{x_d}}\quad\quad\quad  \label{crudesum}
\end{eqnarray}
where: 
\begin{itemize}
\item the multiplicative factor $\mathcal{O}(r^{2d-1}) $ is an upper bound estimate (see Remark \ref{shapes})
to the number of rectangles $J_{\bold{k}', \bold{q}'}\subset J_{\bold{r}, \bold{i}}$ such that  $[ J_{\bold{k}',\bold{q}'} \cup J_{\bold{k},\bold{q}} ]=J_{\bold{r},\bold{i}}$.
\item by construction $k_*=\lfloor r^\frac{1}{4}\rfloor$;
\end{itemize}
Now, for any $s$ with $\lfloor r^\frac{1}{4} \rfloor\leq s\leq r-\lfloor r^\frac{1}{4} \rfloor$,
\begin{eqnarray}
&&\sum_{s_{1}=0}^{s}\sum_{s_{2}=0}^{s-s_{1}}\dots \sum_{s_{d}=0}^{s-s_{1}-\dots -s_{d-1}}\delta_{s_1+s_2+\dots +s_d-s}\cdot  c_d \cdot r^{2d-1} \cdot t\cdot \frac{t^{\frac{s-1}{3}}\cdot }{s^{x_d}}\frac{t^{\frac{r-s-1}{3}} }{(r-s)^{x_d}}\\
&\leq & s^d \cdot c_d \cdot r^{2d-1}\cdot  t^\frac{2}{3}\cdot \frac{t^{\frac{r-1}{3}}}{s^{x_d}\cdot (r-s)^{x_d}}\\
&\leq & r^d \cdot c_d \cdot r^{2d-1}\cdot t^\frac{2}{3}\cdot \frac{t^{\frac{r-1}{3}}}{s^{x_d} \cdot (r-s)^{x_d}}\\
&\leq & 2^{x_d} \cdot c_d \cdot r^{2d-1}\cdot  t^\frac{2}{3}\cdot \frac{t^{\frac{r-1}{3}}}{r^{x_d} \cdot r^{x_d/4}}
\end{eqnarray}
as 
$$\max_{\lfloor r^\frac{1}{4} \rfloor\leq s\leq r-\lfloor r^\frac{1}{4} \rfloor}\, \frac{1}{s^{x_d}\cdot (r-s)^{x_d}} \leq \frac{1}{r^{x_d/4}\cdot (r-r^\frac{1}{4})^{x_d}}\leq \frac{2^{x_d}}{r^{x_d}\cdot r^{x_d/4}}$$
since $r-\lfloor r^\frac{1}{4} \rfloor\geq\frac{r}{2}$. Finally, using the inductive hypothesis for $\|V^{(\bold{k},\bold{q})_*}_{J_{\bold{r},\bold{i}}}\|$,
\begin{eqnarray}
\|V^{(\bold{k},\bold{q})}_{J_{\bold{r},\bold{i}}}\|&\leq &\|V^{(\bold{k},\bold{q})_*}_{J_{\bold{r},\bold{i}}}\|+\sum_{s=\lfloor r^\frac{1}{4} \rfloor}^{r-\lfloor r^\frac{1}{4} \rfloor} r^d \cdot 2^{x_d} \cdot c_d \cdot r^{2d-1}\cdot  t^\frac{2}{3}\cdot \frac{t^{\frac{r-1}{3}}}{r^{x_d} \cdot r^{x_d/4}}\\
&\leq &\frac{t^\frac{r-1}{3}}{r^{x_d+2d}} +2^{x_d} \cdot c_d \cdot t^\frac{2}{3} \cdot \frac{t^\frac{r-1}{3}}{r^{\frac{5x_d}{4}-3d}}\\
&\leq & 2\cdot \frac{t^\frac{r-1}{3}}{r^{x_d+2d}}
\end{eqnarray}
since  $x_d= 20d$ and $t\geq 0$ is small enough.

\noindent
\underline{\emph{Regime $\mathcal{R}3)$}}
\\

We recall that, for the potential associated to the given rectangle $J_{\bold{r},\bold{i}}$ and for all those associated to rectangles of smaller size, we assume that (\ref{inductive-reg-1}), (\ref{inductive-reg-2}), (\ref{R3-1}), and (\ref{R3-2}) hold for all (corresponding) steps before $(\bold{k}, \bold{q})$. Then, we prove  that (\ref{R3-1}) holds  in step $(\bold{k}, \bold{q})$, and, consequently,  also (\ref{R3-2}) is true (in step $(\bold{k}, \bold{q})$)\,. In the study of this regime, in order to streamline the notation we use the following definitions
\begin{equation}
(\frac{1}{[\pi_{J_{\bold{r},\bold{i}}}]_1})^{\frac{1}{2}}
:=(\frac{1}{\pi_{J_{\bold{r},\bold{i}}}+1})^{\frac{1}{2}}:=(\frac{1}{\sum_{\bold{j}\in J_{\bold{r},\bold{i}}} P^{\perp}_{\Omega_{\bold{j}}}+1})^{\frac{1}{2}}\quad 
\end{equation}
\begin{equation}
 (\frac{1}{[H^0_{J_{\bold{r},\bold{i}}}]_1})^{\frac{1}{2}}
:=(\frac{1}{\sum_{\bold{j}\in J_{\bold{r},\bold{i}}} H_{\bold{j}}+1})^{\frac{1}{2}}\,.
\end{equation}
We warn the reader that a similar notation with the lower index $1$ is used for the first term in the Lie Schwinger series defining the operators $V^{(\bold{k},\bold{q})}_{J_{\bold{k},\bold{q}}}$ and $S_{J_{\bold{k},\bold{q}}}$.
\\

\noindent
\underline{\emph{Proof of  (\ref{R3-1})}}
\\

\noindent
We first consider 
\begin{eqnarray}
& &(\frac{1}{[\pi_{J_{\bold{r},\bold{i}}}]_1})^{\frac{1}{2}}\, (\frac{1}{[H^0_{J_{\bold{r},\bold{i}}}]_1})^{\frac{1}{2}}\,P^{(+)}_{J_{\bold{r},\bold{i}}}\,V^{(\bold{k},\bold{q})}_{J_{\bold{r},\bold{i}}}\,P^{(-)}_{J_{\bold{r},\bold{i}}}\, (\frac{1}{[H^0_{J_{\bold{r},\bold{i}}}]_1})^{\frac{1}{2}}\, (\frac{1}{[\pi_{J_{\bold{r},\bold{i}}}]_1})^{\frac{1}{2}}\quad\quad\quad \\
&=&(\frac{1}{[\pi_{J_{\bold{r},\bold{i}}}]_1})^{\frac{1}{2}}\,(\frac{1}{[H^0_{J_{\bold{r},\bold{i}}}]_1})^{\frac{1}{2}}\,P^{(+)}_{J_{\bold{r},\bold{i}}}\,V^{(\bold{k},\bold{q})}_{J_{\bold{r},\bold{i}}}\,P^{(-)}_{J_{\bold{r},\bold{i}}}\,.
\end{eqnarray}

Recall that for $(\bold{k},\bold{q})\prec (\bold{r},\bold{i})$ the types of re-expansion that have to be considered correspond to a) and c) in Definition \ref{def-interactions-multi}. The re-expansion of type a) is trivial since it leaves potential unchanged. Using the re-expansion of type c), i.e.,  (\ref{main-def-V-multi}), we obtain
\begin{eqnarray}
& &(\frac{1}{[\pi_{J_{\bold{r},\bold{i}}}]_1})^{\frac{1}{2}}\, (\frac{1}{[H^0_{J_{\bold{r},\bold{i}}}]_1})^{\frac{1}{2}}\,P^{(+)}_{J_{\bold{r},\bold{i}}}\,V^{(\bold{k},\bold{q})}_{J_{\bold{r},\bold{i}}}\,P^{(-)}_{J_{\bold{r},\bold{i}}}\\
&=&(\frac{1}{[\pi_{J_{\bold{r},\bold{i}}}]_1})^{\frac{1}{2}}\, (\frac{1}{[H^0_{J_{\bold{r},\bold{i}}}]_1})^{\frac{1}{2}}P^{(+)}_{J_{\bold{r},\bold{i}}}\,V^{(\bold{k},\bold{q})_{-1}}_{J_{\bold{r},\bold{i}}}\,P^{(-)}_{J_{\bold{r},\bold{i}}}\label{moving}\\
& &+(\frac{1}{[\pi_{J_{\bold{r},\bold{i}}}]_1})^{\frac{1}{2}}\, (\frac{1}{[H^0_{J_{\bold{r},\bold{i}}}]_1})^{\frac{1}{2}}\,P^{(+)}_{J_{\bold{r},\bold{i}}}\,\Big\{\sum_{J_{\bold{k}',\bold{q}'}\in \mathcal{G}^{(\bold{k},\bold{q})}_{J_{\bold{r},\bold{i}}}}\,\sum_{n=1}^{\infty}\frac{1}{n!}\,ad^{n}S_{J_{\bold{k},\bold{q}}}(V^{(\bold{k},\bold{q})_{-1}}_{J_{\bold{k}',\bold{q}'}})\Big\}\,P^{(-)}_{J_{\bold{r},\bold{i}}}\,.\quad\quad\quad\quad  \label{rest-2}
\end{eqnarray}
Similarly to the bounded case in \cite{DFPR3}, we shall re-expand the terms analogous to $V^{(\bold{k},\bold{q})_{-1}}_{J_{\bold{r},\bold{i}}}$ in (\ref{moving}) from $(\bold{k},\bold{q})_{-1}$ down to $(\bold{k}_{**},\bold{q}_{**})$, which is the index corresponding to the greatest rectangle with respect to the ordering $\succ$ in the Regime $\frak{R}2$, with $k_{**}=r-\lfloor r^\frac{1}{4}\rfloor$ by construction. At each step we estimate the terms of the type  (\ref{rest-2}) that are produced by the iteration. 
\\

\noindent
%
%
\emph{Estimate of (\ref{rest-2})}
\\

\noindent
We split the corresponding term, $(\ref{rest-2})$, into
\begin{eqnarray}
& &(\ref{rest-2})\\
&= &(\frac{1}{[\pi_{J_{\bold{r},\bold{i}}}]_1})^{\frac{1}{2}}\, (\frac{1}{[H^0_{J_{\bold{r},\bold{i}}}]_1})^{\frac{1}{2}}\,\Big\{\sum_{J_{\bold{k}',\bold{q}'}\in \mathcal{G}^{(\bold{k},\bold{q})}_{J_{\bold{r},\bold{i}}}}\,\,ad\,S_{J_{\bold{k},\bold{q}}}(V^{(\bold{k},\bold{q})_{-1}}_{J_{\bold{k}',\bold{q}'}})\Big\}\,P^{(-)}_{J_{\bold{r},\bold{i}}}\label{rest-2-first}\\
& &+(\frac{1}{[\pi_{J_{\bold{r},\bold{i}}}]_1})^{\frac{1}{2}}\, (\frac{1}{[H^0_{J_{\bold{r},\bold{i}}}]_1})^{\frac{1}{2}}\,\Big\{\sum_{J_{\bold{k}',\bold{q}'}\in \mathcal{G}^{(\bold{k},\bold{q})}_{J_{\bold{r},\bold{i}}}}\,\sum_{n=2}^{\infty}\frac{1}{n!}\,ad^{n}S_{J_{\bold{k},\bold{q}}}(V^{(\bold{k},\bold{q})_{-1}}_{J_{\bold{k}',\bold{q}'}})\Big\}\,P^{(-)}_{J_{\bold{r},\bold{i}}}\,.\quad\quad \label{rest-2-second}
\end{eqnarray}

\noindent
In (\ref{rest-2-first}) we separately collect the contributions associated with $J_{\bold{k}',\bold{q}'}$ \emph{small} and \emph{large}, depending on whether $(\bold{k}',\bold{q}')$ is a predecessor or a successor of $(\bold{k},\bold{q})$, and denote by $(\mathcal{G}^{(\bold{k},\bold{q})}_{J_{\bold{r},\bold{i}}})_{small}$ the subset whose elements are the \emph{small} $J_{\bold{k}',\bold{q}'}$ in $\mathcal{G}^{(\bold{k},\bold{q})}_{J_{\bold{r},\bold{i}}}$. We denote the corresponding contributions
$$(\ref{rest-2-first})_{small}\quad \text{and}\quad (\ref{rest-2-first})_{large}\,,$$
respectively. 
Next, we analyze some commutators that enter the expression  $(\ref{rest-2-first})_{small}$ that is estimated below. Note that
\begin{eqnarray}
& &[S_{J_{\bold{k},\bold{q}}}\,,\,V^{(\bold{k},\bold{q})_{-1}}_{J_{\bold{k}',\bold{q}'}}]\\
& =&[S_{J_{\bold{k},\bold{q}}}\,,\,P^{(+)}_{J_{\bold{k}',\bold{q}'}}\,V^{(\bold{k},\bold{q})_{-1}}_{J_{\bold{k}',\bold{q}'}}\,P^{(+)}_{J_{\bold{k}',\bold{q}'}}+P^{(-)}_{J_{\bold{k}',\bold{q}'}}\,V^{(\bold{k},\bold{q})_{-1}}_{J_{\bold{k}',\bold{q}'}}\,P^{(-)}_{J_{\bold{k}',\bold{q}'}}]\,\\
& =&[S_{J_{\bold{k},\bold{q}}}\,,\,P^{(+)}_{J_{\bold{k}',\bold{q}'}}\,V^{(\bold{k},\bold{q})_{-1}}_{J_{\bold{k}',\bold{q}'}}\,P^{(+)}_{J_{\bold{k}',\bold{q}'}}+\langle V^{(\bold{k},\bold{q})_{-1}}_{J_{\bold{k}',\bold{q}'}}\rangle\,P^{(-)}_{J_{\bold{k}',\bold{q}'}}]\,\\
& =&[S_{J_{\bold{k},\bold{q}}}\,,\,P^{(+)}_{J_{\bold{k}',\bold{q}'}}\,V^{(\bold{k},\bold{q})_{-1}}_{J_{\bold{k}',\bold{q}'}}\,P^{(+)}_{J_{\bold{k}',\bold{q}'}}]-[S_{J_{\bold{k},\bold{q}}}\,,\,<V^{(\bold{k},\bold{q})_{-1}}_{J_{\bold{k}',\bold{q}'}}>P^{(+)}_{J_{\bold{k}',\bold{q}'}}]\,,
\end{eqnarray}
where we used that $V^{(\bold{k},\bold{q})_{-1}}_{J_{\bold{k}',\bold{q}'}}$ is block-diagonalized since \emph{small} means    $(\bold{k}',\bold{q}')\prec (\bold{k},\bold{q})$. Also note that $P^{(+)}_{J_{\bold{k}',\bold{q}'}}P^{(-)}_{J_{\bold{r},\bold{i}}}=0$  since  $J_{\bold{k}',\bold{q}'} \subset J_{\bold{r},\bold{i}}$ by construction, thus
 \begin{eqnarray}
& &P^{(+)}_{J_{\bold{r},\bold{i}}}\,[S_{J_{\bold{k},\bold{q}}}\,,\,P^{(+)}_{J_{\bold{k}',\bold{q}'}}\,V^{(\bold{k},\bold{q})_{-1}}_{J_{\bold{k}',\bold{q}'}}\,P^{(+)}_{J_{\bold{k}',\bold{q}'}}] \,P^{(-)}_{J_{\bold{r},\bold{i}}}\\
& &-P^{(+)}_{J_{\bold{r},\bold{i}}}\, [S_{J_{\bold{k},\bold{q}}}\,,\,<V^{(\bold{k},\bold{q})_{-1}}_{J_{\bold{k}',\bold{q}'}}>P^{(+)}_{J_{\bold{k}',\bold{q}'}}]\,P^{(-)}_{J_{\bold{r},\bold{i}}}\\
&=&-P^{(+)}_{J_{\bold{r},\bold{i}}}\,P^{(+)}_{J_{\bold{k}',\bold{q}'}}\,V^{(\bold{k},\bold{q})_{-1}}_{J_{\bold{k}',\bold{q}'}}\,P^{(+)}_{J_{\bold{k}',\bold{q}'}}S_{J_{\bold{k},\bold{q}}}P^{(-)}_{J_{\bold{r},\bold{i}}}\\
& &+P^{(+)}_{J_{\bold{r},\bold{i}}}\,<V^{(\bold{k},\bold{q})_{-1}}_{J_{\bold{k}',\bold{q}'}}>P^{(+)}_{J_{\bold{k}',\bold{q}'}}S_{J_{\bold{k},\bold{q}}}\,P^{(-)}_{J_{\bold{r},\bold{i}}}\,.
\end{eqnarray}

\noindent
Recall that 
\begin{equation}\label{def-Sj}
(S_{J_{\bold{k},\bold{q}}})_j:=\frac{1}{G_{J_{\bold{k},\bold{q}}}-E_{J_{\bold{k},\bold{q}}}}P^{(+)}_{J_{\bold{k},\bold{q}}}\,(V^{(\bold{k},\bold{q})_{-1}}_{J_{\bold{k},\bold{q}}})_j\,P^{(-)}_{J_{\bold{k},\bold{q}}}-h.c.\,;
\end{equation}
from Lemma  \ref{control-LS},  for $j\geq 2$ and $t\geq 0$ sufficiently small we get
\begin{equation}
\|\sum_{j=2}^{\infty}t^{j}(S_{J_{\bold{k},\bold{q}}})_j\|\leq C\cdot t\cdot \|(V^{(\bold{k},\bold{q})_{-1}}_{J_{\bold{k},\bold{q}}})_1\|_{H^0}^2.
\end{equation}
We split  $(\ref{rest-2-first})_{small}$ into two contributions: 

\noindent
1) the leading order term
\begin{eqnarray}
& &-(\frac{1}{[\pi_{J_{\bold{r},\bold{i}}}]_1})^{\frac{1}{2}}\, (\frac{1}{[H^0_{J_{\bold{r},\bold{i}}}]_1})^{\frac{1}{2}}\,P^{(+)}_{J_{\bold{r},\bold{i}}}\times \label{lead-2.115}\\
& &\quad \times\Big\{\sum_{J_{\bold{k}',\bold{q}'}\in (\mathcal{G}^{(\bold{k},\bold{q})}_{J_{\bold{r};\bold{i}}})_{small}   }\,\,\,P^{(+)}_{J_{\bold{k}',\bold{q}'}}\,\Big(V^{(\bold{k},\bold{q})_{-1}}_{J_{\bold{k}',\bold{q}'}}-\langle V^{(\bold{k},\bold{q})_{-1}}_{J_{\bold{k}',\bold{q}'}}\rangle\Big)\,P^{(+)}_{J_{\bold{k}',\bold{q}'}}\,\Big(\frac{t}{G_{J_{\bold{k},\bold{q}}}-E_{J_{\bold{k},\bold{q}}}}P^{(+)}_{J_{\bold{k},\bold{q}}}\,V^{(\bold{k},\bold{q})_{-1}}_{J_{\bold{k},\bold{q}}}\,P^{(-)}_{J_{\bold{k},\bold{q}}}-h.c.\Big)\Big\}\,P^{(-)}_{J_{\bold{r},\bold{i}}}\nonumber \\
&= &-(\frac{1}{[\pi_{J_{\bold{r},\bold{i}}}]_1})^{\frac{1}{2}}\,(\frac{1}{[H^0_{J_{\bold{r},\bold{i}}}]_1})^{\frac{1}{2}}\,P^{(+)}_{J_{\bold{r},\bold{i}}}\times\\
& &\quad\times \Big\{\sum_{J_{\bold{k}',\bold{q}'}\in (\mathcal{G}^{(\bold{k},\bold{q})}_{J_{\bold{r};\bold{i}}})_{small}}\,\,\,P^{(+)}_{J_{\bold{k}',\bold{q}'}}\,\Big(V^{(\bold{k},\bold{q})_{-1}}_{J_{\bold{k}',\bold{q}'}}-\langle V^{(\bold{k},\bold{q})_{-1}}_{J_{\bold{k}',\bold{q}'}}\rangle\Big)\,P^{(+)}_{J_{\bold{k}',\bold{q}'}}\,\Big(\frac{t}{G_{J_{\bold{k},\bold{q}}}-E_{J_{\bold{k},\bold{q}}}}P^{(+)}_{J_{\bold{k},\bold{q}}}\,V^{(\bold{k},\bold{q})_{-1}}_{J_{\bold{k},\bold{q}}}\,P^{(-)}_{J_{\bold{k},\bold{q}}}\Big)\Big\}\,P^{(-)}_{J_{\bold{r},\bold{i}}}\nonumber
\end{eqnarray}
where  we used $P^{(+)}_{J_{\bold{k},\bold{q}}}\,P^{(-)}_{J_{\bold{r},\bold{i}}}=0$; 

\noindent
2) the remainder term
\begin{equation}
-(\frac{1}{[\pi_{J_{\bold{r},\bold{i}}}]_1})^{\frac{1}{2}}\, (\frac{1}{[H^0_{J_{\bold{r},\bold{i}}}]_1})^{\frac{1}{2}}\,P^{(+)}_{J_{\bold{r},\bold{i}}}\,\Big\{\sum_{J_{\bold{k}',\bold{q}'}\in (\mathcal{G}^{(\bold{k},\bold{q})}_{J_{\bold{r};\bold{i}}})_{small}}\,\,\,P^{(+)}_{J_{\bold{k}',\bold{q}'}}\,\Big(V^{(\bold{k},\bold{q})_{-1}}_{J_{\bold{k}',\bold{q}'}}-\langle V^{(\bold{k},\bold{q})_{-1}}_{J_{\bold{k}',\bold{q}'}}\rangle\Big)\,P^{(+)}_{J_{\bold{k}',\bold{q}'}}\,\sum_{j=2}^{\infty}t^{j}(S_{J_{\bold{k},\bold{q}}})_j\,\Big\}\,P^{(-)}_{J_{\bold{r},\bold{i}}}\,.\label{rem-2.115}
\end{equation}
To estimate the leading order term  (\ref{lead-2.115}), we use the identity
\begin{eqnarray}
& &\|(\ref{lead-2.115})\|  \label{2.158}\\
&=&\|\sum_{J_{\bold{k}',\bold{q}'}\in (\mathcal{G}^{(\bold{k},\bold{q})}_{J_{\bold{r};\bold{i}}})_{small}}\,(\frac{1}{[\pi_{J_{\bold{r},\bold{i}}}]_1})^{\frac{1}{2}}\, (\frac{1}{[H^0_{J_{\bold{r},\bold{i}}}]_1})^{\frac{1}{2}}\,\,P^{(+)}_{J_{\bold{r},\bold{i}}}\,\,P^{(+)}_{J_{\bold{k}',\bold{q}'}}\Big(V^{(\bold{k},\bold{q})_{-1}}_{J_{\bold{k}',\bold{q}'}}-\langle V^{(\bold{k},\bold{q})_{-1}}_{J_{\bold{k}',\bold{q}'}}\rangle\Big)\,P^{(+)}_{J_{\bold{k}',\bold{q}'}}  (\frac{1}{[H^0_{J_{\bold{r},\bold{i}}\setminus J_{\bold{k},\bold{q}}}]_1})^{\frac{1}{2}} \nonumber \\
& &\quad \times \frac{t}{G_{J_{\bold{k},\bold{q}}}-E_{J_{\bold{k},\bold{q}}}}\,P^{(+)}_{J_{\bold{k},\bold{q}}}\,V^{(\bold{k},\bold{q})_{-1}}_{J_{\bold{k},\bold{q}}}\,P^{(-)}_{J_{\bold{k},\bold{q}}}\,P^{(-)}_{J_{\bold{r},\bold{i}}}  \|\,\\
&= &\|\sum_{J_{\bold{k}',\bold{q}'}\in (\mathcal{G}^{(\bold{k},\bold{q})}_{J_{\bold{r};\bold{i}}})_{small}}\,(\frac{1}{[\pi_{J_{\bold{r},\bold{i}}}]_1})^{\frac{1}{2}}\, (\frac{1}{[H^0_{J_{\bold{r},\bold{i}}}]_1})^{\frac{1}{2}}\,\,P^{(+)}_{J_{\bold{r},\bold{i}}}\,\,P^{(+)}_{J_{\bold{k}',\bold{q}'}}\Big(V^{(\bold{k},\bold{q})_{-1}}_{J_{\bold{k}',\bold{q}'}}-\langle V^{(\bold{k},\bold{q})_{-1}}_{J_{\bold{k}',\bold{q}'}}\rangle\Big)\,P^{(+)}_{J_{\bold{k}',\bold{q}'}} (\frac{1}{[H^0_{J_{\bold{k}',\bold{q}'}}]_1})^{\frac{1}{2}}  \,\nonumber \\
& &\quad\times  (\frac{[H^0_{J_{\bold{k}',\bold{q}'}}]_1}{[H^0_{J_{\bold{r},\bold{i}}\setminus J_{\bold{k},\bold{q}}}]_1})^{\frac{1}{2}}  (\frac{[\pi_{J_{\bold{k},\bold{q}}}]_1}{[H^0_{J_{\bold{k},\bold{q}}}]_1})^{\frac{1}{2}} \,([H^0_{J_{\bold{k},\bold{q}}}]_1)^{\frac{1}{2}}\, (\frac{1}{[\pi_{J_{\bold{k},\bold{q}}}]_1})^{\frac{1}{2}}\frac{t}{G_{J_{\bold{k},\bold{q}}}-E_{J_{\bold{k},\bold{q}}}}\,P^{(+)}_{J_{\bold{k},\bold{q}}}\,V^{(\bold{k},\bold{q})_{-1}}_{J_{\bold{k},\bold{q}}}\,P^{(-)}_{J_{\bold{k},\bold{q}}}P^{(-)}_{J_{\bold{r},\bold{i}}}\|\quad \quad \quad 
\end{eqnarray}
where we have inserted $ (\frac{1}{[H^0_{J_{\bold{r},\bold{i}}\setminus J_{\bold{k},\bold{q}}}]_1})^{\frac{1}{2}}$ for free since
\begin{equation}
[ (\frac{1}{[H^0_{J_{\bold{r},\bold{i}}\setminus J_{\bold{k},\bold{q}}}]_1})^{\frac{1}{2}}\,,\,\frac{t}{G_{J_{\bold{k},\bold{q}}}-E_{J_{\bold{k},\bold{q}}}}]=[ (\frac{1}{[H^0_{J_{\bold{r},\bold{i}}\setminus J_{\bold{k},\bold{q}}}]_1})^{\frac{1}{2}}\,,\,P^{(+)}_{J_{\bold{k},\bold{q}}}\,V^{(\bold{k},\bold{q})_{-1}}_{J_{\bold{k},\bold{q}}}\,P^{(-)}_{J_{\bold{k},\bold{q}}}]=0
\end{equation}
and $ (\frac{1}{[H^0_{J_{\bold{r},\bold{i}}\setminus J_{\bold{k},\bold{q}}}]_1})^{\frac{1}{2}}\,P^{(-)}_{J_{\bold{r},\bold{i}}}=P^{(-)}_{J_{\bold{r},\bold{i}}}$. 
In Lemma \ref{inductive-aux} we derive the key estimate
\begin{eqnarray}
& &\|([H^0_{J_{\bold{k},\bold{q}}}]_1)^{\frac{1}{2}}\, (\frac{1}{[\pi_{J_{\bold{k},\bold{q}}}]_1})^{\frac{1}{2}}\frac{1}{G_{J_{\bold{k},\bold{q}}}-E_{J_{\bold{k},\bold{q}}}}\,P^{(+)}_{J_{\bold{k},\bold{q}}}\,V^{(\bold{k},\bold{q})_{-1}}_{J_{\bold{k},\bold{q}}}\,P^{(-)}_{J_{\bold{k},\bold{q}}}P^{(-)}_{J_{\bold{r},\bold{i}}}\|\label{ineq-lemma-hs}\\
&\leq &
 C\cdot  \| (\frac{1}{[\pi_{J_{\bold{k},\bold{q}}}]_1})^{\frac{1}{2}}\,\,(\frac{1}{[H^0_{J_{\bold{k},\bold{q}}}]_1})^{\frac{1}{2}}\,P^{(+)}_{J_{\bold{k},\bold{q}}}\,V^{(\bold{k},\bold{q})_{-1}}_{J_{\bold{k},\bold{q}}}\,P^{(-)}_{J_{\bold{k},\bold{q}}}\|\,.
\end{eqnarray}

For the rest of the expression, first, in order to streamline the notation, we define
\begin{equation}
T_{J_{\bold{r},\bold{i}}}^{J_{\bold{k},\bold{q}},J_{\bold{k}',\bold{q}'}}:= (\frac{[H^0_{J_{\bold{k}',\bold{q}'}}]_1}{[H^0_{J_{\bold{r},\bold{i}}\setminus J_{\bold{k},\bold{q}}}]_1})^{\frac{1}{2}}(\frac{[\pi_{J_{\bold{k},\bold{q}}}]_1}{[H^0_{J_{\bold{k},\bold{q}}}]_1})^{\frac{1}{2}}
\end{equation}
and  we introduce the symbols
  \begin{equation}
\overline{\sum}_{J_{\bold{k}',\bold{q}' }\,,\,J_{\bold{k}'',\bold{q}''}} =\sum_{J_{\bold{k}',\bold{q}'}\,,\,J_{\bold{k}'',\bold{q}''}\,\in (\mathcal{G}^{(\bold{k},\bold{q})}_{J_{\bold{r};\bold{i}}})_{small}\,;\, J_{\bold{k}',\bold{q}'}\cap J_{\bold{k}'',\bold{q}''}=\emptyset}
  \end{equation}
  and
   \begin{equation}
  \sum'_{J_{\bold{k}',\bold{q}' }\,,\,J_{\bold{k}'',\bold{q}''}}  =\sum_{J_{\bold{k}',\bold{q}'}\,,\,J_{\bold{k}'',\bold{q}''}\,\in (\mathcal{G}^{(\bold{k},\bold{q})}_{J_{\bold{r};\bold{i}}})_{small}\,;\, J_{\bold{k}',\bold{q}'}\cap J_{\bold{k}'',\bold{q}''}\neq\emptyset}\,.
  \end{equation}
We can write
  \begin{eqnarray}
& &\|\sum_{J_{\bold{k}',\bold{q}'}\in (\mathcal{G}^{(\bold{k},\bold{q})}_{J_{\bold{r};\bold{i}}})_{small}}\,(\frac{1}{[\pi_{J_{\bold{r},\bold{i}}}]_1})^{\frac{1}{2}}\, (\frac{1}{[H^0_{J_{\bold{r},\bold{i}}}]_1})^{\frac{1}{2}}\,\,P^{(+)}_{J_{\bold{r},\bold{i}}}\,\,P^{(+)}_{J_{\bold{k}',\bold{q}'}}\Big(V^{(\bold{k},\bold{q})_{-1}}_{J_{\bold{k}',\bold{q}'}}-\langle V^{(\bold{k},\bold{q})_{-1}}_{J_{\bold{k}',\bold{q}'}}\rangle\Big)\,P^{(+)}_{J_{\bold{k}',\bold{q}'}}   (\frac{1}{[H^0_{J_{\bold{k}',\bold{q}'}}]_1})^{\frac{1}{2}}\,T_{J_{\bold{r},\bold{i}}}^{J_{\bold{k},\bold{q}},J_{\bold{k}',\bold{q}'}}\|^2 \nonumber \\
& \le&\sup_{\| \psi\|=1}\,\overline{\sum}_{J_{\bold{k}',\bold{q}' \,,\,J_{\bold{k}'',\bold{q}''}}}  \,\Big|\langle (\frac{1}{[\pi_{J_{\bold{r},\bold{i}}}]_1})^{\frac{1}{2}}\, (\frac{1}{[H^0_{J_{\bold{r},\bold{i}}}]_1})^{\frac{1}{2}}\,\psi\,,\label{nonitersect}\,\\
& &\quad\quad P^{(+)}_{J_{\bold{k}',\bold{q}'}}\Big(V^{(\bold{k},\bold{q})_{-1}}_{J_{\bold{k}',\bold{q}'}}-\langle V^{(\bold{k},\bold{q})_{-1}}_{J_{\bold{k}',\bold{q}'}}\rangle\Big)\,P^{(+)}_{J_{\bold{k}',\bold{q}'}}(\frac{1}{[H^0_{J_{\bold{k}',\bold{q}'}}]_1})^{\frac{1}{2}}\,T_{J_{\bold{r},\bold{i}}}^{J_{\bold{k},\bold{q}},J_{\bold{k}',\bold{q}'}}
\nonumber \\
& &\quad\quad\quad\times \,T_{J_{\bold{r},\bold{i}}}^{J_{\bold{k},\bold{q}},J_{\bold{k}'',\bold{q}''}}\,(\frac{1}{[H^0_{J_{\bold{k}'',\bold{q}''}}]_1})^{\frac{1}{2}}\, 
P^{(+)}_{J_{\bold{k}'',\bold{q}''}}\Big(V^{(\bold{k},\bold{q})_{-1}}_{J_{\bold{k}'',\bold{q}''}}-\langle V^{(\bold{k},\bold{q})_{-1}}_{J_{\bold{k}'',\bold{q}''}}\rangle\Big)\,P^{(+)}_{J_{\bold{k}'',\bold{q}''}}\,(\frac{1}{[\pi_{J_{\bold{r},\bold{i}}}]_1})^{\frac{1}{2}}\, (\frac{1}{[H^0_{J_{\bold{r},\bold{i}}}]_1})^{\frac{1}{2}}\,\psi \rangle\Big| \nonumber \\
& &+\sup_{\| \psi\|=1}\,  \sum'_{J_{\bold{k}',\bold{q}' }\,,\,J_{\bold{k}'',\bold{q}''}} \,\Big|\langle (\frac{1}{[\pi_{J_{\bold{r},\bold{i}}}]_1})^{\frac{1}{2}}\, (\frac{1}{[H^0_{J_{\bold{r},\bold{i}}}]_1})^{\frac{1}{2}}\,\psi\,,\label{intersect}\,\\
& &\quad\quad P^{(+)}_{J_{\bold{k}',\bold{q}'}}\Big(V^{(\bold{k},\bold{q})_{-1}}_{J_{\bold{k}',\bold{q}'}}-\langle V^{(\bold{k},\bold{q})_{-1}}_{J_{\bold{k}',\bold{q}'}}\rangle\Big)\,P^{(+)}_{J_{\bold{k}',\bold{q}'}} (\frac{1}{[H^0_{J_{\bold{k}',\bold{q}'}}]_1})^{\frac{1}{2}}\,T_{J_{\bold{r},\bold{i}}}^{J_{\bold{k},\bold{q}},J_{\bold{k}',\bold{q}'}}
\nonumber\\
& &\quad\quad\quad\times \,T_{J_{\bold{r},\bold{i}}}^{J_{\bold{k},\bold{q}},J_{\bold{k}'',\bold{q}''}}\,(\frac{1}{[H^0_{J_{\bold{k}'',\bold{q}''}}]_1})^{\frac{1}{2}}\, 
P^{(+)}_{J_{\bold{k}'',\bold{q}''}}\Big(V^{(\bold{k},\bold{q})_{-1}}_{J_{\bold{k}'',\bold{q}''}}-\langle V^{(\bold{k},\bold{q})_{-1}}_{J_{\bold{k}'',\bold{q}''}}\rangle\Big)\,P^{(+)}_{J_{\bold{k}'',\bold{q}''}}\,(\frac{1}{[\pi_{J_{\bold{r},\bold{i}}}]_1})^{\frac{1}{2}}\, (\frac{1}{[H^0_{J_{\bold{r},\bold{i}}}]_1})^{\frac{1}{2}}\,\psi \rangle\Big| \nonumber 
\end{eqnarray}
    
\noindent
\emph{Leading terms in $(\ref{rest-2})$: Contribution proportional to (\ref{nonitersect})}

\noindent
We observe that by exploiting the identities
\begin{equation}
P^{(+)}_{J_{\bold{k}',\bold{q}'}} \,= \, P^{(+)}_{J_{\bold{k}',\bold{q}'}} (\frac{1}{[H^0_{J_{\bold{k}',\bold{q}'}}]_1})^{\frac{1}{2}}\,(\frac{[H^0_{J_{\bold{k}',\bold{q}'}}]_1}{H^0_{J_{\bold{k}',\bold{q}'}}})^{\frac{1}{2}}\,  (H^0_{J_{\bold{k}',\bold{q}'}})^{\frac{1}{2}}
\end{equation}
\begin{equation}
P^{(+)}_{J_{\bold{k}'',\bold{q}'}} \,= \, P^{(+)}_{J_{\bold{k}'',\bold{q}'}} \, (\frac{1}{[H^0_{J_{\bold{k}'',\bold{q}''}}]_1})^{\frac{1}{2}}\,(\frac{[H^0_{J_{\bold{k}'',\bold{q}''}}]_1}{H^0_{J_{\bold{k}'',\bold{q}''}}})^{\frac{1}{2}}\,  (H^0_{J_{\bold{k}'',\bold{q}''}})^{\frac{1}{2}}
\end{equation}
we can write
\begin{eqnarray}
& &\sup_{\| \psi\|=1}\,\overline{\sum}_{J_{\bold{k}',\bold{q}' \,,\,J_{\bold{k}'',\bold{q}''}}}  \,\Big|\langle (\frac{1}{[\pi_{J_{\bold{r},\bold{i}}}]_1})^{\frac{1}{2}}\,(\frac{1}{[H^0_{J_{\bold{r},\bold{i}}}]_1})^{\frac{1}{2}}\,\psi\,,\label{nonitersect-bis}\,\\
& &\quad\quad P^{(+)}_{J_{\bold{k}',\bold{q}'}}\Big(V^{(\bold{k},\bold{q})_{-1}}_{J_{\bold{k}',\bold{q}'}}-\langle V^{(\bold{k},\bold{q})_{-1}}_{J_{\bold{k}',\bold{q}'}}\rangle\Big)\,P^{(+)}_{J_{\bold{k}',\bold{q}'}} (\frac{1}{[H^0_{J_{\bold{k}',\bold{q}'}}]_1})^{\frac{1}{2}}\,T_{J_{\bold{r},\bold{i}}}^{J_{\bold{k},\bold{q}},J_{\bold{k}',\bold{q}'}}
\\
& &\quad\quad\quad\times \,T_{J_{\bold{r},\bold{i}}}^{J_{\bold{k},\bold{q}},J_{\bold{k}'',\bold{q}''}}\,(\frac{1}{[H^0_{J_{\bold{k}'',\bold{q}''}}]_1})^{\frac{1}{2}}\, 
P^{(+)}_{J_{\bold{k}'',\bold{q}''}}\Big(V^{(\bold{k},\bold{q})_{-1}}_{J_{\bold{k}'',\bold{q}''}}-\langle V^{(\bold{k},\bold{q})_{-1}}_{J_{\bold{k}'',\bold{q}''}}\rangle\Big)\,P^{(+)}_{J_{\bold{k}'',\bold{q}''}}\,(\frac{1}{[\pi_{J_{\bold{r},\bold{i}}}]_1})^{\frac{1}{2}}\, (\frac{1}{[H^0_{J_{\bold{r},\bold{i}}}]_1})^{\frac{1}{2}}\,\psi \rangle\Big| \nonumber  \\
&=&\sup_{\| \psi\|=1}\,\overline{\sum}_{J_{\bold{k}',\bold{q}' \,,\,J_{\bold{k}'',\bold{q}''}}}  \,\Big|\langle (H^0_{J_{\bold{k}',\bold{q}'}})^{\frac{1}{2}}\,(\frac{1}{[\pi_{J_{\bold{r},\bold{i}}}]_1})^{\frac{1}{2}}\, (\frac{1}{[H^0_{J_{\bold{r},\bold{i}}}]_1})^{\frac{1}{2}}\,\psi\,,\label{nonitersect-bis-bis}\,\\
& &\quad\quad  (\frac{[H^0_{J_{\bold{k}',\bold{q}'}}]_1}{H^0_{J_{\bold{k}',\bold{q}'}}})^{\frac{1}{2}}\, (\frac{1}{[H^0_{J_{\bold{k}',\bold{q}'}}]_1})^{\frac{1}{2}}\,P^{(+)}_{J_{\bold{k}',\bold{q}'}}\Big(V^{(\bold{k},\bold{q})_{-1}}_{J_{\bold{k}',\bold{q}'}}-\langle V^{(\bold{k},\bold{q})_{-1}}_{J_{\bold{k}',\bold{q}'}}\rangle\Big)\,P^{(+)}_{J_{\bold{k}',\bold{q}'}} (\frac{1}{[H^0_{J_{\bold{k}',\bold{q}'}}]_1})^{\frac{1}{2}}\,\nonumber \\
& &\quad\quad\quad  \times T_{J_{\bold{r},\bold{i}}}^{J_{\bold{k},\bold{q}},J_{\bold{k}',\bold{q}'}} \,T_{J_{\bold{r},\bold{i}}}^{J_{\bold{k},\bold{q}},J_{\bold{k}'',\bold{q}''}} \nonumber \\
& &\times \,(\frac{1}{[H^0_{J_{\bold{k}'',\bold{q}''}}]_1})^{\frac{1}{2}}\, 
P^{(+)}_{J_{\bold{k}'',\bold{q}''}}\Big(V^{(\bold{k},\bold{q})_{-1}}_{J_{\bold{k}'',\bold{q}''}}-\langle V^{(\bold{k},\bold{q})_{-1}}_{J_{\bold{k}'',\bold{q}''}}\rangle\Big)\,P^{(+)}_{J_{\bold{k}'',\bold{q}''}}\,(\frac{1}{[H^0_{J_{\bold{k}'',\bold{q}''}}]_1})^{\frac{1}{2}}\, (\frac{[H^0_{J_{\bold{k}'',\bold{q}''}}]_1}{H^0_{J_{\bold{k}'',\bold{q}''}}})^{\frac{1}{2}}\nonumber  \\
& &\times  (H^0_{J_{\bold{k}'',\bold{q}''}})^{\frac{1}{2}}\,(\frac{1}{[\pi_{J_{\bold{r},\bold{i}}}]_1})^{\frac{1}{2}}\, (\frac{1}{[H^0_{J_{\bold{r},\bold{i}}}]_1})^{\frac{1}{2}}\,\psi \rangle\Big| . \nonumber
\end{eqnarray}
Next we observe that for $J_{\bold{k}',\bold{q}'}\cap J_{\bold{k}'',\bold{q}''}=\emptyset$, since 
 $$[P^{(+)}_{J_{\bold{k}',\bold{q}'}}\,,\,\Big(V^{(\bold{k},\bold{q})_{-1}}_{J_{\bold{k}'',\bold{q}''}}-\langle V^{(\bold{k},\bold{q})_{-1}}_{J_{\bold{k}'',\bold{q}''}}\rangle\Big)]=[\Big(V^{(\bold{k},\bold{q})_{-1}}_{J_{\bold{k}',\bold{q}'}}-\langle V^{(\bold{k},\bold{q})_{-1}}_{J_{\bold{k}',\bold{q}'}}\rangle\Big)\,,\,P^{(+)}_{J_{\bold{k}'',\bold{q}''}}]=0\,,$$
  we have
\begin{eqnarray}
& &P^{(+)}_{J_{\bold{k}',\bold{q}'}}\Big(V^{(\bold{k},\bold{q})_{-1}}_{J_{\bold{k}',\bold{q}'}}-\langle V^{(\bold{k},\bold{q})_{-1}}_{J_{\bold{k}',\bold{q}'}}\rangle\Big)\,P^{(+)}_{J_{\bold{k}',\bold{q}'}}P^{(+)}_{J_{\bold{k}'',\bold{q}''}}\Big(V^{(\bold{k},\bold{q})_{-1}}_{J_{\bold{k}'',\bold{q}''}}-\langle V^{(\bold{k},\bold{q})_{-1}}_{J_{\bold{k}'',\bold{q}''}}\rangle\Big)\,P^{(+)}_{J_{\bold{k}'',\bold{q}''}}\\
&=&P^{(+)}_{J_{\bold{k}',\bold{q}'}}P^{(+)}_{J_{\bold{k}'',\bold{q}''}}\\
& & \times \Big(V^{(\bold{k},\bold{q})_{-1}}_{J_{\bold{k}',\bold{q}'}}-\langle V^{(\bold{k},\bold{q})_{-1}}_{J_{\bold{k}',\bold{q}'}}\rangle\Big)\,P^{(+)}_{J_{\bold{k}',\bold{q}'}}P^{(+)}_{J_{\bold{k}'',\bold{q}''}}\Big(V^{(\bold{k},\bold{q})_{-1}}_{J_{\bold{k}'',\bold{q}''}}-\langle V^{(\bold{k},\bold{q})_{-1}}_{J_{\bold{k}'',\bold{q}''}}\rangle\Big)\,\\
& & \times  \,P^{(+)}_{J_{\bold{k}',\bold{q}'}}P^{(+)}_{J_{\bold{k}'',\bold{q}''}}\,.  \end{eqnarray}
From this observation we deduce that
\begin{eqnarray}
& &(\ref{nonitersect})\\
&= &\sup_{\| \psi\|=1}\,\overline{\sum}_{J_{\bold{k}',\bold{q}' \,,\,J_{\bold{k}'',\bold{q}''}}}  \,\Big|\langle P^{(+)}_{J_{\bold{k}'',\bold{q}''}}(H^0_{J_{\bold{k}',\bold{q}'}})^{\frac{1}{2}}\,(\frac{1}{[\pi_{J_{\bold{r},\bold{i}}}]_1})^{\frac{1}{2}}\, (\frac{1}{[H^0_{J_{\bold{r},\bold{i}}}]_1})^{\frac{1}{2}}\,\psi\,,\\
& &\quad\quad  (\frac{[H^0_{J_{\bold{k}',\bold{q}'}}]_1}{H^0_{J_{\bold{k}',\bold{q}'}}})^{\frac{1}{2}}\, (\frac{1}{[H^0_{J_{\bold{k}',\bold{q}'}}]_1})^{\frac{1}{2}}\,P^{(+)}_{J_{\bold{k}',\bold{q}'}}\Big(V^{(\bold{k},\bold{q})_{-1}}_{J_{\bold{k}',\bold{q}'}}-\langle V^{(\bold{k},\bold{q})_{-1}}_{J_{\bold{k}',\bold{q}'}}\rangle\Big)\,P^{(+)}_{J_{\bold{k}',\bold{q}'}} (\frac{1}{[H^0_{J_{\bold{k}',\bold{q}'}}]_1})^{\frac{1}{2}}\,\nonumber \\
& &\quad\quad\quad  \times T_{J_{\bold{r},\bold{i}}}^{J_{\bold{k},\bold{q}},J_{\bold{k}',\bold{q}'}} \,T_{J_{\bold{r},\bold{i}}}^{J_{\bold{k},\bold{q}},J_{\bold{k}'',\bold{q}''}} \nonumber \\
& &\times \,(\frac{1}{[H^0_{J_{\bold{k}'',\bold{q}''}}]_1})^{\frac{1}{2}}\, 
P^{(+)}_{J_{\bold{k}'',\bold{q}''}}\Big(V^{(\bold{k},\bold{q})_{-1}}_{J_{\bold{k}'',\bold{q}''}}-\langle V^{(\bold{k},\bold{q})_{-1}}_{J_{\bold{k}'',\bold{q}''}}\rangle\Big)\,P^{(+)}_{J_{\bold{k}'',\bold{q}''}}\,(\frac{1}{[H^0_{J_{\bold{k}'',\bold{q}''}}]_1})^{\frac{1}{2}}\, (\frac{[H^0_{J_{\bold{k}'',\bold{q}''}}]_1}{H^0_{J_{\bold{k}'',\bold{q}''}}})^{\frac{1}{2}}\nonumber  \\
& &\times  P^{(+)}_{J_{\bold{k}',\bold{q}'}}\,(H^0_{J_{\bold{k}'',\bold{q}''}})^{\frac{1}{2}}\,(\frac{1}{[\pi_{J_{\bold{r},\bold{i}}}]_1})^{\frac{1}{2}}\, (\frac{1}{[H^0_{J_{\bold{r},\bold{i}}}]_1})^{\frac{1}{2}}\,\psi \rangle\Big|\,. \nonumber
\end{eqnarray}
Hence, we can estimate
\begin{eqnarray}
& &(\ref{nonitersect})\\
&= &\sup_{\| \psi\|=1}\,\overline{\sum}_{J_{\bold{k}',\bold{q}' \,,\,J_{\bold{k}'',\bold{q}''}}}  \,\|P^{(+)}_{J_{\bold{k}'',\bold{q}''}}(H^0_{J_{\bold{k}',\bold{q}'}})^{\frac{1}{2}}\,(\frac{1}{[\pi_{J_{\bold{r},\bold{i}}}]_1})^{\frac{1}{2}}\, (\frac{1}{[H^0_{J_{\bold{r},\bold{i}}}]_1})^{\frac{1}{2}}\,\psi \| \\
& &\quad\quad \times \| (\frac{[H^0_{J_{\bold{k}',\bold{q}'}}]_1}{H^0_{J_{\bold{k}',\bold{q}'}}})^{\frac{1}{2}}\,P^{(+)}_{J_{\bold{k}',\bold{q}'}}\|\cdot \| (\frac{1}{[H^0_{J_{\bold{k}',\bold{q}'}}]_1})^{\frac{1}{2}}\,P^{(+)}_{J_{\bold{k}',\bold{q}'}}\Big(V^{(\bold{k},\bold{q})_{-1}}_{J_{\bold{k}',\bold{q}'}}-\langle V^{(\bold{k},\bold{q})_{-1}}_{J_{\bold{k}',\bold{q}'}}\rangle\Big)\,P^{(+)}_{J_{\bold{k}',\bold{q}'}} (\frac{1}{[H^0_{J_{\bold{k}',\bold{q}'}}]_1})^{\frac{1}{2}}\|\,\nonumber \\
& &\quad\quad\quad  \times \|T_{J_{\bold{r},\bold{i}}}^{J_{\bold{k},\bold{q}},J_{\bold{k}',\bold{q}'}} \,T_{J_{\bold{r},\bold{i}}}^{J_{\bold{k},\bold{q}},J_{\bold{k}'',\bold{q}''}}\| \nonumber \\
& &\times \, \|(\frac{1}{[H^0_{J_{\bold{k}'',\bold{q}''}}]_1})^{\frac{1}{2}}\, 
P^{(+)}_{J_{\bold{k}'',\bold{q}''}}\Big(V^{(\bold{k},\bold{q})_{-1}}_{J_{\bold{k}'',\bold{q}''}}-\langle V^{(\bold{k},\bold{q})_{-1}}_{J_{\bold{k}'',\bold{q}''}}\rangle\Big)\,P^{(+)}_{J_{\bold{k}'',\bold{q}''}}\,(\frac{1}{[H^0_{J_{\bold{k}'',\bold{q}''}}]_1})^{\frac{1}{2}}\,\|\,  \|(\frac{[H^0_{J_{\bold{k}'',\bold{q}''}}]_1}{H^0_{J_{\bold{k}'',\bold{q}''}}})^{\frac{1}{2}}\, P^{(+)}_{J_{\bold{k}'',\bold{q}''}}\| \nonumber  \\
& &\times \|P^{(+)}_{J_{\bold{k}',\bold{q}'}}\,(H^0_{J_{\bold{k}'',\bold{q}''}})^{\frac{1}{2}}\,(\frac{1}{[\pi_{J_{\bold{r},\bold{i}}}]_1})^{\frac{1}{2}}\, (\frac{1}{[H^0_{J_{\bold{r},\bold{i}}}]_1})^{\frac{1}{2}}\,\psi \| \nonumber \\
&\leq &\sup_{\| \psi\|=1}\,\sum_{J_{\bold{k}',\bold{q}' }\,,\,J_{\bold{k}'',\bold{q}''} \in  \,\mathcal{G}^{(\bold{k},\bold{q})}_{J_{\bold{r},\bold{i}}}}  \, 8 \|V^{(\bold{k},\bold{q})_{-1}}_{J_{\bold{k}',\bold{q}'}}\|_{H^0}\cdot \|V^{(\bold{k},\bold{q})_{-1}}_{J_{\bold{k}'',\bold{q}''}}\|_{H^0}
\cdot \| T_{J_{\bold{r},\bold{i}}}^{J_{\bold{k},\bold{q}},J_{\bold{k}',\bold{q}'}} \| \cdot \| T_{J_{\bold{r},\bold{i}}}^{J_{\bold{k},\bold{q}},J_{\bold{k}'',\bold{q}''}}\| \\
& &\quad \times \Big\{ \frac{1}{2}\Big\|\frac{(H^0_{J_{\bold{k}',\bold{q}'}})^{\frac{1}{2}}\,P^{(+)}_{J_{\bold{k}'',\bold{q}''}}}{(\sum_{\bold{j}\in J_{\bold{r},\bold{i}}} P^{\perp}_{\Omega_{\bold{j}}}+1)^{\frac{1}{2}}(\sum_{\bold{j}\in J_{\bold{r},\bold{i}}} H_{\bold{j}}+1)^{\frac{1}{2}}}\,\psi\Big\|^2+ \frac{1}{2}\Big\|\frac{(H^0_{J_{\bold{k}'',\bold{q}''}})^{\frac{1}{2}}\,P^{(+)}_{J_{\bold{k}',\bold{q}'}}}{(\sum_{\bold{j}\in J_{\bold{r},\bold{i}}} P^{\perp}_{\Omega_{\bold{j}}}+1)^{\frac{1}{2}}(\sum_{\bold{j}\in J_{\bold{r},\bold{i}}} H_{\bold{j}}+1)^{\frac{1}{2}}}\,\psi\Big\|^2\Big\} \nonumber \,
\end{eqnarray}
where we have used that
\begin{equation}
 \|(\frac{[H^0_{J_{\bold{k}'',\bold{q}''}}]_1}{H^0_{J_{\bold{k}'',\bold{q}''}}})^{\frac{1}{2}}\, P^{(+)}_{J_{\bold{k}'',\bold{q}''}}\| \leq \sqrt{2}\,.
\end{equation}
It is enough to study
\begin{eqnarray}
& & \sup_{\| \psi\|=1}\,\sum_{J_{\bold{k}',\bold{q}' }\,,\,J_{\bold{k}'',\bold{q}''} \in  \,\mathcal{G}^{(\bold{k},\bold{q})}_{J_{\bold{r},\bold{i}}}} 
 \, 4 \|V^{(\bold{k},\bold{q})_{-1}}_{J_{\bold{k}',\bold{q}'}}\|_{H^0}
\cdot \| T_{J_{\bold{r},\bold{i}}}^{J_{\bold{k},\bold{q}},J_{\bold{k}',\bold{q}'}}\|\cdot \|V^{(\bold{k},\bold{q})_{-1}}_{J_{\bold{k}'',\bold{q}''}}\|_{H^0}
\cdot \| T_{J_{\bold{r},\bold{i}}}^{J_{\bold{k},\bold{q}},J_{\bold{k}'',\bold{q}''}}\| \quad \label{nonintersect-norms}\\
& & \times  \langle \psi\,,\, \frac{H^0_{J_{\bold{k}'',\bold{q}''}}\,P^{(+)}_{J_{\bold{k}',\bold{q}'}}}{(\sum_{\bold{j}\in J_{\bold{r},\bold{i}}} H_{\bold{j}}+1)\,(\sum_{\bold{j}\in J_{\bold{r},\bold{i}}} P^{\perp}_{\Omega_{\bold{j}}}+1)}\,\psi \rangle . \,
\end{eqnarray}
We observe that 
\begin{itemize}
\item
\begin{equation}
 \langle \psi\,,\, \frac{H^0_{J_{\bold{k}'',\bold{q}''}}\,P^{(+)}_{J_{\bold{k}',\bold{q}'}}}{(\sum_{\bold{j}\in J_{\bold{r},\bold{i}}} H_{\bold{j}}+1)\,(\sum_{\bold{j}\in J_{\bold{r},\bold{i}}} P^{\perp}_{\Omega_{\bold{j}}}+1)}\,\psi \rangle\geq 0
\end{equation}
\item
by (inductive) hypothesis the estimate, $\mathcal{E}^{(\bold{k},\bold{q})_{-1}}_{k'}$,  of the norm $\|V^{(\bold{k},\bold{q})_{-1}}_{J_{\bold{k}',\bold{q}'}}\|_{H^0}$ does not depend on $\bold{q}'$, i.e.,
\begin{equation}
\|V^{(\bold{k},\bold{q})_{-1}}_{J_{\bold{k}',\bold{q}'}}\|_{H^0}\leq \mathcal{E}^{(\bold{k},\bold{q})_{-1}}_{k'}\,,
\end{equation}
 furthermore 
\begin{equation}
\| T_{J_{\bold{r},\bold{i}}}^{J_{\bold{k},\bold{q}},J_{\bold{k}',\bold{q}'}}\|\leq c_d \cdot (n_{J_{\bold{k},\bold{q}}\cap J_{\bold{k}',\bold{q}'}})^{\frac{1}{2}}
\end{equation} where $n_{J_{\bold{k},\bold{q}}\cap J_{\bold{k}',\bold{q}'}}$ is the number of sites in $J_{\bold{k},\bold{q}}\cap J_{\bold{k}',\bold{q}'}$. 
We define the subset
\begin{equation}
[\mathcal{G}^{(\bold{k},\bold{q})}_{J_{\bold{r},\bold{i}}}]_n:=\{J_{\bold{k}',\bold{q}'}\in \mathcal{G}^{(\bold{k},\bold{q})}_{J_{\bold{r},\bold{i}}}\quad : \quad n_{J_{\bold{k},\bold{q}}\cap J_{\bold{k}',\bold{q}'}}=n\}
\end{equation}
and $n_{\bold{k},\bold{q}}$ the number of sites in $J_{\bold{k},\bold{q}}$.
\end{itemize}
Then it is convenient to re-write
\begin{eqnarray}
& & \sup_{\| \psi\|=1}\,\sum_{J_{\bold{k}',\bold{q}' }\,,\,J_{\bold{k}'',\bold{q}''} \in  \,\mathcal{G}^{(\bold{k},\bold{q})}_{J_{\bold{r},\bold{i}}}} 
 \, 4 \|V^{(\bold{k},\bold{q})_{-1}}_{J_{\bold{k}',\bold{q}'}}\|_{H^0}
\cdot \| T_{J_{\bold{r},\bold{i}}}^{J_{\bold{k},\bold{q}},J_{\bold{k}',\bold{q}'}}\|\cdot \|V^{(\bold{k},\bold{q})_{-1}}_{J_{\bold{k}'',\bold{q}''}}\|_{H^0}
\cdot \| T_{J_{\bold{r},\bold{i}}}^{J_{\bold{k},\bold{q}},J_{\bold{k}'',\bold{q}''}}\|\quad \nonumber\\
& & \times  \langle \psi\,,\, \frac{H^0_{J_{\bold{k}'',\bold{q}''}}\,P^{(+)}_{J_{\bold{k}',\bold{q}'}}}{(\sum_{\bold{j}\in J_{\bold{r},\bold{i}}} H_{\bold{j}}+1)\,(\sum_{\bold{j}\in J_{\bold{r},\bold{i}}} P^{\perp}_{\Omega_{\bold{j}}}+1)}\,\psi \rangle \,\\
&\leq& \sup_{\| \psi\|=1}\,\sum_{n=1}^{n_{\bold{k,\bold{q}}}} \,\sum_{\bold{s}:\exists \bold{u} \text{ with }J_{\bold{s},\bold{u}} \,\in \,[\mathcal{G}^{(\bold{k},\bold{q})}_{J_{\bold{r},\bold{i}}}]_n}\,\,\sum_{m=1}^{n_{\bold{k,\bold{q}}}}\,\sum_{\bold{h}:\exists \bold{p} \text{ with }J_{\bold{h},\bold{p}} \,\in \,[\mathcal{G}^{(\bold{k},\bold{q})}_{J_{\bold{r},\bold{i}}}]_m}\, \\
& &\quad \times (c_d\cdot 2\cdot  \mathcal{E}^{(\bold{k},\bold{q})_{-1}}_{s}\cdot n^{\frac{1}{2}})\cdot (c_d\cdot 2\cdot  \mathcal{E}^{(\bold{k},\bold{q})_{-1}}_{h}\cdot m^{\frac{1}{2}})\\
& &\quad \times  \langle \psi\,,\, \frac{\Big(\sum_{\bold{p}\,:\,\, J_{\bold{h},\bold{p}} \,\in \,[\mathcal{G}^{(\bold{k},\bold{q})}_{J_{\bold{r},\bold{i}}}]_m}H^0_{J_{\bold{h},\bold{p}}}\Big)\, \Big(\sum_{\bold{u}\,:\,\, J_{\bold{s},\bold{u}} \,\in \,[\mathcal{G}^{(\bold{k},\bold{q})}_{J_{\bold{r},\bold{i}}}]_n}P^{(+)}_{J_{\bold{s},\bold{u}}}\Big)}{(\sum_{\bold{j}\in J_{\bold{r},\bold{i}}} H_{\bold{j}}+1)\,(\sum_{\bold{j}\in J_{\bold{r},\bold{i}}} P^{\perp}_{\Omega_{\bold{j}}}+1)}\,\psi \rangle\,.
\end{eqnarray}

Now assume that there are $1\leq l\leq d$ components of $\bold{k}$ different from the corresponding ones in $\bold{r}$, with no loss of generality we assume that these are the first $l$ components; we have
\begin{eqnarray}
& & \sup_{\| \psi\|=1}\,\sum_{n=1}^{n_{\bold{k,\bold{q}}}} \,\sum_{\bold{s}:\exists \bold{u} \text{ with }J_{\bold{s},\bold{u}} \,\in \,[\mathcal{G}^{(\bold{k},\bold{q})}_{J_{\bold{r},\bold{i}}}]_n}\,\,\sum_{m=1}^{n_{\bold{k,\bold{q}}}}\,\sum_{\bold{h}:\exists \bold{p} \text{ with }J_{\bold{h},\bold{p}} \,\in \,[\mathcal{G}^{(\bold{k},\bold{q})}_{J_{\bold{r},\bold{i}}}]_m}\, \\
& &\quad \times (c_d\cdot 2\cdot  \mathcal{E}^{(\bold{k},\bold{q})_{-1}}_{s}\cdot n^{\frac{1}{2}})\cdot (c_d\cdot 2\cdot  \mathcal{E}^{(\bold{k},\bold{q})_{-1}}_{h}\cdot m^{\frac{1}{2}})\\
& &\quad \times  \langle \psi\,,\, \frac{\Big(\sum_{\bold{p}\,:\,\, J_{\bold{h},\bold{p}} \,\in \,[\mathcal{G}^{(\bold{k},\bold{q})}_{J_{\bold{r},\bold{i}}}]_m}H^0_{J_{\bold{h},\bold{p}}}\Big)\, \Big(\sum_{\bold{u}\,:\,\, J_{\bold{s},\bold{u}} \,\in \,[\mathcal{G}^{(\bold{k},\bold{q})}_{J_{\bold{r},\bold{i}}}]_n}P^{(+)}_{J_{\bold{s},\bold{u}}}\Big)}{(\sum_{\bold{j}\in J_{\bold{r},\bold{i}}} H_{\bold{j}}+1)\,(\sum_{\bold{j}\in J_{\bold{r},\bold{i}}} P^{\perp}_{\Omega_{\bold{j}}}+1)}\,\psi \rangle\,\\
&\leq &C_d\Big\{  \sum_{m=1}^{n_{\bold{k,\bold{q}}}}m^{\frac{1}{2}}\,\Big\{ \sum_{s_{1}=r_1-k_1}^{r-1} \dots \sum_{s_{l}=r_l - k_l}^{r-1} \sum_{s_{l+1}=0}^{r}\dots \sum_{s_{d}=0}^{r} \,\theta
_{m}^{^{\bold{k},\bold{q}}}(\bold{s})\,\frac{t^{\left(\sum_{j=1}^{d}\frac{s_j}{3}\right)-\frac{1}{3}}}{(s_1+\dots +s_d)^{x_{d}}}\cdot \Big[  \prod_{j=l+1}^{d}(s_j+1)\Big]\Big\}\,\Big\}^2, \quad\quad\quad  \label{final}
\end{eqnarray}
where $\theta
_{m}^{^{\bold{k},\bold{q}}}(\bold{s})$ constraints the sums over $s_1,\dots ,s_d$ to the rectangles $J_{\bold{s},\bold{u}}$ such that $n_{J_{\bold{s},\bold{u}}\cap J_{\bold{k},\bold{q}}}=m$, and  in the step leading to  (\ref{final}) we use:
\begin{itemize}
\item
the "weight" 
 \begin{equation}
 \|V^{(\bold{k},\bold{q})_{-1}}_{J_{\bold{k}',\bold{q}'}}\|_{H^0}\leq 96\cdot \frac{t^{\frac{k'-1}{3}}}{(k')^{x_{d}}}
 \end{equation}
 
  by the inductive hypotheses (\ref{R3-2}) and (\ref{R3-3});
\item
for fixed $\bold{k}^\prime$, if $k_j\neq r_j$ for $j=1,\cdots ,l$ then $q^\prime_1,\cdots, q^\prime_l$ are uniquely determined by the condition $[ J_{\bold{k}',\bold{q}'} \cup J_{\bold{k},\bold{q}} ]=J_{\bold{r},\bold{i}}$; thus we have
$$
 \sum_{\bold{u}\,;\,u_{1},\dots,u_l=\text{fixed}\,,\, J_{\bold{s},\bold{u}} \,\in \,[\mathcal{G}^{(\bold{k},\bold{q})}_{J_{\bold{r},\bold{i}}}]_m}\,\,\,P^{(+)}_{J_{\bold{s},\bold{u}}}\,\leq \Big\{\prod_{j=l+1}^{d}(s_j+1)\Big\}\,\sum_{\bold{j}\in J_{\bold{r},\bold{i}}}  P^{\perp}_{\Omega_{\bold{j}}}$$
 $$
 \sum_{\bold{u}\,;\,u_{1},\dots,u_l=\text{fixed}\,,\, J_{\bold{s},\bold{u}} \,\in \,[\mathcal{G}^{(\bold{k},\bold{q})}_{J_{\bold{r},\bold{i}}}]_m}\,\,\,H^{0}_{J_{\bold{s},\bold{u}}}\,\leq \Big\{\prod_{j=l+1}^{d}(s_j+1)\Big\}\,\sum_{\bold{j}\in J_{\bold{r},\bold{i}}}  H_{\Omega_{\bold{j}}}$$

which can be obtained using the same reasoning of Corollary \ref{op-ineq-2}.
\end{itemize}

\noindent
Next, for $j=1,\dots,l$, set 
\begin{equation}\label{def-rho} 
\rho_j:=s_j-(r_j-k_j)\quad \rightarrow \quad s_j=\rho_j+(r_j-k_j)\,,
\end{equation}
and note that since $s_j\geq r_j-k_j$ for  $j=1,\dots,l$, and $s_j\geq 0$ for $j=l+1,\dots,d$, 
\begin{equation}
(s_1+\dots +s_d)^{x_{d}}\geq (r_1-k_1+\dots +r_l-k_l)^{x_{d}}\,.
\end{equation}
Next, we point out that if $J_{\bold{s},\bold{u}}\in [\mathcal{G}^{(\bold{k},\bold{q})}_{J_{\bold{r},\bold{i}}}]_m$ then, since 
\begin{equation}
m=n_{J_{\bold{s},\bold{u}}\cap J_{\bold{k},\bold{q}}}=\prod_{j=1}^l(\rho_j+1)\,\prod_{j=l+1}^d(s_j+1),
\end{equation}
we have
\begin{equation}
\sum_{j=1}^l\rho_j+\sum_{j=l+1}^ds_j+1\geq  m^{1/d},
\end{equation}
and using $(m^{1/d}-1)\geq a^\prime (m-1)^{1/d}$ where $a^\prime>0$ is a constant (dependent on $d$), we get (recalling that $t\in[0,1)$)
\begin{equation}
t^{\sum_{j=1}^{d}\frac{s_{j}}{3}}  \leq t^{\sum_{j=1}^{l}\frac{(r_j-k_j)}{3}}\cdot t^{\sum_{j=l+1}^{d}\frac{s_{j}}{6}}\cdot  t^{\sum_{j=1}^{l}\frac{\rho_{j}}{6}} \cdot t^{a \cdot (m-1)^{\frac{1}{d}}}
\end{equation}
where $a$ is a constant (dependent on $d$).

Hence we can estimate
\begin{eqnarray}
& &\Big( \sup_{\| \psi\|=1}\,\sum_{J_{\bold{k}',\bold{q}' }\,,\,J_{\bold{k}'',\bold{q}''} \in  \,\mathcal{G}^{(\bold{k},\bold{q})}_{J_{\bold{r},\bold{i}}}} 
 \, 4 \|V^{(\bold{k},\bold{q})_{-1}}_{J_{\bold{k}',\bold{q}'}}\|_{H^0}
\cdot \| T_{J_{\bold{r},\bold{i}}}^{J_{\bold{k},\bold{q}},J_{\bold{k}',\bold{q}'}}\|\cdot \|V^{(\bold{k},\bold{q})_{-1}}_{J_{\bold{k}'',\bold{q}''}}\|_{H^0}
\cdot \| T_{J_{\bold{r},\bold{i}}}^{J_{\bold{k},\bold{q}},J_{\bold{k}'',\bold{q}''}}\|  \label{3-1}\\
& &\quad\quad\quad \times  \langle \psi\,,\, \frac{H^0_{J_{\bold{k}'',\bold{q}''}}\,P^{(+)}_{J_{\bold{k}',\bold{q}'}}}{(\sum_{\bold{j}\in J_{\bold{r},\bold{i}}} H_{\bold{j}}+1)\,(\sum_{\bold{j}\in J_{\bold{r},\bold{i}}} P^{\perp}_{\Omega_{\bold{j}}}+1)}\,\psi \rangle\Big)^{\frac{1}{2}} \,\nonumber \\
& \leq &C\cdot t^{-1/3}\,t^{\sum_{j=1}^{l}\frac{(r_j-k_j)}{3}}\cdot \sum_{m=1}^{n_{\bold{k,\bold{q}}}} m^{\frac{1}{2}}\, t^{a \cdot (m-1)^{\frac{1}{d}}}\\
& &\quad \times\Big\{ \sum_{s_{l+1}=0}^{r}\dots \sum_{s_{d}=0}^{r} \frac{1}{(r_1-k_1+\dots +r_l-k_l)^{x_{d}}}\cdot \Big[ t^{\sum_{j=l+1}^{d}\frac{s_{j}}{6}}\cdot \prod_{j=l+1}^{d}(s_j+1)\Big]  \sum_{\rho_{1}=0}^{\infty}\dots \sum_{\rho_{l}=0}^{\infty} t^{\sum_{j=1}^{l}\frac{\rho_{j}}{6}} \Big\}\nonumber \\
& \leq & C\cdot t^{-1/3} t^{\sum_{j=1}^{l}\frac{(r_j-k_j)}{3}} \cdot \sum_{m=1}^{n_{\bold{k,\bold{q}}}} m^{\frac{1}{2}}\, t^{a \cdot (m-1)^{\frac{1}{d}}} \cdot  \frac{1}{(r_1-k_1+\dots +r_l-k_l)^{x_{d}}}\times \label{3-2,5}\\
& &\quad\quad\quad \times  \Big\{\sum_{s_{l+1}=0}^{r}\dots \sum_{s_{d}=0}^{r} \Big[ t^{\sum_{j=l+1}^{d}\frac{s_{j}}{6}}\cdot \prod_{j=l+1}^{d}(s_j+1)\Big] \sum_{\rho_{1}=0}^{\infty}\dots \sum_{\rho_{l}=0}^{\infty} t^{\sum_{j=1}^{l}\frac{\rho_{j}}{6}}  \Big\}\nonumber \\
&\leq &C_d\cdot  t^{-1/3} (\frac{t^{\frac{r-1}{3}}}{t^{\frac{k-1}{3}}}) \cdot \frac{1}{(r_1-k_1+\dots +r_l-k_l)^{x_{d}}} \label{3-3}
\end{eqnarray}
 where  for the step from (\ref{3-2,5}) to (\ref{3-3})
 we have exploited:
\begin{itemize}
\item
\begin{equation}
\sum_{s_{l+1}=0}^{r}\dots \sum_{s_{d}=0}^{r} \Big[ t^{\sum_{j=l+1}^{d}\frac{s_{j}}{6}}\cdot \prod_{j=l+1}^{d}(s_j+1)\Big] \sum_{\rho_{1}=0}^{\infty}\dots \sum_{\rho_{l}=0}^{\infty} t^{\sum_{j=1}^{l}\frac{\rho_{j}}{6}}
\end{equation}
is bounded from above by a constant (which depends on $d$);
\item
for the considered $\bold{k}$ 
 \begin{equation}
t^{\sum_{j=1}^{l}\frac{(r_j-k_j)}{3}}=t^{\frac{r-k}{3}}
 \end{equation}
since by assumption $k_{j}=r_j$ for $j=l+1,\dots,d$.
\end{itemize}

\noindent
\emph{Leading terms in $(\ref{rest-2})$: Contribution proportional to (\ref{intersect})}
\\

\noindent
We estimate
\begin{eqnarray}
& &\sup_{\| \psi\|=1}\,  \sum'_{J_{\bold{k}',\bold{q}' }\,,\,J_{\bold{k}'',\bold{q}''}} \,\Big|\langle (\frac{1}{[\pi_{J_{\bold{r},\bold{i}}}]_1})^{\frac{1}{2}}\, (\frac{1}{[H^0_{J_{\bold{r},\bold{i}}}]_1})^{\frac{1}{2}}\,\psi\,,\label{intersect-bis}\,\\
& &\quad\quad P^{(+)}_{J_{\bold{k}',\bold{q}'}}\Big(V^{(\bold{k},\bold{q})_{-1}}_{J_{\bold{k}',\bold{q}'}}-\langle V^{(\bold{k},\bold{q})_{-1}}_{J_{\bold{k}',\bold{q}'}}\rangle\Big)\,P^{(+)}_{J_{\bold{k}',\bold{q}'}}(\frac{1}{[H^0_{J_{\bold{k}',\bold{q}'}}]_1})^{\frac{1}{2}}\,T_{J_{\bold{r},\bold{i}}}^{J_{\bold{k},\bold{q}},J_{\bold{k}',\bold{q}'}}
\nonumber\\
& &\quad\quad\quad\times \,T_{J_{\bold{r},\bold{i}}}^{J_{\bold{k},\bold{q}},J_{\bold{k}'',\bold{q}''}}\,(\frac{1}{[H^0_{J_{\bold{k}'',\bold{q}''}}]_1})^{\frac{1}{2}}\, 
P^{(+)}_{J_{\bold{k}'',\bold{q}''}}\Big(V^{(\bold{k},\bold{q})_{-1}}_{J_{\bold{k}'',\bold{q}''}}-\langle V^{(\bold{k},\bold{q})_{-1}}_{J_{\bold{k}'',\bold{q}''}}\rangle\Big)\,P^{(+)}_{J_{\bold{k}'',\bold{q}''}}\,(\frac{1}{[\pi_{J_{\bold{r},\bold{i}}}]_1})^{\frac{1}{2}}\, (\frac{1}{[H^0_{J_{\bold{r},\bold{i}}}]_1})^{\frac{1}{2}}\,\psi \rangle\Big| \quad\quad\quad \nonumber\\
 &\leq &\sup_{\| \psi\|=1}\,  \sum'_{J_{\bold{k}',\bold{q}' }\,,\,J_{\bold{k}'',\bold{q}''}}  \, \, 8 \|V^{(\bold{k},\bold{q})_{-1}}_{J_{\bold{k}',\bold{q}'}}\|_{H^0}\cdot \|V^{(\bold{k},\bold{q})_{-1}}_{J_{\bold{k}'',\bold{q}''}}\|_{H^0}
\cdot \| T_{J_{\bold{r},\bold{i}}}^{J_{\bold{k},\bold{q}},J_{\bold{k}',\bold{q}'}} \| \cdot \| T_{J_{\bold{r},\bold{i}}}^{J_{\bold{k},\bold{q}},J_{\bold{k}'',\bold{q}''}}\|
 \label{fin-intersect} \\
& & \times \Big\{ \frac{1}{2}\Big\|\frac{(H^0_{J_{\bold{k}',\bold{q}'}})^{\frac{1}{2}}\,}{(\sum_{\bold{j}\in J_{\bold{r},\bold{i}}} P^{\perp}_{\Omega_{\bold{j}}}+1)^{\frac{1}{2}}(\sum_{\bold{j}\in J_{\bold{r},\bold{i}}} H_{\bold{j}}+1)^{\frac{1}{2}}}\,\psi\Big\|^2+ \frac{1}{2}\Big\|\frac{(H^0_{J_{\bold{k}'',\bold{q}''}})^{\frac{1}{2}}\,}{(\sum_{\bold{j}\in J_{\bold{r},\bold{i}}} P^{\perp}_{\Omega_{\bold{j}}}+1)^{\frac{1}{2}}(\sum_{\bold{j}\in J_{\bold{r},\bold{i}}} H_{\bold{j}}+1)^{\frac{1}{2}}}\,\psi\Big\|^2\Big\} \nonumber
\end{eqnarray}
Since (\ref{fin-intersect}) is symmetric under the permutaton of $J_{\bold{k}',\bold{q}'}$ with $J_{\bold{k}'',\bold{q}''}$, we further get
\begin{eqnarray}
&& (\ref{intersect-bis})\\
&\leq &\sup_{\| \psi\|=1}\,  \sum'_{J_{\bold{k}',\bold{q}' }\,,\,J_{\bold{k}'',\bold{q}''}}  \, \, 8 \|V^{(\bold{k},\bold{q})_{-1}}_{J_{\bold{k}',\bold{q}'}}\|_{H^0}\cdot \|V^{(\bold{k},\bold{q})_{-1}}_{J_{\bold{k}'',\bold{q}''}}\|_{H^0}\cdot \| T_{J_{\bold{r},\bold{i}}}^{J_{\bold{k},\bold{q}},J_{\bold{k}',\bold{q}'}} \| \cdot \| T_{J_{\bold{r},\bold{i}}}^{J_{\bold{k},\bold{q}},J_{\bold{k}'',\bold{q}''}}\|\label{3-4}\\
&&\quad\quad \times \langle \psi\,,\frac{H^0_{J_{\bold{k}',\bold{q}'}}\,}{(\sum_{\bold{j}\in J_{\bold{r},\bold{i}}} P^{\perp}_{\Omega_{\bold{j}}}+1)(\sum_{\bold{j}\in J_{\bold{r},\bold{i}}} H_{\bold{j}}+1)}\,\psi \rangle. \nonumber
\end{eqnarray}

Similarly to (\ref{3-1})-(\ref{3-3}), if we suppose that there are $1\leq l\leq d$ components of $\bold{k}$ different from the corresponding ones in $\bold{r}$ (and without loss of generality we identify them with the first $l$ components) we have the bound
 \begin{eqnarray}
& &(\ref{3-4})\\
&\leq&  \sum_{J_{\bold{k}',\bold{q}'}\,\in \,\mathcal{G}^{(\bold{k},\bold{q})}_{J_{\bold{r},\bold{i}}}}  \,\, 8 \|V^{(\bold{k},\bold{q})_{-1}}_{J_{\bold{k}',\bold{q}'}}\|_{H^0}\cdot  \| T_{J_{\bold{r},\bold{i}}}^{J_{\bold{k},\bold{q}},J_{\bold{k}',\bold{q}'}} \| \cdot \langle \psi\,,\frac{H^0_{J_{\bold{k}',\bold{q}'}}\,}{(\sum_{\bold{j}\in J_{\bold{r},\bold{i}}} P^{\perp}_{\Omega_{\bold{j}}}+1)(\sum_{\bold{j}\in J_{\bold{r},\bold{i}}} H_{\bold{j}}+1)}\,\psi \rangle \quad \nonumber\\
& &\quad \times \sum_{J_{\bold{k}'',\bold{q}''}\,\in \,\mathcal{G}^{(\bold{k},\bold{q})}_{J_{\bold{r},\bold{i}}}\,\, :\,\,J_{\bold{k}',\bold{q}'}\cap J_{\bold{k}'',\bold{q}''}\neq \emptyset} \|V^{(\bold{k},\bold{q})_{-1}}_{J_{\bold{k}'',\bold{q}''}}\|_{H^0}\cdot  \| T_{J_{\bold{r},\bold{i}}}^{J_{\bold{k},\bold{q}},J_{\bold{k}'',\bold{q}''}}\|\, \nonumber\\
&\leq &C \cdot \Big\{ \sum_{m=1}^{n_{\bold{k,\bold{q}}}} m^{\frac{1}{2}}\, \sum_{s_{1}=r_1-k_1}^{r-1} \dots \sum_{s_{l}=r_l - k_l}^{r-1} \sum_{s_{l+1}=0}^{r}\dots \sum_{s_{d}=0}^{r} \,\theta
_{m}^{^{\bold{k},\bold{q}}}(\bold{s})\,\frac{t^{(\sum_{j=1}^{d}\frac{s_j}{3})-\frac{1}{3}}}{(s_1+\dots +s_d)^{x_{d}}}\cdot \Big[  \prod_{j=l+1}^{d}(s_j+1)\Big]\Big\} \nonumber\\
& &\quad\times \sum_{n=1}^{n_{\bold{k,\bold{q}}}} n^{\frac{1}{2}}\cdot\,  \Big(\sum_{w=r-k}^r \frac{t^{\frac{w-1}{3}}}{w^{x_d}}\,\,\theta
_{n}^{^{\bold{k},\bold{q}}}(\bold{w})\,\cdot \Big(\prod_{j=1}^{d}s_j\Big)\cdot w^{d-1}\Big)\label{3-2-1} 
\end{eqnarray}
which follows from the estimate
$$\sum_{J_{\bold{k}'',\bold{q}''}\,\in \,[\mathcal{G}^{(\bold{k},\bold{q})}_{J_{\bold{r},\bold{i}}}]_n\,\, :\,\,J_{\bold{k}',\bold{q}'}\cap J_{\bold{k}'',\bold{q}''}\neq \emptyset} \|V^{(\bold{k},\bold{q})_{-1}}_{J_{\bold{k}'',\bold{q}''}}\|_{H^0}\cdot  \| T_{J_{\bold{r},\bold{i}}}^{J_{\bold{k},\bold{q}},J_{\bold{k}'',\bold{q}''}}\| \leq \mathcal{O}\Big(n^{\frac{1}{2}}\cdot \sum_{w=r-k}^r \frac{t^{\frac{w-1}{3}}}{w^{x_d}}\cdot \Big(\prod_{j=1}^{d}s_j\Big)\cdot w^{d-1}\Big)$$
where:
\begin{itemize}
\item[i)]
$\mathcal{O}(\Big(\prod_{j=1}^{d}s_j\Big)\cdot w^{d-1})$ is a bound from above of the number of rectangles $J_{\bold{w}, \bold{q}''}$ overlapping with the rectangle $J_{\bold{s}, \bold{q}'}$;
\item[ii)]
$ \mathcal{O}( \frac{t^{\frac{w-1}{3}}}{w^{x_d}})$ is the bound to $ \|V^{(\bold{k},\bold{q})_{-1}}_{J_{\bold{w},\bold{q}''}}\|_{H^0} $ from the inductive hypotheses;
\item[iii)]
$\mathcal{O}(n^{\frac{1}{2}})$ is the bound to $\| T_{J_{\bold{r},\bold{i}}}^{J_{\bold{k},\bold{q}},J_{\bold{k}'',\bold{q}''}}\| $\,.
\end{itemize}
Next, using the notation in (\ref{def-rho})  and arguments similar to the ones used in (\ref{3-2,5})-(\ref{3-3}), we write
\begin{eqnarray}
& &(\ref{3-2-1})\nonumber \\
&\leq &C\cdot  (t^{-1/3} \frac{t^{\frac{r-1}{3}}}{t^{\frac{k-1}{3}}})^2 \cdot  \sum_{m=1}^{n_{\bold{k,\bold{q}}}} m^{\frac{1}{2}}\, t^{a \cdot (m-1)^{\frac{1}{d}}}\Big\{ \frac{ 1}{(r_1-k_1+\dots +r_l-k_l)^{x_{d}}}\\
& &\quad \times  \sum_{s_{l+1}=0}^{r}\dots \sum_{s_{d}=0}^{r}\, \sum_{n=1}^{n_{\bold{k,\bold{q}}}} n^{\frac{1}{2}}\, t^{a \cdot (n-1)^{\frac{1}{d}}}\,   \sum_{w=r-k}^{r}\big[w^{d-1}\frac{1}{w^{x_{d}}}\big]\label{inters-bis}\\
& &\quad \times  \Big[  \Big(\prod_{j=l+1}^{d}s_j\Big)\cdot t^{\sum_{j=l+1}^{d}s_{j}/6}\cdot \prod_{j=l+1}^{d}(s_j+1)\Big]\sum_{\rho_{1}=0}^{\infty}\dots \sum_{\rho_{l}=0}^{\infty} t^{\sum_{j=1}^{l}\frac{\rho_{j}}{6}} \prod_{j=1}^l\Big(\rho_j+r_j-k_j\Big)\Big\} \nonumber 
\end{eqnarray}
Now we multiply (\ref{inters-bis}) by
$$\frac{ (r_1-k_1+\dots +r_l-k_l)^d}{(r_1-k_1+\dots +r_l-k_l)^{l}}\geq 1\,$$
and we obtain
\begin{eqnarray}
& &(\ref{3-2-1})\\
&\leq &C\cdot  (t^{-1/3} \frac{t^{\frac{r-1}{3}}}{t^{\frac{k-1}{3}}})^2 \cdot  \sum_{m=1}^{n_{\bold{k,\bold{q}}}} m^{\frac{1}{2}}\, t^{a \cdot (m-1)^{\frac{1}{d}}} \Big\{ \frac{ (r_1-k_1+\dots +r_l-k_l)^d}{(r_1-k_1+\dots +r_l-k_l)^{x_{d}}}\\
& &\quad \times \sum_{s_{l+1}=0}^{r}\dots \sum_{s_{d}=0}^{r}  \sum_{n=1}^{n_{\bold{k,\bold{q}}}} n^{\frac{1}{2}}\, t^{a \cdot (n-1)^{\frac{1}{d}}} \sum_{w=r-k}^{r}\big[w^{d-1}\frac{1}{w^{x_{d}}}\big] \label{inters-bisbis}\\
&&\quad \times  \Big[  \Big(\prod_{j=l+1}^{d}s_j\Big)\cdot t^{\sum_{j=l+1}^{d}s_{j}/6}\cdot \prod_{j=l+1}^{d}(s_j+1)\Big]\sum_{\rho_{1}=0}^{\infty}\dots \sum_{\rho_{l}=0}^{\infty} t^{\sum_{j=1}^{l}\frac{\rho_{j}}{6}} \prod_{j=1}^l\Big(\frac{\rho_j+r_j-k_j}{(r_1-k_1+\dots +r_l-k_l)}\Big)\Big\}\nonumber \\
& \leq &C_d\cdot  (t^{-1/3}\frac{t^{\frac{r-1}{3}}}{t^{\frac{k-1}{3}}})^2 \cdot( \frac{1}{(r_1-k_1+\dots +r_l-k_l)^{x_{d}-d}})^2\label{inters-bisbisbis} 
 \end{eqnarray}
where in the step from (\ref{inters-bisbis}) to (\ref{inters-bisbisbis})  we have used  $x_{d}\geq d+1$, and that the following quantities are bounded from above by a $d$-dependent constant:\\
\begin{itemize}
\item
$$ \sum_{n=1}^{n_{\bold{k,\bold{q}}}} n^{\frac{1}{2}}\, t^{a \cdot (n-1)^{\frac{1}{d}}}$$
\item
$$(r_1-k_1+\dots +r_l-k_l)^{x_{d}-d} \sum_{w=r-k}^{r}\,w^{d-1}\frac{1}{w^{x_{d}}}$$
\item
$$\sum_{s_{l+1}=0}^{r}\dots \sum_{s_{d}=0}^{r}  \Big(\prod_{j=l+1}^{d}s_j\Big)\cdot t^{\sum_{j=l+1}^{d}s_{j}/6}\cdot \prod_{j=l+1}^{d}(s_j+1)$$
\item
$$\sum_{\rho_{1}=0}^{\infty}\dots \sum_{\rho_{l}=0}^{\infty} t^{\sum_{j=1}^{l}\frac{\rho_{j}}{6}} \prod_{j=1}^l\Big(\frac{\rho_j+r_j-k_j}{(r_1-k_1+\dots +r_l-k_l)}\Big)\,.$$
\end{itemize}

\noindent
\emph{Leading terms in  $(\ref{rest-2})$: Contribution proportional to  (\ref{2.158})}
\\

\noindent
Finally, making use of 
\begin{eqnarray}
&(\ref{2.158})&\\
&\leq&\|\sum_{J_{\bold{k}',\bold{q}'}\in (\mathcal{G}^{(\bold{k},\bold{q})}_{J_{\bold{r};\bold{i}}})_{small}}\,(\frac{1}{[\pi_{J_{\bold{r},\bold{i}}}]_1})^{\frac{1}{2}}\, (\frac{1}{[H^0_{J_{\bold{r},\bold{i}}}]_1})^{\frac{1}{2}}\,\,P^{(+)}_{J_{\bold{r},\bold{i}}}\,\,P^{(+)}_{J_{\bold{k}',\bold{q}'}}\Big(V^{(\bold{k},\bold{q})_{-1}}_{J_{\bold{k}',\bold{q}'}}-\langle V^{(\bold{k},\bold{q})_{-1}}_{J_{\bold{k}',\bold{q}'}}\rangle\Big)\,P^{(+)}_{J_{\bold{k}',\bold{q}'}}   (\frac{1}{[H^0_{J_{\bold{k}',\bold{q}'}}]_1})^{\frac{1}{2}}  \,\nonumber \\
& &\quad\times (\frac{[H^0_{J_{\bold{k}',\bold{q}'}}]_1}{[H^0_{J_{\bold{r},\bold{i}}\setminus J_{\bold{k},\bold{q}}}]_1})^{\frac{1}{2}}   (\frac{[\pi_{J_{\bold{k},\bold{q}}}]_1}{[H^0_{J_{\bold{k},\bold{q}}}]_1})^{\frac{1}{2}} \|\cdot \|([H^0_{J_{\bold{k},\bold{q}}}]_1)^{\frac{1}{2}}\, (\frac{1}{[\pi_{J_{\bold{k},\bold{q}}}]_1})^{\frac{1}{2}}\frac{t}{G_{J_{\bold{k},\bold{q}}}-E_{J_{\bold{k},\bold{q}}}}\,P^{(+)}_{J_{\bold{k},\bold{q}}}\,V^{(\bold{k},\bold{q})_{-1}}_{J_{\bold{k},\bold{q}}}\,P^{(-)}_{J_{\bold{k},\bold{q}}}P^{(-)}_{J_{\bold{r},\bold{i}}}\|\quad \quad \quad 
\end{eqnarray}
together with  (\ref{ineq-lemma-hs}),
we obtain that at fixed $\bold{k}$ with $l$ components being different from the corresponding components of $\bold{r}$,
\begin{equation}\label{contributo-small-leading}
\|(\ref{lead-2.115})\| \leq C_d\cdot t^{2/3}\cdot \frac{ t^{\frac{r-1}{3}}}{ (r_1-k_1+\dots +r_l-k_l)^{x_{d}-d} \cdot k^{x_{d}+2d}}.
\end{equation}
\\

\noindent
\emph{Higher order terms in  $(\ref{rest-2})$}

\noindent
In order to obtain (\ref{R3-1}), with regard to $(\ref{rest-2})$ we must still estimate:
\begin{itemize}
\item the  remainder  (\ref{rem-2.115})  arising from the study of $(\ref{rest-2-first})_{small}$; 
\item the terms corresponding to  $(\ref{rest-2-first})_{large}$, i.e., proportional to terms with $J_{\bold{k}',\bold{q}'}$ such that  $(\bold{k}',\bold{q}')\succ (\bold{k},\bold{q})$;
\item  the contribution from (\ref{rest-2-second}).
\end{itemize}
We observe that:

\noindent
i) in all these terms there are either two factors $S_{\bold{k},\bold{q}}$ or two factors $\|(V^{(k,q-1)}_{J_{k,q}})_1\|_{H^0}$  or $J_{\bold{k}',\bold{q}'}$ is large such that $(\bold{k}',\bold{q}')\succ (\bold{k},\bold{q})$,  thus we get at least  an extra factor $\mathcal{O}(t^{\frac{r-2  r^{1/4}-1}{3}})$;

\noindent
ii) by Remark \ref{shapes}, a bound from above of the total number of the elements of $\mathcal{G}^{(\bold{k},\bold{q})}_{J_{\bold{r},\bold{i}}}$ is
\begin{equation}
\mathcal{O}(r^{d-1}\cdot \sum_{l=1}^{r}l^{d-1})\leq \mathcal{O}(r^{2d-1})\,.
\end{equation}

\noindent
Thus, from the inductive hypotheses (\ref{R3-2}) and (\ref{R3-3}) we get
\begin{eqnarray}
& &\|(\ref{rem-2.115}) \|+\|(\ref{rest-2-first})_{large}\|+\|(\ref{rest-2-second})\| \label{sum-rem}\\
&\leq&C_d\cdot t\cdot r^{4d-1}\cdot t^{\frac{r-2r^{1/4}-1}{3}} \cdot \frac{ t^{\frac{r-1}{3}}}{ (r-k)^{x_{d}} \cdot k^{x_{d}}}.
\end{eqnarray}

When $k$ is fixed, the number of contributions of the type (\ref{sum-rem}) is at most $\mathcal{O}(r^d\cdot k^{d-1})$ .
\\

\noindent
\emph{Complete estimate of  (\ref{R3-1})}

\noindent
Summing up, by the estimates of (\ref{lead-2.115}), (\ref{rem-2.115}), $(\ref{rest-2-first})_{large}$, and $(\ref{rest-2-second})$ that have been derived, we can conclude that 

\begin{eqnarray}
& &\|(\frac{1}{\sum_{\bold{j}\in J_{\bold{r},\bold{i}}} P^{\perp}_{\Omega_{\bold{j}}}+1})^{\frac{1}{2}}\, (\frac{1}{H^0_{ J_{\bold{r},\bold{i}}}+1})^{\frac{1}{2}}\,P^{(+)}_{J_{\bold{r},\bold{i}}}\,V^{(\bold{k},\bold{q})}_{J_{\bold{r},\bold{i}}}\,P^{(-)}_{J_{\bold{r},\bold{i}}}\|\\
&\leq &\|P^{(+)}_{J_{\bold{r},\bold{i}}}\,V^{(\bold{k},\bold{q})_{**}}_{J_{\bold{r},\bold{i}}}\,P^{(-)}_{J_{\bold{r},\bold{i}}}\|_{H^0} \label{first-summand}\\
& &+C_d\cdot t\cdot \sum_{l=1}^{d-1}\binom{d}{l}\cdot \sum_{k_1=0}^{r_1-1}\dots \sum_{k_l=0}^{r_{l}-1}\,\Theta(\sum_{j=1}^{l}k_j-r+\lfloor r^\frac{1}{4}\rfloor )\,\times \label{th-in-0}\\
& &\quad\quad\quad\quad\quad\quad \times \frac{ t^{\frac{r-1}{3}}}{ (r_1-k_1+\dots +r_l-k_l)^{x_{d}-d} \cdot k^{x_{d}+2d}}\label{th-in}
  \\
& &+\sum_{k=r-\lfloor r^\frac{1}{4}\rfloor }^{k= r-1 }\,\Big\{  
C_d\cdot t\cdot r^{5d-1}\cdot k^{d-1}\cdot t^{\frac{r-2r^\frac{1}{4}-1}{3}} \cdot \frac{ t^{\frac{r-1}{3}}}{ (r-k)^{x_{d}} \cdot k^{x_{d}}}\Big\} \label{th-fin}
\end{eqnarray}
where $\Theta$ is the characteristic function of $\mathbb{R}^+$.

To estimate (\ref{th-in-0})-(\ref{th-in})
we observe that we have
\begin{eqnarray}
&&\sum_{k_1=0}^{r_1-1}\dots \sum_{k_l=0}^{r_{l}-1}\,\Theta(\sum_{j=1}^{l}k_j-r+r^\frac{1}{4})\,\times\frac{ t^{\frac{r-1}{3}}}{ (r_1-k_1+\dots +r_l-k_l)^{x_{d}-d} \cdot k^{x_{d}+2d}}\\
&\leq &C_d\cdot  \sum_{s_1=1}^{r_1}\dots \sum_{s_l=1}^{r_{l}}\frac{ t^{\frac{r-1}{3}}}{ (s_1+s_2+\ldots +s_l)^{x_{d}-d} \cdot r^{x_{d}+2d}}\\
&\leq &C_d\cdot \frac{t^\frac{r-1}{3}}{r^{x_{d}+2d}}\sum_{s_1=1}^{\infty}\dots \sum_{s_l=1}^{\infty}\frac{ 1}{ (s_1+s_2+\ldots +s_l)^{x_{d}-d}}\\
&= & C_d\cdot \frac{t^\frac{r-1}{3}}{r^{x_{d}+2d}}
\end{eqnarray}
where we used that since $x_d-d>l$,  $\sum_{s_1=1}^{\infty}\dots \sum_{s_l=1}^{\infty}\frac{ 1}{ (s_1+s_2+\ldots +s_l)^{x_{d}-d}}$ is bounded by a $d$-dependent costant. Therefore we see that the overall quantity in (\ref{th-in-0})-(\ref{th-in}) can be made less than
$\frac{1}{2}\cdot \frac{t^\frac{r-1}{3}}{r^{x_{d}+2d}}$ provided $t\geq 0$ is small enough.\\
As for (\ref{th-fin}), this quantity can be estimated in the following way:
\begin{eqnarray}
&&\sum_{k=r-\lfloor r^\frac{1}{4}\rfloor}^{r-1}C_d\cdot t\cdot r^{5d-1}\cdot k^{d-1}\cdot t^{\frac{r-2r^\frac{1}{4}-1}{3}} \cdot \frac{ t^{\frac{r-1}{3}}}{ (r-k)^{x_{d}} \cdot k^{x_{d}}}\\\
&\leq &
r^\frac{1}{4}\cdot 2^{x_d} \cdot C_d \cdot t \cdot t^{\frac{r-2r^\frac{1}{4}-1}{3}}\frac{t^\frac{r-1}{3}}{r^{x_{d}-6d+2}}\\
&=&2^{x_d}\cdot C_d \cdot t\cdot  t^{\frac{r-2r^\frac{1}{4}-1}{3}}\cdot \frac{t^\frac{r-1}{3}}{r^{x_{d}-6d+\frac{7}{4}}}\\
&\leq &\frac{1}{2}\cdot \frac{t^\frac{r-1}{3}}{r^{x_{d}+2d}}
\end{eqnarray}
where the last inequality holds provided we take $t\geq 0$ to be so small to fulfill the inequality
$$2^{x_d}\cdot C_d \cdot t\cdot t^\frac{r-2r^\frac{1}{4}-1}{3}\leq \frac{1}{3\cdot r^{8d-\frac{7}{4}}}$$
 uniformly in $r$. 

\noindent
Finally, for $t\geq 0$ small enough, we obtain
\begin{eqnarray}
& &\|(\frac{1}{\sum_{\bold{j}\in J_{\bold{r},\bold{i}}} P^{\perp}_{\Omega_{\bold{j}}}+1})^{\frac{1}{2}}\, (\frac{1}{H^0_{ J_{\bold{r},\bold{i}}}+1})^{\frac{1}{2}}\,P^{(+)}_{J_{\bold{r},\bold{i}}}\,V^{(\bold{k},\bold{q})}_{J_{\bold{r},\bold{i}}}\,P^{(-)}_{J_{\bold{r},\bold{i}}}\|\\
&\leq &2\cdot\frac{t^{\frac{r-1}{3}}}{r^{x_{d}+2d}} +2\cdot\frac{1}{2}\cdot \frac{t^{\frac{r-1}{3}}}{r^{x_{d}+2d}} \label{secondo}\\
&=&3\cdot \frac{t^{\frac{r-1}{3}}}{r^{x_{d}+2d}}
\end{eqnarray}
as claimed.

The estimate of
\begin{eqnarray*}
& &(\frac{1}{\sum_{\bold{j}\in J_{\bold{r},\bold{i}}} P^{\perp}_{\Omega_{\bold{j}}}+1})^{\frac{1}{2}}\, (\frac{1}{H^0_{ J_{\bold{r},\bold{i}}}+1})^{\frac{1}{2}}\, P^{(+)}_{J_{\bold{r},\bold{i}}}\,V^{(\bold{k},\bold{q})}_{J_{\bold{r},\bold{i}}}\,P^{(+)}_{J_{\bold{r},\bold{i}}}\, (\frac{1}{H^0_{ J_{\bold{r},\bold{i}}}+1})^{\frac{1}{2}}\,(\frac{1}{\sum_{\bold{j}\in J_{\bold{r},\bold{i}}} P^{\perp}_{\Omega_{\bold{j}}}+1})^{\frac{1}{2}}\,
\end{eqnarray*}
follows the same procedure. The estimate of
\begin{eqnarray}
& &(\frac{1}{\sum_{\bold{j}\in J_{\bold{r},\bold{i}}} P^{\perp}_{\Omega_{\bold{j}}}+1})^{\frac{1}{2}}\, (\frac{1}{H^0_{ J_{\bold{r},\bold{i}}}+1})^{\frac{1}{2}}\, P^{(-)}_{J_{\bold{r},\bold{i}}}\,V^{(\bold{k},\bold{q})}_{J_{\bold{r},\bold{i}}}\,P^{(-)}_{J_{\bold{r},\bold{i}}}\, (\frac{1}{H^0_{ J_{\bold{r},\bold{i}}}+1})^{\frac{1}{2}}\,(\frac{1}{\sum_{\bold{j}\in J_{\bold{r},\bold{i}}} P^{\perp}_{\Omega_{\bold{j}}}+1})^{\frac{1}{2}} \quad\quad \\
&= & P^{(-)}_{J_{\bold{r},\bold{i}}}\,V^{(\bold{k},\bold{q})}_{J_{\bold{r},\bold{i}}}\,P^{(-)}_{J_{\bold{r},\bold{i}}}\end{eqnarray}
can be also be performed in the same manner and is actually simpler since the terms proportional to $J_{\bold{k}',\bold{q}'}$ \emph{small} in the expansion are identically  zero.

\noindent
\underline{\emph{Proof of  (\ref{R3-2})}}

\noindent
We observe that
\begin{eqnarray}
& &\|P^{(\#)}_{J_{\bold{r},\bold{i}}}\,V^{(\bold{k},\bold{q})}_{J_{\bold{r},\bold{i}}}\,P^{(\hat{\#})}_{J_{\bold{r},\bold{i}}}\|_{H^0}\\
&\leq &\|(\sum_{\bold{j}\in J_{\bold{r},\bold{i}}} P^{\perp}_{\Omega_{\bold{j}}}+1)^{\frac{1}{2}}\|^2\\
& &\quad \times\|(\frac{1}{[\pi_{J_{\bold{r},\bold{i}}}]_1})^{\frac{1}{2}}\,(\frac{1}{[H^0_{J_{\bold{r},\bold{i}}}]_1})^{\frac{1}{2}}\,P^{(\#)}_{J_{\bold{r},\bold{i}}}\,V^{(\bold{k},\bold{q})}_{J_{\bold{r},\bold{i}}}\,P^{(\hat{\#})}_{J_{\bold{r},\bold{i}}}\, (\frac{1}{[H^0_{J_{\bold{r},\bold{i}}}]_1})^{\frac{1}{2}}\,(\frac{1}{[\pi_{J_{\bold{r},\bold{i}}}]_1})^{\frac{1}{2}}\,\|
\end{eqnarray}
 where $\#,\hat{\#}=\pm $, and $\|(\sum_{\bold{j}\in J_{\bold{r},\bold{i}}} P^{\perp}_{\Omega_{\bold{j}}}+1)^{\frac{1}{2}}\|^2\leq r^d+1\leq 2 \cdot r^{d}$ and thus we can use (\ref{R3-1}) which is proven above.
\\
In order to prove (\ref{R3-3}), it is enough to combine (\ref{bound-V}) and (\ref{R3-2}).\\
\\

\noindent
\emph{Induction step to prove S2)}

\noindent
Since we have already proven S1), we can now apply Lemma \ref{gap} and Corollary \ref{cor-gap}. Thus S2) holds for $t\geq 0$ sufficiently small but independent of $N$, $\bold{k}$, and $\bold{q}$. \qed

We can now derive the main result of the paper.

\begin{thm}\label{main-res}
Under the assumption that (\ref{gaps}), (\ref{potential}) and (\ref{klmn-cond}) hold, the Hamiltonian $K_{\Lambda_N^d}$ defined in (\ref{Hamiltonian}) has the following properties: There exists some $t_d > 0$ such that, for any $t\in \mathbb{R}$ with 
$\vert t \vert < t_d$, and for all $N < \infty$,
\begin{enumerate}
\item[(i)]{ $K_{\Lambda_N^d}\equiv K_{\Lambda_N^d}(t)$ has a unique ground-state; and}
\item[(ii)]{ the energy spectrum of $K_{\Lambda_N^d}$ has a strictly positive gap, $\Delta_{N}(t) \geq \frac{1}{2}$, above the ground-state energy.}
\end{enumerate}
\end{thm}

\noindent
\emph{Proof.}
The effective Hamiltonian at the final step is $K_{\Lambda_N^d}^{(\bold{N-1},\bold{1})} \equiv G_{J_{\bold{N-1},\bold{1}}}+tV^{(\bold{N-1},\bold{1})}_{J_{\bold{N-1},\bold{1}}}$. The composition of the unitary operators associated with each block-diagonalization step yields a unitary operator 
$U_N(t)$, such that  
$$U_N(t)K_{\Lambda_N^d}(t) U_N(t)^*=G_{J_{\bold{N-1},\bold{1}}}+tV^{(\bold{N-1},\bold{1})}_{J_{\bold{N-1},\bold{1}}}=: \widetilde{K}_{\Lambda_N^d}(t).$$  
The statement of the Theorem follows from (\ref{gap-1})-(\ref{gap-2}), for $(\bold{k},\bold{q})=(\bold{N-1},\bold{1})$, where we also include the block-diagonalized potential $V^{(\bold{N-1},\bold{1})}_{J_{\bold{N-1},\bold{1}}}$, that we control by Theorem \ref{th-norms}. \qed

\section{Some Technical Lemmas}\label{teclem}

We here prove some technical Lemmas that are part of the inductive proof of Theorem \ref{th-norms}.  All the estimates performed in the Lemmas never depend on $\bold{k}, \bold{q}$ or $\bold{N}$. 

\begin{lem}\label{control-LS}
Assume that there exists $t_d>0$ (which does not depend on $\bold{k}, \bold{q}$ or $\bold{N}$) such that for every $|t|<t_d$ we have $\|V^{(\bold{k},\bold{q})_{-1}}_{J_{\bold{k},\bold{q}}}\| \leq 96  \cdot \frac{t^{\frac{k-1}{3}}}{k^{x_d}}$ with $x_d=20d$, and $\Delta_{J_{k,q}}\geq \frac{1}{2}$. Then for any $N$ and $(\bold{k},\bold{q})$  the inequalities
\begin{equation}\label{bound-V}
\|V^{(\bold{k},\bold{q})}_{J_{\bold{k},\bold{q}}}\|_{H_0}\leq 2\|V^{(\bold{k},\bold{q})_{-1}}_{J_{\bold{k},\bold{q}}}\|_{H_0}\,,
\end{equation}
\begin{equation}\label{bound-S}
\|S_{J_{\bold{k}, \bold{q}}}\|\leq C\cdot t \cdot \|V^{(\bold{k},\bold{q})_{-1}}_{J_{\bold{k},\bold{q}}}\|_{H_0}\,,
\end{equation}
\begin{equation}\label{Vsquare}
\|\sum_{j=2}^{\infty}t^{j}(S_{J_{\bold{k},\bold{q}}})_j\|\leq C\cdot t \cdot\|V^{(\bold{k},\bold{q})_{-1}}_{J_{\bold{k},\bold{q}}}\|_{H_0}^2,
\end{equation}
and
\begin{equation}\label{S-Hest}
\|S_{J_{\bold{k},\bold{q}}}(H^0_{J_{\bold{k},\bold{q}}}+1)^{\frac{1}{2}}\|=\|(H^0_{J_{\bold{k},\bold{q}}}+1)^{\frac{1}{2}}S_{J_{\bold{k},\bold{q}}}\| \leq C\, t\,
 \| V^{(\bold{k},\bold{q})_{-1}}_{J_{\bold{k},\bold{q}}}\|_{H_0}
\end{equation}
hold true for a universal constant $C$.
\end{lem}

\noindent
\emph{Proof}

\noindent
Let us recall that
\begin{equation}
V^{(\bold{k},\bold{q})}_{J_{\bold{k},\bold{q}}}:= \sum_{j=1}^{\infty}t^{j-1}(V^{(\bold{k},\bold{q})_{-1}}_{J_{\bold{k},\bold{q}}})^{diag}_j \,
\end{equation}
and
\begin{equation}
S_{J_{\bold{k},\bold{q}}}:=\sum_{j=1}^{\infty}t^j(S_{J_{\bold{k},\bold{q}}})_j\
\end{equation}
where
\begin{eqnarray}\label{defpotentials}
& &(V^{(\bold{k},\bold{q})_{-1}}_{J_{\bold{k},\bold{q}}})^{diag}_j \,\label{def-Vj}:=\\
& &\sum_{p\geq 2, v_1\geq 1 \dots, v_p\geq 1\, ; \, v_1+\dots+v_p=j}\frac{1}{p!}\text{ad}\,(S_{J_{\bold{k},\bold{q}}})_{v_1}\Big(\text{ad}\,(S_{J_{\bold{k},\bold{q}}})_{v_2}\dots (\text{ad}\,(S_{I_{\bold{k},\bold{q}}})_{v_p}(G_{J_{\bold{k},\bold{q}}}))\dots \Big)\\
& &+\sum_{p\geq 1, v_1\geq 1 \dots, v_p\geq 1\, ; \, v_1+\dots+v_p=j-1}\frac{1}{p!}\text{ad}\,(S_{J_{\bold{k},\bold{q}}})_{v_1}\Big(\text{ad}\,(S_{J_{\bold{k},\bold{q}}})_{v_2}\dots (\text{ad}\,(S_{J_{\bold{k},\bold{q}}})_{v_p}(V^{(\bold{k},\bold{q})_{-1}}_{J_{\bold{k},\bold{q}}}))\dots \Big)\quad\quad\quad\quad\,.
\end{eqnarray}
and
\begin{equation}\label{formula-Sj-bis}
(S_{J_{\bold{k},\bold{q}}})_j:=ad^{-1}\,G_{J_{\bold{k},\bold{q}}}\,((V^{(\bold{k},\bold{q})_{-1}}_{J_{\bold{k},\bold{q}}})^{od}_j):=\frac{1}{G_{J_{\bold{k},\bold{q}}}-E_{J_{\bold{k},\bold{q}}}}P^{(+)}_{J_{\bold{k},\bold{q}}}\,(V^{(\bold{k},\bold{q})_{-1}}_{J_{\bold{k},\bold{q}}})_j\,P^{(-)}_{J_{\bold{k},\bold{q}}}-h.c.\,.
\end{equation}
Hence we derive
\begin{eqnarray}
& &\text{ad}\,(S_{J_{\bold{k},\bold{q}}})_{r_p}(G_{J_{\bold{k},\bold{q}}})\\
&=&\text{ad}\,(S_{J_{\bold{k},\bold{q}}})_{r_p}(G_{J_{\bold{k},\bold{q}}}-E_{J_{\bold{k},\bold{q}}})\\
&=&\,[\frac{1}{G_{J_{\bold{k},\bold{q}}}-E_{J_{\bold{k},\bold{q}}}}P^{(+)}_{J_{\bold{k},\bold{q}}}\,(V^{(\bold{k},\bold{q})_{-1}}_{J_{\bold{k},\bold{q}}})_j\,P^{(-)}_{J_{\bold{k},\bold{q}}} \,,\,G_{J_{\bold{k},\bold{q}}}-E_{J_{\bold{k},\bold{q}}}]+h.c.\\
&=&P^{(+)}_{J_{\bold{k},\bold{q}}}\,(V^{(\bold{k},\bold{q})_{-1}}_{J_{\bold{k},\bold{q}}})_{r_p}\,P^{(-)}_{J_{\bold{k},\bold{q}}}+P^{(-)}_{J_{\bold{k},\bold{q}}}\,(V^{(\bold{k},\bold{q})_{-1}}_{J_{\bold{k},\bold{q}}})_{r_p}\,P^{(+)}_{J_{\bold{k},\bold{q}}}\,.
\end{eqnarray}

\noindent
The first thing we need to do is show the following inequality
\begin{equation}\label{Snorm}
\|(S_{J_{\bold{k},\bold{q}}})_j\|\leq \frac{2\sqrt{2}}{\Delta_{I_{k,q}}} \| (V_{J_{\bold{k},\bold{q}}}^{(\bold{k}, \bold{q})_{-1}})_j\|_{H^0}\,,
\end{equation}

\noindent
where  $ \| (V_{J_{\bold{k},\bold{q}}}^{(\bold{k}, \bold{q})_{-1}})_j\|_{H_0}$ will be proved to be finite in the next step.
As for the estimate in (\ref{Snorm}), it is enough to make the following computations:
\begin{eqnarray}
& &\|(S_{J_{\bold{k},\bold{q}}})_j\|\label{S-in}\\
&\leq& 2\left\|\frac{1}{G_{J_{\bold{k},\bold{q}}}-E_{J_{\bold{k},\bold{q}}}}P^{(+)}_{J_{\bold{k},\bold{q}}}\,(V^{(\bold{k},\bold{q})_{-1}}_{J_{\bold{k},\bold{q}}})_j\,P^{(-)}_{J_{\bold{k},\bold{q}}}\right\|\\
&=&2\left\|\frac{1}{G_{J_{\bold{k},\bold{q}}}-E_{J_{\bold{k},\bold{q}}}}P^{(+)}_{J_{\bold{k},\bold{q}}}(H_{J_{\bold{k},\bold{q}}}^0+1)^{\frac{1}{2}}(H_{J_{\bold{k},\bold{q}}}^0+1)^{-\frac{1}{2}}(V^{(\bold{k},\bold{q})_{-1}}_{J_{\bold{k},\bold{q}}})_j(H_{J_{\bold{k},\bold{q}}}^0+1)^{-\frac{1}{2}}P^{(-)}_{J_{\bold{k},\bold{q}}}\right\| \\
&\leq&2\left\|\frac{1}{G_{J_{\bold{k},\bold{q}}}-E_{J_{\bold{k},\bold{q}}}}P^{(+)}_{J_{\bold{k},\bold{q}}}(H_{J_{\bold{k},\bold{q}}}^0+1)^{\frac{1}{2}}\right\|   \|(V_{J_{\bold{k},\bold{q}}}^{(\bold{k},\bold{q})_{-1}})_j\|_{H^0}\\
&\leq & \frac{2\sqrt{2}}{\Delta_{J_{\bold{k},\bold{q}}}}\|(V_{J_{\bold{k},\bold{q}}}^{(\bold{k},\bold{q})_{-1}})_j\|_{H^0}\,, \label{S-fin}
\end{eqnarray}
where we have used (\ref{op-norm-G-2}) for the last inequality.

\noindent
Making use of (\ref{op-norm-G}) and $(H_{J_{\bold{k},\bold{q}}}^0+1)^{\frac{1}{2}}P^{(-)}_{J_{\bold{k},\bold{q}}}=P^{(-)}_{J_{\bold{k},\bold{q}}}$, we can similarly estimate
\begin{equation}\label{S-norm-2}
\|(S_{J_{\bold{k},\bold{q}}})_{j}(H^0_{J_{\bold{k},\bold{q}}}+1)^{\frac{1}{2}}\|=\|(H^0_{J_{\bold{k},\bold{q}}}+1)^{\frac{1}{2}}(S_{J_{\bold{k},\bold{q}}})_{j}\|\leq \frac{2+\sqrt{2}}{\Delta_{J_{\bold{k},\bold{q}}}} \|(V_{J_{\bold{k},\bold{q}}}^{(\bold{k}, \bold{q})_{-1}})_{j}\|_{H^0}\,.
\end{equation}
We now want to show that
\begin{eqnarray}
\|(V^{(\bold{k},\bold{q})_{-1}}_{J_{\bold{k},\bold{q}}})_j\|_{H^0}&\leq&\label{V-ineq}\\
\sum_{p=2}^{j}\,\frac{(2c)^p}{p!}&&\sum_{ r_1\geq 1 \dots, r_p\geq 1\, ; \, r_1+\dots+r_p=j}\,\|\,(V^{(\bold{k},\bold{q})_{-1}}_{J_{\bold{k},\bold{q}}})_{r_1}\|_{H^0}\|\,(V^{(\bold{k},\bold{q})_{-1}}_{J_{\bold{k},\bold{q}}})_{r_2}\|_{H^0}\dots \|\,(V^{(\bold{k},\bold{q})_{-1}}_{J_{\bold{k},\bold{q}}})_{r_p}\|_{H^0}\nonumber \\
+2\|V^{(\bold{k},\bold{q})_{-1}}_{J_{\bold{k},\bold{q}}}\|_{H^0}&& \sum_{p=1}^{j-1}\,\frac{(2c)^p}{p!}\,\sum_{ r_1\geq 1 \dots, r_p\geq 1\, ; \, r_1+\dots+r_p=j-1}\,\|\,(V^{(\bold{k},\bold{q})_{-1}}_{J_{\bold{k},\bold{q}}})_{r_1}\|_{H^0}\|\,(V^{(\bold{k},\bold{q})_{-1}}_{J_{\bold{k},\bold{q}}})_{r_2}\|_{H^0}\dots \|\,(V^{(\bold{k},\bold{q})_{-1}}_{J_{\bold{k},\bold{q}}})_{r_p}\|_{H^0}\,,\nonumber 
\end{eqnarray}
where $c:= \frac{2+\sqrt{2}}{\Delta_{J_{\bold{k},\bold{q}}}}(> \frac{2\sqrt{2}}{\Delta_{J_{\bold{k},\bold{q}}}})$.
To this end, we note that formula (\ref{defpotentials}) yielding $(V_{J_{\bold{k},\bold{q}}}^{(\bold{k}, \bold{q})_{-1}})_j$ contains two sums. We first work on the second, namely
$$\sum_{p\geq 1, r_1\geq 1 \dots, r_p\geq 1\, ; \, r_1+\dots+r_p=j-1}\frac{1}{p!}\text{ad}\,(S_{J_{\bold{k},\bold{q}}})_{r_1}\Big(\text{ad}\,(S_{J_{\bold{k},\bold{q}}})_{r_2}\dots (\text{ad}\,.(S_{J_{\bold{k},\bold{q}}})_{r_p}(V^{(\bold{k},\bold{q})_{-1}}_{J_{\bold{k},\bold{q}}})\Big)\,. $$
Each summand of the above sum is in turn a sum of $2^p$ terms which, up to a sign,  are permutations of
$$(S_{J_{\bold{k},\bold{q}}})_{r_1}(S_{J_{\bold{k},\bold{q}}})_{r_2}\ldots (S_{J_{\bold{k},\bold{q}}})_{r_p}V_{J_{\bold{k},\bold{q}}}^{(\bold{k}, \bold{q})_{-1}}\,,$$
with the potential $V_{J_{\bold{k},\bold{q}}}^{(\bold{k}, \bold{q})_{-1}}$ allowed to appear at any position.
We treat only one of these terms, as the others can be studied in the same way. For instance, we can 
study $$(S_{J_{\bold{k},\bold{q}}})_{r_1} V_{J_{\bold{k},\bold{q}}}^{(\bold{k}, \bold{q})_{-1}}   (S_{J_{\bold{k},\bold{q}}})_{r_2}\ldots (S_{J_{\bold{k},\bold{q}}})_{r_p}.$$
Note that
\begin{eqnarray}
&&\|(S_{J_{\bold{k},\bold{q}}})_{r_1} V_{J_{\bold{k},\bold{q}}}^{(\bold{k},\bold{q})_{-1}}   (S_{J_{\bold{k},\bold{q}}})_{r_2}\ldots (S_{J_{\bold{k},\bold{q}}})_{r_p}\|_{H^0} \nonumber \\
&=&\|(H_{J_{\bold{k},\bold{q}}}^0+1)^{-\frac{1}{2}} (S_{J_{\bold{k},\bold{q}}})_{r_1}  (H_{J_{\bold{k},\bold{q}}}^0+1)^{\frac{1}{2}} \\
&\cdot&  (H_{J_{\bold{k},\bold{q}}}^0+1)^{-\frac{1}{2}} V_{J_{\bold{k},\bold{q}}}^{(\bold{k},\bold{q})_{-1}}(H_{J_{\bold{k},\bold{q}}}^0+1)^{-\frac{1}{2}}(H_{J_{\bold{k},\bold{q}}}^0+1)^{\frac{1}{2}}    (S_{J_{\bold{k},\bold{q}}})_{r_2}\ldots (S_{J_{\bold{k},\bold{q}}})_{r_p}  (H_{J_{\bold{k},\bold{q}}}^0+1)^{-\frac{1}{2}}\|  \nonumber \\
&\leq& \|V_{J_{\bold{k},\bold{q}}}^{(\bold{k},\bold{q})_{-1}} \|_{H^0} \|(S_{J_{\bold{k},\bold{q}}})_{r_1}(H_{J_{\bold{k},\bold{q}}}^0+1)^{\frac{1}{2}} \|\, \|(H_{J_{\bold{k},\bold{q}}}^0+1)^{\frac{1}{2}}    (S_{J_{\bold{k},\bold{q}}})_{r_2}\| \ldots \|(S_{J_{\bold{k},\bold{q}}})_{r_p}\| \nonumber\\
&\leq &c^p \|V^{(\bold{k},\bold{q})_{-1}}_{J_{\bold{k},\bold{q}}}\|_{H^0} \|\,(V^{(\bold{k},\bold{q})_{-1}}_{J_{\bold{k},\bold{q}}})_{r_1}\|_{H^0}\|\,(V^{(\bold{k},\bold{q})_{-1}}_{J_{\bold{k},\bold{q}}})_{r_2}\|_{H^0}\dots \|\,(V^{(\bold{k},\bold{q})_{-1}}_{J_{\bold{k},\bold{q}}})_{r_p}\|_{H^0}\,,\nonumber
\end{eqnarray}
where we made use of (\ref{Snorm}) and (\ref{S-norm-2}). 
\noindent
Collecting these terms together, we get to the second sum of (\ref{V-ineq}).\\
Regarding the first sum in (\ref{defpotentials}), i.e.,

$$\sum_{p\geq 2, r_1\geq 1 \dots, r_p\geq 1\, ; \, r_1+\dots+r_p=j}\frac{1}{p!}\text{ad}\,(S_{J_{\bold{k},\bold{q}}})_{r_1}\Big(\text{ad}\,(S_{J_{\bold{k},\bold{q}}})_{r_2}\dots (\text{ad}\,(S_{J_{\bold{k},\bold{q}}})_{r_p}(G_{J_{\bold{k},\bold{q}}}) \Big)\,,$$
note that each summand is in turn the sum (up to a sign) of permutations of
$$(S_{J_{\bold{k},\bold{q}}})_{r_1}(S_{J_{\bold{k},\bold{q}}})_{r_2}\ldots(S_{J_{\bold{k},\bold{q}}})_{r_{p-1}}[-P_{J_{\bold{k},\bold{q}}}^{(+)}(V_{J_{\bold{k},\bold{q}}}^{(\bold{k},\bold{q})_{-1}})_{r_p} P_{J_{\bold{k},\bold{q}}}^{(-)} - P_{J_{\bold{k},\bold{q}}}^{(-)}(V_{J_{\bold{k},\bold{q}}}^{(\bold{k},\bold{q})_{-1}})_{r_p}P_{J_{\bold{k},\bold{q}}}^{(+)}]\,.$$ 
Now with a slight variation of the computations above we see that the $\|\cdot\|_{H^0}$-norm of the first sum  in  (\ref{defpotentials}) is bounded
from above by 
$$\sum_{p=2}^{j}\,\frac{(2c)^p}{p!}\sum_{ r_1\geq 1 \dots, r_p\geq 1\, ; \, r_1+\dots+r_p=j}\,\|\,(V^{(\bold{k},\bold{q})_{-1}}_{J_{\bold{k},\bold{q}}})_{r_1}\|_{H^0}\|\,(V^{(\bold{k},\bold{q})_{-1}}_{J_{\bold{k},\bold{q}}})_{r_2}\|_{H^0}\dots \|\,(V^{(\bold{k},\bold{q})_{-1}}_{J_{\bold{k},\bold{q}}})_{r_p}\|_{H^0}\,;
$$
where we have assumed that $c>1$, without harming the generality.

From now on,  we closely follow the proof of Theorem 3.2 in \cite{DFFR}; that is, assuming $\|V^{(\bold{k},\bold{q})_{-1}}_{J_{\bold{k},\bold{q}}}\|_{H_0}\neq 0$, we recursively define numbers $B_j$, $j\geq 1$, by the equations
\begin{eqnarray}
B_1&:= &\|V^{(\bold{k},\bold{q})_{-1}}_{J_{\bold{k},\bold{q}}}\|_{H^0}=\|(V^{(\bold{k},\bold{q})_{-1}}_{J_{\bold{k},\bold{q}}})_1\|_{H^0} \label{B1} \,,\\
B_j&:=&\frac{1}{a}\sum_{k=1}^{j-1}B_{j-k}B_k\,,\quad j\geq 2\,, \label{def-Bj}
\end{eqnarray}
with \,$a>0$\, satisfying the equation
\begin{equation}\label{a-eq}
e^{2c a}-1+ \left( \frac{e^{2c a}-2c a -1}{a}\right)-1=0
\end{equation}
Using (\ref{B1}), (\ref{def-Bj}), (\ref{V-ineq}),  an easy induction shows that (see Theorem 3.2 in \cite{DFFR}) for $j\geq 2$
\begin{equation}\label{bound-V-B}
\|(V^{(\bold{k},\bold{q})_{-1}}_{J_{\bold{k},\bold{q}}})_j\|_{H^0}\leq B_j\,\Big(\frac{e^{2c a}-2c a-1}{a}\Big)+2\|V^{(\bold{k},\bold{q})_{-1}}_{J_{\bold{k},\bold{q}}}\|_{H^0}\,B_{j-1}\Big(\frac{e^{2c a}-1}{a}\Big)\,.
\end{equation}
From (\ref{B1}) and (\ref{def-Bj}) it also follows that
\begin{equation}\label{bound-b}
  B_j\geq \frac{2B_{j-1}\|\,V^{(\bold{k},\bold{q})_{-1}}_{J_{\bold{k},\bold{q}}}\,\|_{H^0}}{a}\,\quad \Rightarrow\quad B_{j-1}\leq a\frac{B_j}{2\|\,V^{(\bold{k},\bold{q})_{-1}}_{J_{\bold{k},\bold{q}}}\,\|_{H^0}}\,,
\end{equation} 
which, along with (\ref{bound-V-B}) and (\ref{a-eq}), yields
\begin{equation}\label{bound-b-bis}
B_j\geq \|\,(V^{(\bold{k},\bold{q})_{-1}}_{J_{\bold{k},\bold{q}}})_j\|_{H^0}\,.
\end{equation} 
The numbers $B_j$ are the Taylor coefficients of the function
\begin{equation}
f(x):=\frac{a}{2}\cdot \left(\,1-\sqrt{1- (\frac{4}{a}\cdot \|V^{(\bold{k},\bold{q})_{-1}}_{J_{\bold{k},\bold{q}}}\|_{H^0}) \,x }\,\right)\,,
\end{equation}
(see  \cite{DFFR}). We observe that
\begin{eqnarray}
\|(V^{(\bold{k},\bold{q})_{-1}}_{J_{\bold{k},\bold{q}}})^{diag}_j \|_{H^0}&=& \max\{\|P^{(+)}_{J_{\bold{k},\bold{q}}}(V^{(\bold{k},\bold{q})_{-1}}_{J_{\bold{k},\bold{q}}})_j P^{(+)}_{J_{\bold{k},\bold{q}}}\|_{H^0}\,,\,\|P^{(-)}_{J_{\bold{k},\bold{q}}}(V^{(\bold{k},\bold{q})_{-1}}_{J_{\bold{k},\bold{q}}})_j P^{(-)}_{J_{\bold{k},\bold{q}}}\|_{H^0}\}\label{diag-est-in}\\
& =&\max_{\#=\pm }\,\|(\frac{1}{H^0_{J_{\bold{k},\bold{q}}}+1})^{\frac{1}{2}}P^{(\#)}_{J_{\bold{k},\bold{q}}}(V^{(\bold{k},\bold{q})_{-1}}_{J_{\bold{k},\bold{q}}})_j P^{(\#)}_{J_{\bold{k},\bold{q}}}(\frac{1}{H^0_{J_{\bold{k},\bold{q}}}+1})^{\frac{1}{2}}\|\\
&=&\max_{\#=\pm }\,\|P^{(\#)}_{J_{\bold{k},\bold{q}}}(\frac{1}{H^0_{J_{\bold{k},\bold{q}}}+1})^{\frac{1}{2}}(V^{(\bold{k},\bold{q})_{-1}}_{J_{\bold{k},\bold{q}}})_j (\frac{1}{H^0_{J_{\bold{k},\bold{q}}}+1})^{\frac{1}{2}}P^{(\#)}_{J_{\bold{k},\bold{q}}}\|\\
&\leq & \|(V^{(\bold{k},\bold{q})_{-1}}_{J_{\bold{k},\bold{q}}})_j \|_{H^0}\,.\label{diag-est-fin}
\end{eqnarray}
Therefore, the radius of analyticity, $t_0$,  of 
\begin{equation}
\sum_{j=1}^{\infty}t^{j-1}\|(V^{(\bold{k},\bold{q})_{-1}}_{J_{\bold{k},\bold{q}}})^{diag}_j \|_{H^0}=\frac{d}{dt}\,\Big(\sum_{j=1}^{\infty}\frac{t^{j}}{j}\|(V^{(\bold{k},\bold{q})_{-1}}_{J_{\bold{k},\bold{q}}})^{diag}_j \|_{H^0}\Big)
\end{equation}
is bounded from below by the radius of analyticity of $\sum_{j=1}^{\infty}x^jB_j$, i.e.,
\begin{equation}\label{radius}
t_0\geq \frac{a}{4\|V^{(\bold{k},\bold{q})_{-1}}_{J_{\bold{k},\bold{q}}}\|_{H^0}}\geq \frac{a}{4}
\end{equation}
where we have assumed $0<t<1$ and invoked  the assumption $\|V^{(\bold{k},\bold{q})_{-1}}_{I_{r,i}}\|_{H^0}\leq t^{\frac{r-1}{4}}$.
By virtue of the inequality in (\ref{Snorm}), the same bound holds for the radius of convergence of the series $S_{J_{\bold{k},\bold{q}}}:=\sum_{j=1}^{\infty}t^j(S_{J_{\bold{k},\bold{q}}})_j\,$\,.
For $0<t<1$ and in the interval $(0,\frac{a}{8})$, by using (\ref{B1}) and  (\ref{bound-b-bis}) we can estimate
\begin{eqnarray}
\sum_{j=1}^{\infty}t^{j-1}\|(V^{(\bold{k},\bold{q})_{-1}}_{J_{\bold{k},\bold{q}}})^{diag}_j \|_{H^0}&\leq &\frac{1}{t}\sum_{j=1}^{\infty}t^jB_j\\
&=&\frac{1}{t}\cdot \frac{a}{2}\cdot \left(\,1-\sqrt{1- (\frac{4}{a}\cdot \|V^{(\bold{k},\bold{q})_{-1}}_{J_{\bold{k},\bold{q}}}\|_{H^0}) \,t }\,\right)\\
&\leq &(1+C_a \cdot t )\,\|V^{(\bold{k},\bold{q})_{-1}}_{J_{\bold{k},\bold{q}}}\|_{H^0}
\end{eqnarray}
for some $a$-dependent constant $C_a>0$.
Hence the inequality in (\ref{bound-V})  holds true, as long as $t$ is sufficiently small, independently of $N$, $k$, and $q$. 
Inequality  (\ref{bound-S}) and (\ref{S-Hest}) can be derived in a similar way, using  (\ref{S-in})-(\ref{S-fin}) and (\ref{S-norm-2}), respectively. 

\noindent
As far as (\ref{Vsquare}) is concerned, we start from 
\begin{equation}
\|\sum_{j=2}^{\infty}t^{j}(S_{J_{\bold{k},\bold{q}}})_j\|\leq\sum_{j=2}^\infty t^j\|(S_{J_{\bold{k},\bold{q}}})_j\|
\leq 4\sum_{j=2}^\infty t^j \|(V^{(\bold{k},\bold{q})_{-1}}_{J_{\bold{k},\bold{q}}})_j\|_{H_0}\leq 4\sum_{j=2}^{\infty}t^jB_j\\
\end{equation}
then, using $B_1\equiv \|V^{(\bold{k},\bold{q})_{-1}}_{J_{\bold{k},\bold{q}}}\|_{H_0}$ and a Taylor expansion, we estimate
\begin{eqnarray}
\sum_{j=2}^\infty t^jB_{j}&=&\frac{a}{2}\cdot \Big(\,1-\sqrt{1-\frac{4}{a}\cdot \|V^{(\bold{k},\bold{q})_{-1}}_{J_{\bold{k},\bold{q}}}\|_{H_0}\,t }\,\Big)\,-t\cdot \|V^{(\bold{k},\bold{q})_{-1}}_{J_{\bold{k},\bold{q}}}\|_{H_0}\\
&\leq&D_a\cdot t\cdot \|V^{(\bold{k},\bold{q})_{-1}}_{J_{\bold{k},\bold{q}}}\|_{H_0}^2
\end{eqnarray}
where $D_a$ depends only on $a$.
\qed

\begin{lem}\label{weightcontrol}
Assume that $t\geq 0$ is sufficiently small so that (\ref{inductive-reg-1}), (\ref{inductive-reg-2}), (\ref{R3-2}), and (\ref{R3-3}) hold for the potentials associated with any rectangle $J_{\bold{l}',\bold{i}'}$, with $(\bold{l}', \bold{i}')\preceq (\bold{r}, \bold{i})$,  in steps $(\bold{k}', \bold{q}')\prec (\bold{k}, \bold{q})$.
Then for $\mathfrak{b}\in\mathcal{B}_{V^{(\bold{k},\bold{q})}_{J_{\bold{r},\bold{i}}}}$ (see Definition \ref{def-tree}) we have
\begin{equation}\label{bnorm}
\|\mathfrak{b}\|_{H^0}:=\|(H^0_{J_{\bold{r},\bold{i}}}+1)^{-\frac{1}{2}}\,\mathfrak{b}\, (H^0_{J_{\bold{r},\bold{i}}}+1)^{-\frac{1}{2}}\|\leq t^{\frac{r-1}{3}}\, \prod_{R\in\mathcal{R}_{\mathfrak{b}}} ((c+1)\frac{t^{\frac{1}{3}}}{(\rho(R))^{x_d}})
\end{equation}
where $c$ is a universal constant and $\rho(R)$ is the size of $R\in\mathcal{R}_{\mathfrak{b}}$, i.e. $R=J_{\bold{s},\bold{u}}$ for some ${\bold{s},\bold{u}}$ and $\rho(R)=s$.
\end{lem}
\emph{Proof}

We first show that
\begin{eqnarray}
& &\|\mathfrak{b}\|_{H^0}:=\|(H^0_{J_{\bold{r},\bold{i}}}+1)^{-\frac{1}{2}}\,\mathfrak{b}\, (H^0_{J_{\bold{r},\bold{i}}}+1)^{-\frac{1}{2}}\|\label{R1normest}\\
&=&\|(\frac{1}{H^0_{J_{\bold{r},\bold{i}}}+1})^{\frac{1}{2}}\mathcal{A}_{J_{\bold{k}^{(1)},\bold{q}^{(1)}}}(\,(\cdots  \mathcal{A}_{J_{\bold{k}^{(\vert\mathcal{R}_b\vert -1)},\bold{q}^{(\vert\mathcal{R}_b\vert -1)}}}(V_{\mathcal{L}_\mathfrak{b}}))\cdots )\, )(\frac{1}{H^0_{J_{\bold{r},\bold{i}}}+1})^{\frac{1}{2}}\|\nonumber\\
&\leq &(C\cdot t)^{\vert \mathcal{R}_b\vert -1}\,\|V_{\mathcal{L}_{\mathfrak{b}}}\|_{H^0} \prod_{i\in \{1,\cdots,\vert\mathcal{R}_b\vert -1\}}\|V^{(\bold{k}^{(i)},\bold{q}^{(i)})_{-1}}_{J_{\bold{k}^{(i)},\bold{q}^{(i)}}} \|_{H^0},\label{intermest}
\end{eqnarray}
where $V_{\mathcal{L}_{\mathfrak{b}}}$ is the potential associated with the leaf of $\mathfrak{b}$ (point 8. of Definition \ref{def-tree}).
The estimate in (\ref{intermest}) follows directly from iterating the inequality
\begin{eqnarray}
& &\|(\frac{1}{H^0_{\mathcal{J}^{(j)}}+1})^{\frac{1}{2}} \mathcal{A}_{J_{\bold{k}^{(j)},\bold{q}^{(j)}}}(B) (\frac{1}{H^0_{\mathcal{J}^{(j)}}+1})^{\frac{1}{2}}\| \label{intermest-bis}\\
&\leq&(C\cdot t) \,\|V^{(\bold{k}^{(j)},\bold{q}^{(j)})_{-1}}_{J_{\bold{k}^{(j)},\bold{q}^{(j)}}} \|_{H^0}\, \|(\frac{1}{H^0_{\mathcal{J}^{(j+1)}}+1})^{\frac{1}{2}} B\, (\frac{1}{H^0_{\mathcal{J}^{(j+1)}}+1})^{\frac{1}{2}} \|\label{intermest-bis-bis}
\end{eqnarray}
where 
$$B:=\mathcal{A}_{J_{\bold{k}^{(j)},\bold{q}^{(j)}}}(\,(\cdots  \mathcal{A}_{J_{\bold{k}^{(\vert\mathcal{R}_b\vert-1)},\bold{q}^{(\vert\mathcal{R}_b\vert-1)}}}(V_{\mathcal{L}_{\mathfrak{b}}}))\cdots )\, )$$
 and
\begin{equation}
\mathcal{J}^{(j)}=[\cup_{i\in \{j,j+1,\cdots,\vert\mathcal{R}_b\vert\}}J_{\bold{k}^{(i)},\bold{q}^{(i)}}]
\end{equation}
is the minimal rectangle associated with the connected set $\cup_{i\in \{l,l+1,\cdots,\vert\mathcal{R}_b\vert\}}J_{\bold{k}^{(i)},\bold{q}^{(i)}}$.

\noindent
In order to show the inequality in (\ref{intermest-bis})-(\ref{intermest-bis-bis}) we have to control
\begin{equation}
\|(\frac{1}{H^0_{\mathcal{J}^{(j)}}+1})^{\frac{1}{2}} \mathcal{A}_{J_{\bold{k}^{(j)},\bold{q}^{(j)}}}(B) (\frac{1}{H^0_{\mathcal{J}^{(j)}}+1})^{\frac{1}{2}}\|=\|\sum_{n=1}^{\infty}\frac{1}{n!}\,(H_{\mathcal{J}^{(j)}}^0+1)^{-\frac{1}{2}}\,ad^{n}S_{J_{\bold{k}^{(j)},\bold{q}^{(j)}}}(B)\,(H_{\mathcal{J}^{(j)}}^0+1)^{-\frac{1}{2}}\|\,.
\end{equation}
This amounts to study terms of the type
\begin{equation}
(H_{\mathcal{J}^{(j)}}^0+1)^{-\frac{1}{2}}\,S_{J_{\bold{k}^{(j)},\bold{q}^{(j)}}}\dots S_{J_{\bold{k}^{(j)},\bold{q}^{(j)}}} B\, S_{J_{\bold{k}^{(j)},\bold{q}^{(j)}}}\dots S_{J_{\bold{k}^{(j)},\bold{q}^{(j)}}}\,(H_{\mathcal{J}^{(j)}}^0+1)^{-\frac{1}{2}}
\end{equation}
that we re-write as
\begin{equation}
(H_{\mathcal{J}^{(j)}}^0+1)^{-\frac{1}{2}}\,S_{J_{\bold{k}^{(j)},\bold{q}^{(j)}}}\dots S_{J_{\bold{k}^{(j)},\bold{q}^{(j)}}}(H_{\mathcal{J}^{(j+1)}}^0+1)^{\frac{1}{2}}(H_{\mathcal{J}^{(j+1)}}^0+1)^{-\frac{1}{2}} B\, S_{J_{\bold{k}^{(j)},\bold{q}^{(j)}}}\dots S_{J_{\bold{k}^{(j)},\bold{q}^{(j)}}}\,(H_{\mathcal{J}^{(j)}}^0+1)^{-\frac{1}{2}}\,.
\end{equation}
We estimate the norm of
\begin{equation}
(H_{\mathcal{J}^{(j)}}^0+1)^{-\frac{1}{2}}\,S_{J_{\bold{k}^{(j)},\bold{q}^{(j)}}}\dots S_{J_{\bold{k}^{(j)},\bold{q}^{(j)}}}(H_{\mathcal{J}^{(j+1)}}^0+1)^{\frac{1}{2}}\,.
\end{equation}
by  inserting $\charf=(H_{\mathcal{J}^{(j+1)}\setminus J_{\bold{k}^{(j)},\bold{q}^{(j)}}}^0+1)^{\frac{1}{2}}(H_{\mathcal{J}^{(j+1)}\setminus J_{\bold{k}^{(j)},\bold{q}^{(j)}}}^0+1)^{-\frac{1}{2}}$ and exploiting $$[H_{\mathcal{J}^{(j+1)}\setminus J_{\bold{k}^{(j)},\bold{q}^{(j)}}}^0\,,\,S_{J_{\bold{k}^{(j)},\bold{q}^{(j)}}}]=0$$ that holds since the two supports, $\mathcal{J}^{(j+1)}\setminus J_{\bold{k}^{(j)},\bold{q}^{(j)}}$ and $J_{\bold{k}^{(j)},\bold{q}^{(j)}}$, are nonoverlapping by construction. Here $H_{\mathcal{J}^{(j+1)}\setminus J_{\bold{k}^{(j)},\bold{q}^{(j)}}}$ is of course naturally defined even if $\mathcal{J}^{(j+1)}\setminus J_{\bold{k}^{(j)},\bold{q}^{(j)}}$ is not necessarily a rectangle. Consequently we can write
\begin{eqnarray}
& &(H_{\mathcal{J}^{(j)}}^0+1)^{-\frac{1}{2}}(H_{\mathcal{J}^{(j+1)}\setminus J_{\bold{k}^{(j)},\bold{q}^{(j)}}}^0+1)^{\frac{1}{2}}(H_{\mathcal{J}^{(j+1)}\setminus J_{\bold{k}^{(j)},\bold{q}^{(j)}}}^0+1)^{-\frac{1}{2}}\,S_{J_{\bold{k}^{(j)},\bold{q}^{(j)}}}\dots S_{J_{\bold{k}^{(j)},\bold{q}^{(j)}}}(H_{\mathcal{J}^{(j)}}^0+1)^{\frac{1}{2}}\nonumber \\
&=&(H_{\mathcal{J}^{(j)}}^0+1)^{-\frac{1}{2}}(H_{\mathcal{J}^{(j+1)}\setminus J_{\bold{k}^{(j)},\bold{q}^{(j)}}}^0+1)^{\frac{1}{2}}\,S_{J_{\bold{k}^{(j)},\bold{q}^{(j)}}}\dots S_{J_{\bold{k}^{(j)},\bold{q}^{(j)}}}(H_{\mathcal{J}^{(j+1)}\setminus J_{\bold{k}^{(j)},\bold{q}^{(j)}}}^0+1)^{-\frac{1}{2}}(H_{\mathcal{J}^{(j)}}^0+1)^{\frac{1}{2}}\,.\nonumber
\end{eqnarray}
The key inequality in (\ref{intermest-bis})-(\ref{intermest-bis-bis}) is obtained by making use of:
\begin{itemize}
\item
the results from Lemma \ref{control-LS} that we can exploit due to the inductive hypothesis for S2)
\begin{equation}\label{bound-S}
\|S_{J_{\bold{k}^{(j)},\bold{q}^{(j)}}}\|\leq C\, t\,
 \| V^{(\bold{k}^{(j)},\bold{q}^{(j)})_{-1}}_{J_{\bold{k}^{(j)},\bold{q}^{(j)}}}\|_{H_0}
\end{equation}
\begin{equation}
\|S_{J_{\bold{k}^{(j)},\bold{q}^{(j)}}}(H^0_{J_{\bold{k}^{(j)},\bold{q}^{(j)}}}+1)^{\frac{1}{2}}\| \leq C\,t\,
 \| V^{(\bold{k}^{(j)},\bold{q}^{(j)})_{-1}}_{J_{\bold{k}^{(j)},\bold{q}^{(j)}}}\|_{H_0}\,;
\end{equation}
\item
the operator norm bound
\begin{equation}
\|(H_{\mathcal{J}^{(j)}}^0+1)^{-\frac{1}{2}}(H_{\mathcal{J}^{(j+1)}\setminus J_{\bold{k}^{(j)},\bold{q}^{(j)}}}^0+1)^{\frac{1}{2}}\|\leq 1
\end{equation}
that follows  from the spectral theorem for commuting operators and from the inclusion $\mathcal{J}^{(j+1)}\setminus J_{\bold{k}^{(j)},\bold{q}^{(j)}}\subset \mathcal{J}^{(j)}$;
\item the operator norm bound
\begin{equation}
\|(H^0_{J_{\bold{k}^{(j)},\bold{q}^{(j)}}}+1)^{-\frac{1}{2}}(H_{\mathcal{J}^{(j+1)}\setminus J_{\bold{k}^{(j)},\bold{q}^{(j)}}}^0+1)^{-\frac{1}{2}}(H_{\mathcal{J}^{(j+1)}}^0+1)^{\frac{1}{2}}\|\leq 1
\end{equation}
that follows  from the spectral theorem for commuting operators.
\end{itemize}
Now, starting from (\ref{intermest}), we can use the inductive hypothesis, 
\begin{align*}
 &(C\cdot t)^{\vert \mathcal{R}_b\vert -1}\,\|V_{\mathcal{L}_{\mathfrak{b}}}\|_{H^0} \prod_{i\in \{1,\cdots,\vert\mathcal{R}_b\vert -1\}}\|V^{(\bold{k}^{(i)},\bold{q}^{(i)})_{-1}}_{J_{\bold{k}^{(i)},\bold{q}^{(i)}}} \|_{H^0}\leq  (C\cdot t)^{\vert \mathcal{R}_b\vert -1}\, \prod_{R\in\mathcal{R}_{\mathfrak{b}}} (c^\prime \, \frac{t^{\frac{\rho(R)-1}{3}}}{\rho(R)^{x_d}})\\
\leq& t^{\frac{r-\vert \mathcal{R}_{\mathfrak{b}}\vert}{3}} (C\cdot t)^{\vert \mathcal{R}_b\vert -1}\,\prod_{R\in\mathcal{R}_{\mathfrak{b}}} (c^\prime\, \frac{1}{\rho(R)^{x_d}})\leq  t^{\frac{r- 1}{3}} (C\cdot t^{2/3})^{\vert \mathcal{R}_b\vert -1}\,\prod_{R\in\mathcal{R}_{\mathfrak{b}}} (c^\prime\, \frac{1}{\rho(R)^{x_d}})\leq t^{\frac{r- 1}{3}} \,\prod_{R\in\mathcal{R}_{\mathfrak{b}}} ((c+1)\, \frac{t^{1/3}}{\rho(R)^{x_d}}),
\end{align*}
for some universal constants $c$ and $c^\prime$, where the second inequality is due to the requirement that $J_{\bold{r},\bold{i}}$ is the minimal rectangle associated with $\cup_{i\in  \{1,\cdots,\vert \mathcal{R}_{\mathfrak{b}}\vert\}}J_{\bold{k}^{(i)},\bold{q}^{(i)}}$ and the last inequality uses $\vert \mathcal{R}_{\mathfrak{b}}\vert \geq 2$.

\qed

Recall the notation
\begin{equation}
(\frac{1}{[\pi_{J_{\bold{r},\bold{i}}}]_1})^{\frac{1}{2}}
:=(\frac{1}{\pi_{J_{\bold{r},\bold{i}}}+1})^{\frac{1}{2}}:=(\frac{1}{\sum_{\bold{j}\in J_{\bold{r},\bold{i}}} P^{\perp}_{\Omega_{\bold{j}}}+1})^{\frac{1}{2}}\quad 
\end{equation}
\begin{equation}
 (\frac{1}{[H^0_{J_{\bold{r},\bold{i}}}]_1})^{\frac{1}{2}}
:=(\frac{1}{\sum_{\bold{j}\in J_{\bold{r},\bold{i}}} H_{\bold{j}}+1})^{\frac{1}{2}}\,.
\end{equation}
\begin{lem}\label{inductive-aux}
Assume that $t\geq 0$ is sufficiently small so that (\ref{inductive-reg-1}), (\ref{inductive-reg-2}), (\ref{R3-2}), and (\ref{R3-3}) hold for the potentials $V_{J_{\bold{k}',\bold{q}'}}^{(\bold{k}, \bold{q})_{-1}}$ associated with any rectangle $J_{\bold{k}',\bold{q}'}$, with $(\bold{k}', \bold{q}')\preceq (\bold{k}, \bold{q})$,  in step $(\bold{k}, \bold{q})_{-1}$.
Then the following estimate holds:
\begin{eqnarray}
& &\|([H^0_{J_{\bold{k},\bold{q}}}]_1)^{\frac{1}{2}}\, (\frac{1}{[\pi_{J_{\bold{k},\bold{q}}}]_1})^{\frac{1}{2}}\frac{1}{G_{J_{\bold{k},\bold{q}}}-E_{J_{\bold{k},\bold{q}}}}\,P^{(+)}_{J_{\bold{k},\bold{q}}}\,V^{(\bold{k},\bold{q})_{-1}}_{J_{\bold{k},\bold{q}}}\,P^{(-)}_{J_{\bold{k},\bold{q}}}\|\\
&\leq & C_d\cdot  \| (\frac{1}{[\pi_{J_{\bold{k},\bold{q}}}]_1})^{\frac{1}{2}}\,\,(\frac{1}{[H^0_{J_{\bold{k},\bold{q}}}]_1})^{\frac{1}{2}}\,P^{(+)}_{J_{\bold{k},\bold{q}}}\,V^{(\bold{k},\bold{q})_{-1}}_{J_{\bold{k},\bold{q}}}\,P^{(-)}_{J_{\bold{k},\bold{q}}}\|\,,
\end{eqnarray}
where $C_d>0$ is a $d$-dependent constant (i.e., it does not depend on $(\bold{k}, \bold{q})$ or $\bold{N}$).
\end{lem}

\noindent
\emph{Proof}

\noindent

We introduce the definition
\begin{eqnarray}
\mathcal{V}^{(\bold{k},\bold{q})_{-1}}_{J_{\bold{k},\bold{q}}}&:=&P^{(+)}_{J_{\bold{k},\bold{q}}}\,\Big[t\sum_{J_{\bold{k}_{(1)}',\bold{q}'}\subset J_{\bold{k},\bold{q}}}  V^{(\bold{k},\bold{q})_{-1}}_{J_{\bold{k}_{(1)}',\bold{q}'}} P^{(+)}_{J_{\bold{k}_{(1)}',\bold{q}'}}  +\dots+t\sum_{J_{\bold{k}'_{(|\bold{k}|-1)},\bold{q}'} \subset J_{\bold{k},\bold{q}}} V^{(\bold{k},\bold{q})_{-1}}_{J_{\bold{k}'_{(|\bold{k}|-1)}},\bold{q}'}} P^{(+)}_{J_{\bold{k}_{(|\bold{k}|-1)}',\bold{q}'}} \Big]P^{(+)}_{J_{\bold{k},\bold{q}}\quad \quad \\
& &-P^{(+)}_{J_{\bold{k},\bold{q}}}\,\Big[t\sum_{J_{\bold{k}_{(1)}',\bold{q}'}\subset J_{\bold{k},\bold{q}}} \langle V^{(\bold{k},\bold{q})_{-1}}_{J_{\bold{k}_{(1)}',\bold{q}'}}\rangle P^{(+)}_{J_{\bold{k}_{(1)}',\bold{q}'}}  +\dots+t\sum_{J_{\bold{k}'_{(|\bold{k}|-1)},\bold{q}'} \subset J_{\bold{k},\bold{q}}}\langle V^{(\bold{k},\bold{q})_{-1}}_{J_{\bold{k}'_{(|\bold{k}|-1)}},\bold{q}'}}\rangle P^{(+)}_{J_{\bold{k}_{(|\bold{k}|-1)}',\bold{q}'}} \Big]P^{(+)}_{J_{\bold{k},\bold{q}}\quad \quad \quad
\end{eqnarray}
and make use of the Neumann expansion
\begin{equation}
P^{(+)}_{J_{\bold{k},\bold{q}}}\frac{1}{G_{J_{\bold{k},\bold{q}}}-E_{J_{\bold{k},\bold{q}}}}P^{(+)}_{J_{\bold{k},\bold{q}}}=\sum_{j=0}^{\infty}\frac{1}{P^{(+)}_{J_{\bold{k},\bold{q}}}H^0_{J_{\bold{k},\bold{q}}}P^{(+)}_{J_{\bold{k},\bold{q}}}}\Big\{-\mathcal{V}^{(\bold{k},\bold{q})_{-1}}_{J_{\bold{k},\bold{q}}}\frac{1}{P^{(+)}_{J_{\bold{k},\bold{q}}}H^0_{J_{\bold{k},\bold{q}}}P^{(+)}_{J_{\bold{k},\bold{q}}}}\}^j\,
\end{equation}
(that is justified by assuming (\ref{inductive-reg-1}), (\ref{inductive-reg-2}), (\ref{R3-2}), and (\ref{R3-3})).
We write
\begin{equation}
\left(\frac{1}{\sum_{\bold{j}\in J_{\bold{k},\bold{q}}} P^{\perp}_{\Omega_{\bold{j}}}+1}\right)^{\frac{1}{2}}=f(\pi_{J_{\bold{k},\bold{q}}})
\end{equation}
where $\pi_{J_{\bold{k},\bold{q}}}:=\sum_{\bold{j}\in J_{\bold{k},\bold{q}}} P^{\perp}_{\Omega_{\bold{j}}}$, where $f({\color{blue}x}):=\frac{1}{\sqrt{x+1}}$.\\
Although $f(\pi_{J_{\bold{k},\bold{q}}})$ can be understood as the usual continuous functional calculus
of the self-adjoint operator $\pi_{J_{\bold{k}, \bold{q}}}$, in the following computations it will be more convenient to represent it through the Helffer-Sj{\"o}strand calculus (see \cite{HS}, \cite{D95}), which we rather quickly recall here.\\
Since the operator $\pi_{J_{\bold{k},\bold{q}}}$ is bounded with spectrum in $[0, \|\pi_{J_{\bold{k},\bold{q}}}\|]$, we consider a compactly supported smooth positive  function, which by a slight abuse of notation we still denote by $f$, coinciding with $\frac{1}{\sqrt{x+1}}$ in the interval $[0, \|\pi_{J_{\bold{k},\bold{q}}}\|]$ and being $0$ for $x\leq -\frac{1}{2}$ and $x\geq 2 \|\pi_{J_{\bold{k},\bold{q}}}\|$.\\
We then consider an almost analytic estension $\tilde{f}$ of $f$, which can be obtained as in \cite{D95}, namely
we set, for any $n\geq 0$,
$$\tilde{f}(x, y)=\sum_{r=0}^n f^{(r)}(x)\frac{(iy)^r}{r!}J(x,y)$$
 with $J(x,y)= \tau\left(\frac{y}{(1+x^2)^\frac{1}{2}}\right)$, $f^{(r)}(x):=(\frac{d^r}{dx^r}f)(x)$, where $\tau\in C_c^\infty(\mathbb{R})$ is $1$ in $[-1,1]$ and its support is contained
in $[-2,2]$.\\
As shown in \cite{D95}, $f(\pi_{J_{\bold{k}, \bold{q}}})$ can be represented by the following integral
$$f(\pi_{J_{\bold{k}, \bold{q}}})=-\frac{1}{\pi}\int_{\mathbb{C}}\frac{\partial \tilde{f}}{\partial \bar{z}} \frac{1}{z-\pi_{J_{\bold{k}, \bold{q}}}}\textrm{d}x\textrm{d}y$$
where $\frac{\partial}{\partial \bar{z}}$ is the differential operator $\frac{1}{2}\left(\frac{\partial}{\partial x}+ i\frac{\partial}{\partial y}\right)$ and the integral is norm-convergent. The above
integral representation depends neither on $n$ nor on the cut-off function $\tau$.
Lastly, we need to observe that since 
$$\frac{\partial f}{\partial \bar {z}}=\frac{1}{2}\sum_{r=0}^n f^{(r)}(x)\frac{(iy)^r}{r!}J(x,y)+ f^{(n+1)}(x)\frac{(iy)^n}{n!}\Big(J_x(x, y)+iJ_y(x,y)\Big)$$
for any fixed $x\in\mathbb{R}$ we have  
\begin{equation}\label{resregul}
\left|\frac{\partial f}{\partial \bar{z}}\right|= O(|y|^n)\quad \text{ as }|y|\rightarrow 0.
\end{equation}

\noindent
Now we observe that
\begin{eqnarray}
& &(\frac{1}{\sum_{\bold{j}\in J_{\bold{k},\bold{q}}} P^{\perp}_{\Omega_{\bold{j}}}+1})^{\frac{1}{2}}\,(\frac{1}{P^{(+)}_{J_{\bold{k},\bold{q}}}H^0_{J_{\bold{k},\bold{q}}}P^{(+)}_{J_{\bold{k},\bold{q}}}})^{\frac{1}{2}}\mathcal{V}^{(\bold{k},\bold{q})_{-1}}_{J_{\bold{k},\bold{q}}}(\frac{1}{P^{(+)}_{J_{\bold{k},\bold{q}}}H^0_{J_{\bold{k},\bold{q}}}P^{(+)}_{J_{\bold{k},\bold{q}}}})^{\frac{1}{2}}\\
&=&(\frac{1}{P^{(+)}_{J_{\bold{k},\bold{q}}}H^0_{J_{\bold{k},\bold{q}}}P^{(+)}_{J_{\bold{k},\bold{q}}}})^{\frac{1}{2}}\, f(\pi_{J_{\bold{k},\bold{q}}})\,\mathcal{V}^{(\bold{k},\bold{q})_{-1}}_{J_{\bold{k},\bold{q}}}(\frac{1}{P^{(+)}_{J_{\bold{k},\bold{q}}}H^0_{J_{\bold{k},\bold{q}}}P^{(+)}_{J_{\bold{k},\bold{q}}}})^{\frac{1}{2}}\\
&=&(\frac{1}{P^{(+)}_{J_{\bold{k},\bold{q}}}H^0_{J_{\bold{k},\bold{q}}}P^{(+)}_{J_{\bold{k},\bold{q}}}})^{\frac{1}{2}}\mathcal{V}^{(\bold{k},\bold{q})_{-1}}_{J_{\bold{k},\bold{q}}}f(\pi_{J_{\bold{k},\bold{q}}})\,(\frac{1}{P^{(+)}_{J_{\bold{k},\bold{q}}}H^0_{J_{\bold{k},\bold{q}}}P^{(+)}_{J_{\bold{k},\bold{q}}}})^{\frac{1}{2}}\\
& &+(\frac{1}{P^{(+)}_{J_{\bold{k},\bold{q}}}H^0_{J_{\bold{k},\bold{q}}}P^{(+)}_{J_{\bold{k},\bold{q}}}})^{\frac{1}{2}}\,[f(\pi_{J_{\bold{k},\bold{q}})}\,,\,\mathcal{V}^{(\bold{k},\bold{q})_{-1}}_{J_{\bold{k},\bold{q}}}](\frac{1}{P^{(+)}_{J_{\bold{k},\bold{q}}}H^0_{J_{\bold{k},\bold{q}}}P^{(+)}_{J_{\bold{k},\bold{q}}}})^{\frac{1}{2}}
\end{eqnarray}
with
\begin{eqnarray}
& &[f(\pi_{J_{\bold{k},\bold{q}}})\,,\,\mathcal{V}^{(\bold{k},\bold{q})_{-1}}_{J_{\bold{k},\bold{q}}}]\\
& =&[-\frac{1}{\pi}\int_{\mathbb{C}}\,\frac{\partial \tilde{f}}{\partial \bar{z}}  \frac{1}{z-\pi_{J_{\bold{k},\bold{q}}}}\textrm{d}x\textrm{d}y\,,\,\mathcal{V}^{(\bold{k},\bold{q})_{-1}}_{J_{\bold{k},\bold{q}}}]\\
&=&-\frac{1}{\pi}\int_{\mathbb{C}} \frac{\partial f}{\partial \bar{z}} [\frac{1}{z-\pi_{J_{\bold{k}, \bold{q}}}}, \mathcal{V}^{(\bold{k},\bold{q})_{-1}}_{J_{\bold{k},\bold{q}}}]\textrm{d}x\textrm{d}y\label{firstint}\\
&=&-\frac{1}{\pi}\int_{\mathbb{C}}\,\frac{\partial \tilde{f}}{\partial \bar{z}}  \frac{1}{z-\pi_{J_{\bold{k},\bold{q}}}}\,[\pi_{J_{\bold{k},\bold{q}}}\,,\,\mathcal{V}^{(\bold{k},\bold{q})_{-1}}_{J_{\bold{k},\bold{q}}}]\, \frac{1}{z-\pi_{J_{\bold{k},\bold{q}}}}\textrm{d}x\textrm{d}y\label{secondint}\\
&=&-\frac{1}{\pi}\int_{\mathbb{C}}\,\frac{\partial \tilde{f}}{\partial \bar{z}} \,[\pi_{J_{\bold{k},\bold{q}}}\,,\,\mathcal{V}^{(\bold{k},\bold{q})_{-1}}_{J_{\bold{k},\bold{q}}}]\, \frac{1}{z-\pi_{J_{\bold{k},\bold{q}}}}\, \frac{1}{z-\pi_{J_{\bold{k},\bold{q}}}}\textrm{d}x\textrm{d}y\\
& &-\frac{1}{\pi}\int_{\mathbb{C}}\,\frac{\partial \tilde{f}}{\partial \bar{z}} \,[ \frac{1}{z-\pi_{J_{\bold{k},\bold{q}}}}\,,\,[\pi_{J_{\bold{r},\bold{i}}}\,,\,\mathcal{V}^{(\bold{k},\bold{q})_{-1}}_{J_{\bold{k},\bold{q}}}]\,]\, \frac{1}{z-\pi_{J_{\bold{k},\bold{q}}}}\textrm{d}x\textrm{d}y\label{secondlastint}\\
&=&-\frac{1}{\pi}\int_{\mathbb{C}}\,\frac{\partial \tilde{f}}{\partial \bar{z}} \,[\pi_{J_{\bold{k},\bold{q}}}\,,\,\mathcal{V}^{(\bold{k},\bold{q})_{-1}}_{J_{\bold{k},\bold{q}}}]\, \frac{1}{z-\pi_{J_{\bold{k},\bold{q}}}}\, \frac{1}{z-\pi_{J_{\bold{k},\bold{q}}}}\textrm{d}x\textrm{d}y\\
& &-\frac{1}{\pi}\int_{\mathbb{C}}\,\frac{\partial \tilde{f}}{\partial \bar{z}} \,\frac{1}{z-\pi_{J_{\bold{k},\bold{q}}}}[\pi_{J_{\bold{k},\bold{q}}}\,,\,[\pi_{J_{\bold{k},\bold{q}}}\,,\,\mathcal{V}^{(\bold{k},\bold{q})_{-1}}_{J_{\bold{k},\bold{q}}}]\,]\, \frac{1}{z-\pi_{J_{\bold{k},\bold{q}}}}\frac{1}{z-\pi_{J_{\bold{k},\bold{q}}}}\textrm{d}x\textrm{d}y\label{lastint}
\end{eqnarray}
which is a well defined operator when sandwiched with the weight $(\frac{1}{P^{(+)}_{J_{\bold{k},\bold{q}}}H^0_{J_{\bold{k},\bold{q}}}P^{(+)}_{J_{\bold{k},\bold{q}}}})^{\frac{1}{2}}$. 
Note that the formula displayed in (\ref{secondint}) is obtained out of the formula displayed in (\ref{firstint}) by means of a simple application of the  the general  identity
$$\frac{\pi_{J_{\bold{k}, \bold{q}}}}{z-\pi_{J_{\bold{k}, \bold{q}}}}=\frac{z}{z-\pi_{J_{\bold{k},\bold{q}}}}-1$$
which is similarly used to arrive at (\ref{lastint}) from (\ref{secondlastint}).
$$$$
At this point we need to to show that the following inequality
\begin{eqnarray}\label{delineq}
& &\|(\frac{1}{P^{(+)}_{J_{\bold{k},\bold{q}}}H^0_{J_{\bold{k},\bold{q}}}P^{(+)}_{J_{\bold{k},\bold{q}}}})^{\frac{1}{2}}[f(\pi_{J_{\bold{k},\bold{q}}})\,,\,\mathcal{V}^{(\bold{k},\bold{q})_{-1}}_{J_{\bold{k},\bold{q}}}]\,(\frac{1}{P^{(+)}_{J_{\bold{k},\bold{q}}}H^0_{J_{\bold{k},\bold{q}}}P^{(+)}_{J_{\bold{k},\bold{q}}}})^{\frac{1}{2}}\|\leq
C\cdot t
\end{eqnarray}
 holds for some universal constant $C$. This is seen as follows.\\
As for the commutator
\begin{eqnarray}
& & [\pi_{J_{\bold{k},\bold{q}}}\,,\,\mathcal{V}^{(\bold{k},\bold{q})_{-1}}_{J_{\bold{k},\bold{q}}}]\\
&=&P^{(+)}_{J_{\bold{k},\bold{q}}}\Big(t\sum_{J_{\bold{k}_{(1)}',\bold{q}'}\subset J_{\bold{k},\bold{q}}}  [\pi_{J_{\bold{k},\bold{q}}},\,V^{(\bold{k},\bold{q})_{-1}}_{J_{\bold{k}_{(1)}',\bold{q}'}}] P^{(+)}_{J_{\bold{k}_{(1)}',\bold{q}'}}  +\dots+t\sum_{J_{\bold{k}'_{(|\bold{k}|-1)},\bold{q}'} \subset J_{\bold{k},\bold{q}}} [\pi_{J_{\bold{k},\bold{q}}}\,,\,V^{(\bold{k},\bold{q})_{-1}}_{J_{\bold{k}'_{(|\bold{k}|-1)}},\bold{q}'}}] P^{(+)}_{J_{\bold{k}_{(|\bold{k}|-1)}',\bold{q}'}} \Big) P^{(+)}_{J_{\bold{k},\bold{q}} \quad \quad \quad\quad \\
&=&P^{(+)}_{J_{\bold{k},\bold{q}}}\Big(t\sum_{J_{\bold{k}_{(1)}',\bold{q}'}\subset J_{\bold{k},\bold{q}}}  [\pi_{J_{\bold{k}_{(1)}',\bold{q}'}},\,V^{(\bold{k},\bold{q})_{-1}}_{J_{\bold{k}_{(1)}',\bold{q}'}}] P^{(+)}_{J_{\bold{k}_{(1)}',\bold{q}'}}  +\dots+t\sum_{J_{\bold{k}'_{(|\bold{k}|-1)},\bold{q}'} \subset J_{\bold{k},\bold{q}}} [\pi_{J_{\bold{k}_{(|\bold{k}|-1)}',\bold{q}'}}\,,\,V^{(\bold{k},\bold{q})_{-1}}_{J_{\bold{k}'_{(|\bold{k}|-1)}},\bold{q}'}}] P^{(+)}_{J_{\bold{k}_{(|\bold{k}|-1)}',\bold{q}'}} \Big) P^{(+)}_{J_{\bold{k},\bold{q}} \quad \quad \quad\quad
\end{eqnarray}
we observe that the weighted norm of each of the sums displayed in the overall sum above can in turn be estimated in the following way:

\medskip
\begin{eqnarray}
&&\|(\frac{1}{P^{(+)}_{J_{\bold{k},\bold{q}}}H^0_{J_{\bold{k},\bold{q}}}P^{(+)}_{J_{\bold{k},\bold{q}}}})^{\frac{1}{2}}\,P^{(+)}_{J_{\bold{k}, \bold{q}}}\sum_{J_{\bold{k}_{(j)}', \bold{q}'}\subset J_{\bold{k}, \bold{q}}}[\pi_{J_{\bold{k}_{(j)}',\bold{q}'}},\,V^{(\bold{k},\bold{q})_{-1}}_{J_{\bold{k}_{(j)}',\bold{q}'}}]P^{(+)}_{J_{\bold{k}_{(j)}', \bold{q}'}} P^{(+)}_{J_{\bold{k}, \bold{q}}}(\frac{1}{P^{(+)}_{J_{\bold{k},\bold{q}}}H^0_{J_{\bold{k},\bold{q}}}P^{(+)}_{J_{\bold{k},\bold{q}}}})^{\frac{1}{2}}\|\nonumber\\
&\leq& 2\|(\frac{1}{P^{(+)}_{J_{\bold{k},\bold{q}}}H^0_{J_{\bold{k},\bold{q}}}P^{(+)}_{J_{\bold{k},\bold{q}}}})^{\frac{1}{2}}\,P^{(+)}_{J_{\bold{k}, \bold{q}}}\sum_{J_{\bold{k}_{(j)}', \bold{q}'}\subset J_{\bold{k}, \bold{q}}}\,\pi_{J_{\bold{k}_{(j)}',\bold{q}'}}[P^{(+)}_{J_{\bold{k}_{(j)}',\bold{q}'}}(V^{(\bold{k}, \bold{q})_{-1}}_{J_{\bold{k}_i', \bold{q'}}}-\langle V^{(\bold{k},\bold{q})_{-1}}_{J_{\bold{k}_{(j)}',\bold{q}'}} \rangle) P^{(+)}_{J_{\bold{k}_{(j)'}, \bold{q}'}}]\times  \nonumber\\
& & \quad\quad\quad\quad\quad\quad \times  P^{(+)}_{J_{\bold{k}, \bold{q}}}\,(\frac{1}{P^{(+)}_{J_{\bold{k},\bold{q}}}H^0_{J_{\bold{k},\bold{q}}}P^{(+)}_{J_{\bold{k},\bold{q}}}})^{\frac{1}{2}}\| \quad\quad\quad \nonumber \\
&=&\sup_{\|\varphi\|=\|\psi\|=1}  |\langle \varphi\,,\, (\frac{1}{P^{(+)}_{J_{\bold{k},\bold{q}}}H^0_{J_{\bold{k},\bold{q}}}P^{(+)}_{J_{\bold{k},\bold{q}}}})^{\frac{1}{2}}\,P^{(+)}_{J_{\bold{k}, \bold{q}}}\sum_{J_{\bold{k}_{(j)}', \bold{q}'}\subset J_{\bold{k}, \bold{q}}}\,\pi_{J_{\bold{k}_{(j)}',\bold{q}'}}[P^{(+)}_{J_{\bold{k}_{(j)}',\bold{q}'}}(V^{(\bold{k}, \bold{q})_{-1}}_{J_{\bold{k}_i', \bold{q'}}}-\langle V^{(\bold{k},\bold{q})_{-1}}_{J_{\bold{k}_{(j)}',\bold{q}'}} \rangle)\times \nonumber\\
&& P^{(+)}_{J_{\bold{k}_{(j)'}, \bold{q}'}}] 
 \times P^{(+)}_{J_{\bold{k}, \bold{q}}}\,(\frac{1}{P^{(+)}_{J_{\bold{k},\bold{q}}}H^0_{J_{\bold{k},\bold{q}}}P^{(+)}_{J_{\bold{k},\bold{q}}}})^{\frac{1}{2}}\psi \rangle |\,. \label{normprod1}
\end{eqnarray}
If in the norms above we replace the operator $P^{(+)}_{J_{\bold{k}_{(j)}', \bold{q}'}}$ with
$$(H_{J_{\bold{k}_{(j)}', \bold{q}'}}^0+1)^{-\frac{1}{2}}\left(\frac{H_{J_{\bold{k}_{(j)}', \bold{q}'}}^0+1}{H_{J_{\bold{k}_{(j)}', \bold{q}'}}^0}\right)^{\frac{1}{2}}(H_{J_{\bold{k}_{(j)}', \bold{q}'}}^0)^{\frac{1}{2}}P^{(+)}_{J_{\bold{k}_{(j)}', \bold{q}'}}$$
so as to make the $H^0$- norms of the potentials appear, we can use the inductive control on the $H^0$-norms of the potentials, i.e.
\begin{equation}
 \| \,(H_{J_{\bold{k}_{(j)}',\bold{q}'}}^0+1)^{-\frac{1}{2}}(V^{(\bold{k},\bold{q})_{-1}}_{I_{\bold{k}_{(j)}', \bold{q}'}}-\langle V^{(\bold{k}, \bold{q})_{-1}}_{I_{\bold{k}_{(j)}', \bold{q}'}} \rangle)(H_{J_{\bold{k}_{(j)}', \bold{q}'}}^0+1)^{-\frac{1}{2}}\|\leq 2 |t|^{\frac{j-1}{4}}\,,
\end{equation}
along with the obvious bound $\|\pi_{J_{\bold{k}_{(j)}',\bold{q}'}}\|\leq C\cdot j^d$ 
and the Cauchy-Schwartz inequality in $\mathbb{R}^n$, so as to find the following chain of inequalities

\begin{eqnarray*}
& &(\ref{normprod1})\\
&\leq &4\cdot C\cdot j^d\cdot  |t|^{\frac{j-1}{4}} \cdot \sup_{\|\varphi\|=\|\psi\|=1}\sum_{I_{j,i}\subset I_{k,q}}\,\Big\{\|(H_{J_{\bold{k}_{(j)}',\bold{q}'}}^0)^{\frac{1}{2}}\, P^{(+)}_{J_{\bold{k}_{(j)', \bold{q}'} }}P^{(+)}_{J_{\bold{k},\bold{q}}}\Big[(\frac{1}{P^{(+)}_{J_{\bold{k},\bold{q}}}H^0_{J_{\bold{k},\bold{q}}}P^{(+)}_{J_{\bold{k},\bold{q}}}})^{\frac{1}{2}}\Big]^*\,\varphi\| \times\nonumber  \\
& &\quad \quad\quad\quad\quad \quad\quad\quad\quad\quad  \times \|(H_{J_{\bold{k}_{(j)}',\bold{q}'}}^0)^{\frac{1}{2}}\, P^{(+)}_{J_{\bold{k}_{(j)}' , \bold{q}'}}P^{(+)}_{J_{\bold{k},\bold{q}}}(\frac{1}{P^{(+)}_{J_{\bold{k},\bold{q}}}H^0_{J_{\bold{k},\bold{q}}}P^{(+)}_{J_{\bold{k},\bold{q}}}})^{\frac{1}{2}}\,\psi\|\Big\}\\
& \leq &4\cdot C\cdot j^d\cdot  |t|^{\frac{j-1}{4}} \cdot \sup_{\|\psi\|=1}\langle \psi\,,\,\Big[(\frac{1}{P^{(+)}_{J_{\bold{k},\bold{q}}}H^0_{J_{\bold{k},\bold{q}}}P^{(+)}_{J_{\bold{k},\bold{q}}}})^{\frac{1}{2}}\Big]^*\,P^{(+)}_{J_{\bold{k},\bold{q}}}\,(\sum_{I_{j,i}\subset J_{\bold{k},\bold{q}}}H_{J_{\bold{k}_{(j)}',\bold{q}'}}^0)\,P^{(+)}_{J_{\bold{k},\bold{q}}}(\frac{1}{P^{(+)}_{J_{\bold{k},\bold{q}}}H^0_{J_{\bold{k},\bold{q}}}P^{(+)}_{J_{\bold{k},\bold{q}}}})^{\frac{1}{2}}\psi \rangle \nonumber\\
&=&4\cdot C\cdot j^d\cdot  |t|^{\frac{j-1}{4}} \cdot \|\Big[(\frac{1}{P^{(+)}_{J_{\bold{k},\bold{q}}}H^0_{J_{\bold{k},\bold{q}}}P^{(+)}_{J_{\bold{k},\bold{q}}}})^{\frac{1}{2}}\Big]^*\,P^{(+)}_{J_{\bold{k},\bold{q}}}\,(\sum_{J_{\bold{k}_{(j)}',\bold{q}}\subset J_{\bold{k},\bold{q}}}H_{J_{\bold{k}_{(j)}',\bold{q}'}}^0)\,P^{(+)}_{J_{\bold{k},\bold{q}}}(\frac{1}{P^{(+)}_{J_{\bold{k},\bold{q}}}H^0_{J_{\bold{k},\bold{q}}}P^{(+)}_{J_{\bold{k},\bold{q}}}})^{\frac{1}{2}}\|\\
&\leq& 4\cdot C\cdot j^d\cdot  |t|^{\frac{j-1}{4}}C_d j^{d-1}(j+1)^d 
\end{eqnarray*}
where in the last inequality we have exploited
$\sum_{J_{\bold{k}_{(j)}', \bold{q}'}\subset J_{\bold{k},\bold{q}}} H_{J_{\bold{k}_{(i)}',\bold{q}'}}^0\leq C_d j^{d-1}(j+1)^d H^0_{J_{\bold{k}, \bold{q}}}$ (see e.g. \ref{ineq-inter-00}).\\
But then we have
$$\|[\pi_{J_{\bold{k}, \bold{q}}}, V_{J_{\bold{k}, \bold{q}}}^{(\bold{k}, \bold{q})_{-1}}]\|_{H^0}\leq\\
4 C_dt \sum_{j=1}^\infty j^{2d-1} (j+1)^d t^{\frac{j-1}{4}}\leq A t$$
as long as $(0\leq)t$ is small enough, where $A$ is a suitable constant that depends only on $d$.\\

Then we are in a position to estimate the $H^0$-norm of the commutator 
$[f(\pi_{J_{\bold{k}, \bold{q}}}), \mathcal{V}_{J_{\bold{k}, \bold{q}}}^{(\bold{k}, \bold{q})_{-1}} ]$; indeed, we have
\begin{eqnarray*}
& &\|(\frac{1}{P^{(+)}_{J_{\bold{k},\bold{q}}}H^0_{J_{\bold{k},\bold{q}}}P^{(+)}_{J_{\bold{k},\bold{q}}}})^{\frac{1}{2}}\frac{1}{\pi}\int_{\mathbb{C}}\,\frac{\partial \tilde{f}}{\partial \bar{z}} \,[\pi_{J_{\bold{k},\bold{q}}}\,,\,\mathcal{V}^{(\bold{k},\bold{q})_{-1}}_{J_{\bold{k},\bold{q}}}]\, \frac{1}{z-\pi_{J_{\bold{k},\bold{q}}}}\, \frac{1}{z-\pi_{J_{\bold{k},\bold{q}}}}(\frac{1}{P^{(+)}_{J_{\bold{k},\bold{q}}}H^0_{J_{\bold{k},\bold{q}}}P^{(+)}_{J_{\bold{k},\bold{q}}}})^{\frac{1}{2}}\textrm{d}x\textrm{d}y\| \quad\quad\quad \\
&=&\|\frac{1}{\pi}\int_{\mathbb{C}}\,\frac{\partial \tilde{f}}{\partial \bar{z}}\,(\frac{1}{P^{(+)}_{J_{\bold{k},\bold{q}}}H^0_{J_{\bold{k},\bold{q}}}P^{(+)}_{J_{\bold{k},\bold{q}}}})^{\frac{1}{2}} \,[\pi_{J_{\bold{k},\bold{q}}}\,,\,\mathcal{V}^{(\bold{k},\bold{q})_{-1}}_{J_{\bold{k},\bold{q}}}](\frac{1}{P^{(+)}_{J_{\bold{k},\bold{q}}}H^0_{J_{\bold{k},\bold{q}}}P^{(+)}_{J_{\bold{k},\bold{q}}}})^{\frac{1}{2}}\, \frac{1}{z-\pi_{J_{\bold{k},\bold{q}}}}\, \frac{1}{z-\pi_{J_{\bold{k},\bold{q}}}}\textrm{d}x\textrm{d}y\| \\
&\leq&\|(\frac{1}{P^{(+)}_{J_{\bold{k},\bold{q}}}H^0_{J_{\bold{k},\bold{q}}}P^{(+)}_{J_{\bold{k},\bold{q}}}})^{\frac{1}{2}} \,[\pi_{J_{\bold{k},\bold{q}}}\,,\,\mathcal{V}^{(\bold{k},\bold{q})_{-1}}_{J_{\bold{k},\bold{q}}}](\frac{1}{P^{(+)}_{J_{\bold{k},\bold{q}}}H^0_{J_{\bold{k},\bold{q}}}P^{(+)}_{J_{\bold{k},\bold{q}}}})^{\frac{1}{2}}\|\\
& &\times \frac{1}{\pi}\int_{\mathbb{C}}\,\, \|\frac{\partial \tilde{f}}{\partial \bar{z}}\frac{1}{z-\pi_{J_{\bold{k},\bold{q}}}}\, \frac{1}{z-\pi_{J_{\bold{k},\bold{q}}}}\|\textrm{d}x\textrm{d}y
\end{eqnarray*}
where the last norm is seen to be bounded by a universal constant as well, using the spectral theorem and (\ref{resregul}).
Since the summand coming from (\ref{lastint}) can be dealt with in much the same way, the inequality in
(\ref{delineq}) is finally got to.\\

\medskip
\noindent
We can now move on to prove the inequality in the statement. To this aim, we
rewrite the Neumann expansion of the resolvent in the following way:
\begin{eqnarray*}
& &\Big[f(\pi_{J_{\bold{k},\bold{q}}})\,,\,\sum_{j=0}^{\infty}\Big\{(\frac{1}{P^{(+)}_{J_{\bold{k},\bold{q}}}H^0_{J_{\bold{k},\bold{q}}}P^{(+)}_{J_{\bold{k},\bold{q}}}})^{\frac{1}{2}}(-\mathcal{V}^{(\bold{k},\bold{q})_{-1}}_{J_{\bold{k},\bold{q}}})\,(\frac{1}{P^{(+)}_{J_{\bold{k},\bold{q}}}H^0_{J_{\bold{k},\bold{q}}}P^{(+)}_{J_{\bold{k},\bold{q}}}})^{\frac{1}{2}}\Big\}^j]\quad\quad\quad\quad \\
& =&\sum_{j=0}^{\infty}\sum_{l=0}^{j-1}\Big\{\,\Big(\,(\frac{1}{P^{(+)}_{J_{\bold{k},\bold{q}}}H^0_{J_{\bold{k},\bold{q}}}P^{(+)}_{J_{\bold{k},\bold{q}}}})^{\frac{1}{2}}(-\mathcal{V}^{(\bold{k},\bold{q})_{-1}}_{J_{\bold{k},\bold{q}}})\,(\frac{1}{P^{(+)}_{J_{\bold{k},\bold{q}}}H^0_{J_{\bold{k},\bold{q}}}P^{(+)}_{J_{\bold{k},\bold{q}}}})^{\frac{1}{2}}\,\Big)^l\quad\quad \\ 
& &\quad\quad \times \Big[\, f(\pi_{J_{\bold{k},\bold{q}}})\, , \,(\frac{1}{P^{(+)}_{J_{\bold{k},\bold{q}}}H^0_{J_{\bold{k},\bold{q}}}P^{(+)}_{J_{\bold{k},\bold{q}}}})^{\frac{1}{2}}(-\mathcal{V}^{(\bold{k},\bold{q})_{-1}}_{J_{\bold{k},\bold{q}}})\,(\frac{1}{P^{(+)}_{J_{\bold{k},\bold{q}}}H^0_{J_{\bold{k},\bold{q}}}P^{(+)}_{J_{\bold{k},\bold{q}}}})^{\frac{1}{2}}\,\Big] \\
& &\quad\quad \times \Big(\,(\frac{1}{P^{(+)}_{J_{\bold{k},\bold{q}}}H^0_{J_{\bold{k},\bold{q}}}P^{(+)}_{J_{\bold{k},\bold{q}}}})^{\frac{1}{2}}(-\mathcal{V}^{(\bold{k},\bold{q})_{-1}}_{J_{\bold{k},\bold{q}}})\,(\frac{1}{P^{(+)}_{J_{\bold{k},\bold{q}}}H^0_{J_{\bold{k},\bold{q}}}P^{(+)}_{J_{\bold{k},\bold{q}}}})^{\frac{1}{2}}\,\Big)^{j-l-1}
\end{eqnarray*}
from which we see that
\begin{eqnarray*}
& &\|([H^0_{J_{\bold{k},\bold{q}}}]_1)^{\frac{1}{2}}\, (\frac{1}{[\pi_{J_{\bold{k},\bold{q}}}]_1})^{\frac{1}{2}}\,\frac{1}{G_{J_{\bold{k},\bold{q}}}-E_{J_{\bold{k},\bold{q}}}}\,P^{(+)}_{J_{\bold{k},\bold{q}}}\,V^{(\bold{k},\bold{q})_{-1}}_{J_{\bold{k},\bold{q}}}\,P^{(-)}_{J_{\bold{k},\bold{q}}}\|\\
&= &\|([H^0_{J_{\bold{k},\bold{q}}}]_1)^{\frac{1}{2}}\, (\frac{1}{[\pi_{J_{\bold{k},\bold{q}}}]_1})^{\frac{1}{2}}\,\sum_{j=0}^{\infty}\frac{1}{P^{(+)}_{J_{\bold{k},\bold{q}}}H^0_{J_{\bold{k},\bold{q}}}P^{(+)}_{J_{\bold{k},\bold{q}}}}\Big\{-\mathcal{V}^{(\bold{k},\bold{q})_{-1}}_{J_{\bold{k},\bold{q}}}\frac{1}{P^{(+)}_{J_{\bold{k},\bold{q}}}H^0_{J_{\bold{k},\bold{q}}}P^{(+)}_{J_{\bold{k},\bold{q}}}}\}^j\,P^{(+)}_{J_{\bold{k},\bold{q}}}\,V^{(\bold{k},\bold{q})_{-1}}_{J_{\bold{k},\bold{q}}}\,P^{(-)}_{J_{\bold{k},\bold{q}}}\|\quad\quad\quad\quad \\
&\leq&\|\,(\frac{[H^0_{J_{\bold{k},\bold{q}}}]_1}{P^{(+)}_{J_{\bold{k},\bold{q}}}H^0_{J_{\bold{k},\bold{q}}}P^{(+)}_{J_{\bold{k},\bold{q}}}})^{\frac{1}{2}}\,P^{(+)}_{J_{\bold{k},\bold{q}}}\,\sum_{j=0}^{\infty}\Big\{(\frac{1}{P^{(+)}_{J_{\bold{k},\bold{q}}}H^0_{J_{\bold{k},\bold{q}}}P^{(+)}_{J_{\bold{k},\bold{q}}}})^{\frac{1}{2}}(-\mathcal{V}^{(\bold{k},\bold{q})_{-1}}_{J_{\bold{k},\bold{q}}})\,(\frac{1}{P^{(+)}_{J_{\bold{k},\bold{q}}}H^0_{J_{\bold{k},\bold{q}}}P^{(+)}_{J_{\bold{k},\bold{q}}}})^{\frac{1}{2}}\}^j\\
& &\quad \times  (\frac{1}{[\pi_{J_{\bold{k},\bold{q}}}]_1})^{\frac{1}{2}}\, \,(\frac{1}{P^{(+)}_{J_{\bold{k},\bold{q}}}H^0_{J_{\bold{k},\bold{q}}}P^{(+)}_{J_{\bold{k},\bold{q}}}})^{\frac{1}{2}}\,P^{(+)}_{J_{\bold{k},\bold{q}}}\,V^{(\bold{k},\bold{q})_{-1}}_{J_{\bold{k},\bold{q}}}\,P^{(-)}_{J_{\bold{k},\bold{q}}}||\\
& &+\|(\frac{[H^0_{J_{\bold{k},\bold{q}}}]_1}{P^{(+)}_{J_{\bold{k},\bold{q}}}H^0_{J_{\bold{k},\bold{q}}}P^{(+)}_{J_{\bold{k},\bold{q}}}})^{\frac{1}{2}}\,P^{(+)}_{J_{\bold{k},\bold{q}}}\,\,\sum_{j=0}^{\infty}\sum_{l=0}^{j-1}\Big\{\,\Big(\,(\frac{1}{P^{(+)}_{J_{\bold{k},\bold{q}}}H^0_{J_{\bold{k},\bold{q}}}P^{(+)}_{J_{\bold{k},\bold{q}}}})^{\frac{1}{2}}(-\mathcal{V}^{(\bold{k},\bold{q})_{-1}}_{J_{\bold{k},\bold{q}}})\,(\frac{1}{P^{(+)}_{J_{\bold{k},\bold{q}}}H^0_{J_{\bold{k},\bold{q}}}P^{(+)}_{J_{\bold{k},\bold{q}}}})^{\frac{1}{2}}\,\Big)^l\quad\quad \\ 
& &\quad\quad \times \Big[\, f(\pi_{J_{\bold{k},\bold{q}}})\, , \,(\frac{1}{P^{(+)}_{J_{\bold{k},\bold{q}}}H^0_{J_{\bold{k},\bold{q}}}P^{(+)}_{J_{\bold{k},\bold{q}}}})^{\frac{1}{2}}(-\mathcal{V}^{(\bold{k},\bold{q})_{-1}}_{J_{\bold{k},\bold{q}}})\,(\frac{1}{P^{(+)}_{J_{\bold{k},\bold{q}}}H^0_{J_{\bold{k},\bold{q}}}P^{(+)}_{J_{\bold{k},\bold{q}}}})^{\frac{1}{2}}\,\Big] \\
& &\quad\quad \times \Big(\,(\frac{1}{P^{(+)}_{J_{\bold{k},\bold{q}}}H^0_{J_{\bold{k},\bold{q}}}P^{(+)}_{J_{\bold{k},\bold{q}}}})^{\frac{1}{2}}(-\mathcal{V}^{(\bold{k},\bold{q})_{-1}}_{J_{\bold{k},\bold{q}}})\,(\frac{1}{P^{(+)}_{J_{\bold{k},\bold{q}}}H^0_{J_{\bold{k},\bold{q}}}P^{(+)}_{J_{\bold{k},\bold{q}}}})^{\frac{1}{2}}\,\Big)^{j-l-1}\\
& &\quad\quad \times  \, (\frac{1}{P^{(+)}_{J_{\bold{k},\bold{q}}}H^0_{J_{\bold{k},\bold{q}}}P^{(+)}_{J_{\bold{k},\bold{q}}}})^{\frac{1}{2}}\,P^{(+)}_{J_{\bold{k},\bold{q}}}\,V^{(\bold{k},\bold{q})_{-1}}_{J_{\bold{k},\bold{q}}}\,P^{(-)}_{J_{\bold{k},\bold{q}}}\|
\end{eqnarray*}
\begin{eqnarray*}
&\leq&\|(\frac{[H^0_{J_{\bold{k},\bold{q}}}]_1}{P^{(+)}_{J_{\bold{k},\bold{q}}}H^0_{J_{\bold{k},\bold{q}}}P^{(+)}_{J_{\bold{k},\bold{q}}}})^{\frac{1}{2}}\,P^{(+)}_{J_{\bold{k},\bold{q}}}\|\,\Big\{\sum_{j=0}^{\infty}\|(\frac{1}{P^{(+)}_{J_{\bold{k},\bold{q}}}H^0_{J_{\bold{k},\bold{q}}}P^{(+)}_{J_{\bold{k},\bold{q}}}})^{\frac{1}{2}}\mathcal{V}^{(\bold{k},\bold{q})_{-1}}_{J_{\bold{k},\bold{q}}}\,(\frac{1}{P^{(+)}_{J_{\bold{k},\bold{q}}}H^0_{J_{\bold{k},\bold{q}}}P^{(+)}_{J_{\bold{k},\bold{q}}}})^{\frac{1}{2}}\|^j\Big\}\\ 
& &\quad \times \|  (\frac{1}{[\pi_{J_{\bold{k},\bold{q}}}]_1})^{\frac{1}{2}}\,\,(\frac{1}{P^{(+)}_{J_{\bold{k},\bold{q}}}H^0_{J_{\bold{k},\bold{q}}}P^{(+)}_{J_{\bold{k},\bold{q}}}})^{\frac{1}{2}}\,P^{(+)}_{J_{\bold{k},\bold{q}}}\,V^{(\bold{k},\bold{q})_{-1}}_{J_{\bold{k},\bold{q}}}\,P^{(-)}_{J_{\bold{k},\bold{q}}}||\\
& &+\|(\frac{[H^0_{J_{\bold{k},\bold{q}}}]_1}{P^{(+)}_{J_{\bold{k},\bold{q}}}H^0_{J_{\bold{k},\bold{q}}}P^{(+)}_{J_{\bold{k},\bold{q}}}})^{\frac{1}{2}}\,P^{(+)}_{J_{\bold{k},\bold{q}}}\|\,\\
& &\quad\quad \times \sum_{j=0}^{\infty}\sum_{l=0}^{j-1}\Big\|\,\Big(\,(\frac{1}{P^{(+)}_{J_{\bold{k},\bold{q}}}H^0_{J_{\bold{k},\bold{q}}}P^{(+)}_{J_{\bold{k},\bold{q}}}})^{\frac{1}{2}}(-\mathcal{V}^{(\bold{k},\bold{q})_{-1}}_{J_{\bold{k},\bold{q}}})\,(\frac{1}{P^{(+)}_{J_{\bold{k},\bold{q}}}H^0_{J_{\bold{k},\bold{q}}}P^{(+)}_{J_{\bold{k},\bold{q}}}})^{\frac{1}{2}}\,\Big\|^{j-1}\quad\quad \\ 
& &\quad\quad\quad\quad \quad\quad  \times \Big\|\,\Big[\, f(\pi_{J_{\bold{k},\bold{q}}})\, , \,(\frac{1}{P^{(+)}_{J_{\bold{k},\bold{q}}}H^0_{J_{\bold{k},\bold{q}}}P^{(+)}_{J_{\bold{k},\bold{q}}}})^{\frac{1}{2}}(-\mathcal{V}^{(\bold{k},\bold{q})_{-1}}_{J_{\bold{k},\bold{q}}})\,(\frac{1}{P^{(+)}_{J_{\bold{k},\bold{q}}}H^0_{J_{\bold{k},\bold{q}}}P^{(+)}_{J_{\bold{k},\bold{q}}}})^{\frac{1}{2}}\Big]\,\Big\| \\
& &\quad\quad \times \|  (\frac{1}{[\pi_{J_{\bold{k},\bold{q}}}]_1})^{\frac{1}{2}}\,(\frac{1}{P^{(+)}_{J_{\bold{k},\bold{q}}}H^0_{J_{\bold{k},\bold{q}}}P^{(+)}_{J_{\bold{k},\bold{q}}}})^{\frac{1}{2}}\,P^{(+)}_{J_{\bold{k},\bold{q}}}\,V^{(\bold{k},\bold{q})_{-1}}_{J_{\bold{k},\bold{q}}}\,P^{(-)}_{J_{\bold{k},\bold{q}}}\|\\
&\leq &C\cdot \|  (\frac{1}{[\pi_{J_{\bold{r},\bold{i}}}]_1})^{\frac{1}{2}}\, (\frac{1}{P^{(+)}_{J_{\bold{k},\bold{q}}}H^0_{J_{\bold{k},\bold{q}}}P^{(+)}_{J_{\bold{k},\bold{q}}}})^{\frac{1}{2}}\,P^{(+)}_{J_{\bold{k},\bold{q}}}\,V^{(\bold{k},\bold{q})_{-1}}_{J_{\bold{k},\bold{q}}}\,P^{(-)}_{J_{\bold{k},\bold{q}}}\|\\
&\leq &C\cdot \| (\frac{[H^0_{J_{\bold{k},\bold{q}}}]_1}{P^{(+)}_{J_{\bold{k},\bold{q}}}H^0_{J_{\bold{k},\bold{q}}}P^{(+)}_{J_{\bold{k},\bold{q}}}})^{\frac{1}{2}}\,P^{(+)}_{J_{\bold{k},\bold{q}}}\|\cdot \| (\frac{1}{[\pi_{J_{\bold{k},\bold{q}}}]_1})^{\frac{1}{2}}\,\,(\frac{1}{[H^0_{J_{\bold{k},\bold{q}}}]_1})^{\frac{1}{2}}\,P^{(+)}_{J_{\bold{k},\bold{q}}}\,V^{(\bold{k},\bold{q})_{-1}}_{J_{\bold{k},\bold{q}}}\,P^{(-)}_{J_{\bold{k},\bold{q}}}\|\\
&\leq & C\cdot  \| (\frac{1}{[\pi_{J_{\bold{k},\bold{q}}}]_1})^{\frac{1}{2}}\,\,(\frac{1}{[H^0_{J_{\bold{k},\bold{q}}}]_1})^{\frac{1}{2}}\,P^{(+)}_{J_{\bold{k},\bold{q}}}\,V^{(\bold{k},\bold{q})_{-1}}_{J_{\bold{k},\bold{q}}}\,P^{(-)}_{J_{\bold{k},\bold{q}}}\|
\end{eqnarray*}
where we have used (\ref{delineq}), $\| (\frac{[H^0_{J_{\bold{k},\bold{q}}}]_1}{P^{(+)}_{J_{\bold{k},\bold{q}}}H^0_{J_{\bold{k},\bold{q}}}P^{(+)}_{J_{\bold{k},\bold{q}}}})^{\frac{1}{2}}\,P^{(+)}_{J_{\bold{k},\bold{q}}}\|\leq \sqrt{2}$,  and the constant $C$ changes its value from line to line. The proof is thus complete.
\qed

\setcounter{equation}{0}
\begin{appendix}
\section{Appendix A}\label{appA}
We here collect and prove some elementary estimates which are used thoroughout the article.
\begin{lem}\label{op-ineq-1} 
For any $J_{\bold{l},\bold{i}}$, we define
\begin{equation}\label{pro-plus-multi-bis-bis}
P^{(+)}_{J_{\bold{l},\bold{i}}}:=\charf_{\mathcal{H}^{(N^{d})}\ominus \mathcal{H}_{J_{\bold{l},\bold{i}}}}\bigotimes  \Big(\bigotimes_{\bold{j}\in J_{\bold{l},\bold{i}}}P_{\Omega_{\bold{j}}}\Big)^{\perp}\,.
\end{equation}
Then the inequality
\begin{equation}\label{gen-lemmaA-1-bis}
 \sum_{\bold{j}\in J_{\bold{l},\bold{i}}}  \charf_{\mathcal{H}^{(N^d)}\ominus 	\mathcal{H}_\bold{j}} \otimes P^{\perp}_{\Omega_{\bold{j}}}\geq P^{(+)}_{J_{\bold{l},\bold{i}}} \,
\end{equation}
holds true where $P^{\perp}_{\Omega_{\bold{j}}}:=\charf_{\bold{j}}-P_{\Omega_{\bold{j}}}$.
\end{lem}

\noindent
\emph{Proof}

\noindent
Let $A$ be the self-adjoint operator 

$$A:=\sum_{\bold{j}\in J_{\bold{l},\bold{i}}}  \charf_{\mathcal{H}^{(N^d)}\ominus 	\mathcal{H}_\bold{j}} \otimes P^{\perp}_{\Omega_{\bold{j}}}+\charf_{\mathcal{H}^{(N^{d})}\ominus \mathcal{H}_{J_{\bold{l},\bold{i}}}}\bigotimes  \Big(\bigotimes_{\bold{j}\in J_{\bold{l},\bold{i}}}P_{\Omega_{\bold{j}}}\Big)\,.$$
Observe that $A$ is the sum of $(l_1+1)(l_2+1)\ldots(l_d+1)+1$ orthogonal projections which commute with one another, thus its spectrum 
is contained in the set $$\{0, 1,2,\ldots, (l_1+1)(l_2+1)\ldots(l_d+1)+1  \}\,.$$
In particular, as the spectrum is finite, its points are all isolated and therefore they are all eigenvalues. 
Next we will  prove that 
$A$ is invertible, so that $A\geq \charf_{\mathcal{H}^{(N^d)}}$ will follow, which is exactly
the sought inequality as by definition $$\charf_{\mathcal{H}^{(N^d)}}-\charf_{\mathcal{H}^{(N^{d})}\ominus \mathcal{H}_{J_{\bold{l},\bold{i}}}}\bigotimes  \Big(\bigotimes _{\bold{j}\in J_{\bold{l},\bold{i}}}P_{\Omega_{\bold{j}}}\Big)=P^{(+)}_{J_{\bold{l},\bold{i}}}.$$ 
 In light of the remark we made on the spectrum of $A$, proving its invertibility actually amounts to showing its injectivity.
Decomposing the Hilbert space $\mathcal{H}^{(N^d)}$ as $\mathcal{H}_1\otimes \mathcal{H}_2$, where
$\mathcal{H}_1$ and $\mathcal{H}_2$ are respectively $\bigotimes_{\bold{j}\in J_{\bold{l}, \bold{i}}}\mathcal{H}_\bold{j}$ and $\bigotimes_{\bold{j}\in \Lambda_N^d \setminus J_{\bold{l},\bold{i}}}\mathcal{H}_\bold{j}$, induces a factorization of $A$ as $A_1\otimes A_2$, with $A_1$ and $A_2$ acting respectively on $\mathcal{H}_1$ and $\mathcal{H}_2$. 
The injectivity of $A$ is thus equivalent to the injectivity of both $A_1$ and $A_2$. Only $A_2$ has to be dealt with, as
$A_1$ is  simply a multiple of the identity.
By the definition of $A$, $A_2$  coincides with
$$\sum_{\bold{j}\in J_{\bold{l}, \bold{i}}}P^\perp_{\Omega_\bold{j}}+\prod_{\bold{j}\in J_{\bold{l},\bold{i}}}P_{\Omega_{\bold{j}}}\,,$$ 
thus  also $A_2$ is a sum of projectors which
commute with one another.

\noindent
Let $\Psi\in\mathcal{H}_2$ such that
$A_2\Psi=0$. From $(\Psi, A_2\Psi)=0$, we have
\begin{equation}\label{proj-equals}
(\Psi, P^\perp_{\Omega_\bold{j}}\Psi)=0 \,\,\, \forall \bold{j}\in J_{\bold{l}, \bold{i}}\quad \text{and}\quad \prod_{\bold{j}\in J_{\bold{l},\bold{i}}}P_{\Omega_\bold{j}}\Psi=0\,.
\end{equation}
The first equalities in (\ref{proj-equals}) yield $\Psi=P_{\Omega_\bold{j}}\Psi$ for every $\bold{j}\in J_{\bold{l}, \bold{i}}$.
Consequently the second equality implies $\Psi=\prod_{\bold{j}\in J_{\bold{l},\bold{i}}}P_{\Omega_\bold{j}}\Psi=0$, which concludes the proof.
\qed

From previous Lemma \ref{op-ineq-1} we derive:
\begin{cor}\label{op-ineq-2}
For any $J_{\bold{l},\bold{i}}$, we define
\begin{equation}\label{pro-plus-multi-bis-bis}
P^{(+)}_{J_{\bold{l},\bold{i}}}:=\charf_{\mathcal{H}^{(N^{d})}\ominus \mathcal{H}_{J_{\bold{l},\bold{i}}}}\otimes  \Big(\otimes \prod_{\bold{j}\in J_{\bold{l},\bold{i}}}P_{\Omega_{\bold{j}}}\Big)^{\perp}\,.
\end{equation}
Then, for any $J_{\bold{k},\bold{q}}$ the following inequality holds
\begin{equation}\label{gen-lemmaA-2-bis}
\sum_{\bold{i}\,:\,J_{\bold{l},\bold{i}}\subset  J_{\bold{k},\bold{q}} }P^{(+)}_{J_{\bold{l},\bold{i}}}\leq (l+1)^d \sum_{\bold{j}\in  J_{\bold{k},\bold{q}}}\charf_{ \Lambda^{d}_N \setminus \bold{j}}\otimes P^{\perp}_{\Omega_{\bold{j}}}\,
\end{equation}
where $l=|\bold{l}|$.
\end{cor}

\noindent
\emph{Proof}

\noindent
From Lemma \ref{op-ineq-1} we know that
\begin{equation}\label{gen-lemmaA-1-bis-bis}
P^{(+)}_{J_{\bold{l},\bold{i}}}\leq \sum_{\bold{j}\in  J_{\bold{k},\bold{q}}}\charf_{ \Lambda^{d}_N \setminus \bold{j}}\otimes P^{\perp}_{\Omega_{\bold{j}}}\,.
\end{equation}
By summing  at fixed $\bold{l}$  the l-h-s of (\ref{gen-lemmaA-1-bis-bis}) over all $J_{\bold{l},\bold{i}}$ contained in  $J_{\bold{k},\bold{q}}$, for each site $\bold{j}\in J_{\bold{k},\bold{q}}$  we get not more than $$(l_1+1)(l_2+1)\dots(l_d+1)$$ terms of the type
\begin{equation}
 \charf_{\Lambda^{d}_N\setminus \bold{j}} \otimes P^{\perp}_{\Omega_{\bold{j}}}\,.
\end{equation}
Hence the inequality in (\ref{gen-lemmaA-2-bis}) follows.
\qed

The following Lemma is proved analogously to Lemma A.3 in \cite{DFPR1}.
\begin{lem}\label{formboundedness}
For $t>0$ as small  as stated in Corollary \ref{cor-gap}, the following bound holds
\begin{equation}
(\Phi, P^{(+)}_{J_{\bold{k},\bold{q}}}(G_{J_{\bold{k},\bold{q}}}-E_{J_{\bold{k},\bold{q}}})P^{(+)}_{J_{\bold{k},\bold{q}}}\Phi)\geq \frac{\Delta_{J_{\bold{k},\bold{q}}}}{2} (\Phi, P^{(+)}_{J_{\bold{k},\bold{q}}}(H^0_{J_{\bold{k},\bold{q}}}+1)P^{(+)}_{J_{\bold{k},\bold{q}}}\Phi) \label{bound-G}
\end{equation}
for any vector $\Phi$ in the domain of $H^0_{J_{\bold{k},\bold{q}}}$, where $\Delta_{J_{\bold{k},\bold{q}}}$ is the lower bound of the spectral gap determined in Corollary \ref{cor-gap}. Consequently, 
\begin{equation}
\left\|\frac{1}{(G_{J_{\bold{k},\bold{q}}}-E_{J_{\bold{k},\bold{q}}})^{\frac{1}{2}}}P^{(+)}_{J_{\bold{k},\bold{q}}} (H^0_{J_{\bold{k},\bold{q}}}+1)^{\frac{1}{2}}\right\|\leq \frac{\sqrt{2}}{\Delta_{J_{\bold{k},\bold{q}}}^{\frac{1}{2}}}\label{op-norm-G}
\end{equation}
and
\begin{equation}
\left\|\frac{1}{(G_{J_{\bold{k},\bold{q}}}-E_{J_{\bold{k},\bold{q}}})}P^{(+)}_{J_{\bold{k},\bold{q}}} (H^0_{J_{\bold{k},\bold{q}}}+1)^{\frac{1}{2}}\right\|\leq \frac{\sqrt{2}}{\Delta_{J_{\bold{k},\bold{q}}}}\,.\label{op-norm-G-2}
\end{equation}
\end{lem}

%
%

\section{Appendix B}\label{appB}
We here collect some results that are needed for the proof of Theorem \ref{th-norms}, regime $\mathcal{R}$1. The proofs of the following Lemmas can be found in  \cite{DFPR3}, Section 3 and Appendix A.

The following properties are easily deduced for the elements of $\mathcal{B}_{V^{(\bold{k},\bold{q})}_{J_{\bold{r},\bold{i}}}}$.

\begin{lem}[\cite{DFPR3}, Section 3.1.1]    \label{treeprops}
\begin{itemize}
\item[P-i)]
For $\mathfrak{b}\in \mathcal{B}_{V^{(\bold{k},\bold{q})}_{J_{\bold{r},\bold{i}}}}$, the set 
$$\bigcup_{i\in \big\{1,\cdots,\vert \mathcal{R}_{\mathfrak{b}}\vert \big\}}J_{\bold{k}^{(i)},\bold{q}^{(i)}}$$ 
is connected. Likewise, for any fixed $n\in \big\{1\cdots \vert \mathcal{R}_\mathfrak{b}\vert\big\}$, the set 
$\bigcup_{n\leq i\leq \vert\mathcal{R}_{\mathfrak{b}}\vert}J_{\bold{k}^{(i)},\bold{q}^{(i)}}$  is connected. 

\item[P-ii)]
For $\mathfrak{b}\in \mathcal{B}_{V^{(\bold{k},\bold{q})}_{J_{\bold{r},\bold{i}}}}$, the cardinality, $\vert \mathcal{R}_\mathfrak{b}\vert $, of the set $\mathcal{R}_\mathfrak{b}$ of rectangles 
is such that $\vert \mathcal{R}_\mathfrak{b}\vert \geq \mathcal{O}(\frac{r}{k})\geq \mathcal{O}( r^{\frac{3}{4}})$. 

\item[P-iii)] The set $J_{\bold{r},\bold{i}}$ is the minimal rectangle associated with 
$\bigcup_{i\in \big\{1,\cdots,\vert\mathcal{R}_\mathfrak{b}\vert \big\}}J_{\bold{k}^{(i)},\bold{q}^{(i)}}$, for any branch 
$\mathfrak{b}\in \mathcal{B}_{V^{(\bold{k},\bold{q})}_{J_{\bold{r},\bold{i}}}}$.  Furthermore, if we amputate a branch at some vertex by keeping only the descendants of that vertex (i.e., the lower part only) then the same property holds for the rectangle associated with the potential labelling the (new) root vertex of the amputated branch that has been created. 
\item[P-iv)]
Two different branches $\mathfrak{b}, \mathfrak{b}' \in  \mathcal{B}_{V^{(\bold{k},\bold{q})}_{J_{\bold{r},\bold{i}}}}$ are associated with two different (ordered) sets of rectangles $\mathcal{R}_\mathfrak{b}$ and $\mathcal{R}_{\mathfrak{b}'}$.
\end{itemize}
\end{lem}

 The following collection of definitions is needed to specify precisely what we mean by a path visiting rectangles.
 \begin{defn} \label{pathsdef}
\noindent
\begin{itemize}
\item[i)] A path $\Gamma$ is a finite sequence of rectangles $\{J_{\bold{s}^{(i)},\bold{u}^{(i)}}\}_{i=1}^n$, for some $n\in\mathbb{N}$, with the property that $J_{\bold{s}^{(i)},\bold{u}^{(i)}}\neq  J_{\bold{s}^{(i+1)},\bold{u}^{(i+1)}}$ and  $J_{\bold{s}^{(i)},\bold{u}^{(i)}}\cap J_{\bold{s}^{(i+1)},\bold{u}^{(i+1)}}\neq\emptyset$ for every $i=1\cdots n-1$. 
\item[ii)] The set of steps, $\mathcal{S}_\Gamma$, of the path $\Gamma\equiv\{J_{\bold{s}^{(i)},\bold{u}^{(i)}}\}_{i=1}^n$ is the set of ordered pairs $(J_{\bold{s}^{(i)},\bold{u}^{(i)}},J_{\bold{s}^{(i+1)},\bold{u}^{(i+1)}})$, $i=1\cdots n-1$.
\item[iii)] The length, $l_{\Gamma}$, of a path $\Gamma\equiv \{J_{\bold{s}^{(i)},\bold{u}^{(i)}}\}_{i=1}^n$ is $l_{\Gamma}:=n-1$.
\item[iv)] The support, supp$(\Gamma)$,  of a path $\Gamma\equiv \{J_{\bold{s}^{(i)},\bold{u}^{(i)}}\}_{i=1}^n$ is supp$(\Gamma):=\Big\{ J_{\bold{s}^{(i)},\bold{u}^{(i)}},\, i\in\{1\cdots n\}\Big\}.$
\item[v)] A path $\Gamma\equiv \{J_{\bold{s}^{(i)},\bold{u}^{(i)}}\}_{i=1}^n$, $n\geq 2$,  is closed if $J_{\bold{s}^{(1)},\bold{u}^{(1)}}=J_{\bold{s}^{(n)},\bold{u}^{(n)}}$.
\end{itemize}
\end{defn}

We write  the connected set $\bigcup_{i\in\{1,\cdots,|\mathcal{R}_{\mathfrak{b}}|\}}J_{\bold{k}^{(i)},\bold{q}^{(i)}}$ as the union 
$$\bigcup_{\rho=k_0}^{k} \Big(\bigcup_{j=1}^{j_{\rho}}\mathcal{Z}^{(j)}_{\rho}\Big)\,,$$  
where $\{\mathcal{Z}^{(j)}_{\rho}, \quad j=1,\dots,j_{\rho}\}$  are distinct connected components 
of (unions of) rectangles of a given size $\rho$, $k_0\leq \rho \leq k$, starting from the lowest one $k_0\geq 1$, with the following properties:

\noindent
1) $j_{k_0}=1$ (i.e., there is only one component for $\rho=k_0$);

\noindent
2) rectangles of the same size but belonging to different components do \textit{not} overlap, i.e., for any $\rho $, $\mathcal{Z}^{(j)}_{\rho}\cap \mathcal{Z}^{(j')}_{\rho}=\emptyset$\,, for $j\neq j'$.

\noindent
We call $\text{supp}(\mathcal{Z}^{(j)}_{\rho})$, $\rho=k_0, \dots,k\,,\,j=1,\dots, j_{\rho}$,  the set of rectangles of $\mathcal{Z}^{(j)}_{\rho}$, i.e.,
$$\text{supp}(\mathcal{Z}^{(j)}_{\rho}):=\Big\{J_{\bold{k}^{(i)},\bold{q}^{(i)}}: J_{\bold{k}^{(i)},\bold{q}^{(i)}}\subset \mathcal{Z}^{(j)}_{\rho}, 
i\in \big\{1,\cdots,\vert \mathcal{R}_{\mathfrak{b}}\vert \big\} \Big\}.$$ 
\\

The following lemma specifies the map from the set of $\{\mathcal{R}_{\mathfrak{b}}\}$ to a set of paths $\{\Gamma_{\mathfrak{b}}\}$ that was mentioned above.
\begin{lem}[\cite{DFPR3}, Lemma A.5]\label{conn-rect-2}
For $\mathfrak{b}\in  \mathcal{B}_{V^{(\bold{k},\bold{q})}_{J_{\bold{r},\bold{i}}}}$, let
$$ \cup_{i\in\{1,\cdots,|\mathcal{R}_{\mathfrak{b}}|\}}J_{\bold{k}^{(i)},\bold{q}^{(i)}}=\cup_{\rho=k_0}^{k} \cup_{j=1}^{j_{\rho}}\mathcal{Z}^{(j)}_{\rho}$$  where $\{\mathcal{Z}^{(j)}_{\rho}, \quad j=1,\dots,j_{\rho}\}$  are distinct connected components of (unions of) rectangles of same size $\rho$. Then there is a path, $\Gamma_{\mathfrak{b}}$, of length $l_{\Gamma_\mathfrak{b}}$ such that  $$l_{\Gamma_\mathfrak{b}} \leq 2(n_{k_0}+\sum_{j=1}^{j_2}n_{k_0+1}^{(j)}+\dots +\sum_{j=1}^{j_k}n_k^{(j)})-2$$
with $n_\rho^{(j)}:=\vert \text{supp}(\mathcal{Z}^{(j)}_{\rho})\vert$ with the following properties:
\begin{enumerate}
\item[A)]   the support of $\Gamma_\mathfrak{b}$ is  $\mathcal{R}_\mathfrak{\mathfrak{b}}$;
\item[B)] for each component $\mathcal{Z}^{(j)}_{\rho}$ consisting of the union of $n_{\rho}^{(j)}$ rectangles, at most $2n_{\rho}^{(j)}-2$ steps are implemented (i.e., there are at most $2n_{\rho}^{(j)}-2$ steps $\sigma\in\mathcal{S}_{\Gamma_\mathfrak{b}}$ for which $\sigma\in \text{supp}(\mathcal{Z}^{(j)}_{\rho})\times \text{supp}(\mathcal{Z}^{(j)}_{\rho})$);
\item[C)]  there are at most two steps connecting rectangles in $\text{supp}(\mathcal{Z}_\rho^{(j)})$ with rectangles
of lower size: more precisely, for every connected component $\mathcal{Z}_\rho^{(j)}$ there is at most one $J_{\bold{s},\bold{u}}$ in $\text{supp}(\mathcal{Z}_\rho^{(j)})$  such that $(J_{\bold{s}',\bold{u}'},J_{\bold{s},\bold{u}})\in\mathcal{S}_{\Gamma_\mathfrak{b}}$ with $s'<s$, and one $J_{\bold{s},\bold{u}}$ such that $(J_{\bold{s},\bold{u}},J_{\bold{s}',\bold{u}'})\in\mathcal{S}_{\Gamma_\mathfrak{b}}$ with $s<s'$.
\end{enumerate}
\end{lem}

\begin{lem}\label{pathweightest}
Let $\mathfrak{b}\in\mathcal{B}_{V^{(\bold{k},\bold{q})}_{J_{\bold{r},\bold{i}}}}$, then
\begin{equation}
\|\mathfrak{b}\|_{H^0} \leq t^{\frac{r-1}{3}}\cdot \prod_{\sigma\in\mathcal{S}_{\Gamma_{\mathfrak{b}}}}w_\sigma ,
\end{equation}
where $\Gamma_{\mathfrak{b}}$ is the path associated with $\mathfrak{b}$ constructed in Lemma \ref{conn-rect-2}, $\mathcal{S}_{\Gamma_{\mathfrak{b}}}$ is the set of steps of $\Gamma_{\mathfrak{b}}$ and $w_{\sigma}$ is the weight in Eq. \ref{stepweight}.
\end{lem}
\emph{Proof}
%
%
%
\noindent

Consider the rectangles of the set $\text{supp}(\mathcal{Z}^{(j)}_{\rho})$: by definition there are $n_{\rho}^{(j)}$ such rectangles, and, for the paths $\Gamma_{\mathfrak{b}}$, there are at most $2n_{\rho}^{(j)}-2$ steps between them. In addition there are at most $2$ steps, from rectangles of lower size and back,  to be taken into account. 

\noindent
By Lemma \ref{weightcontrol}, we have
\begin{equation}\label{weight-path}
\|\mathfrak{b}\|_{H^0} \leq  t^{\frac{r-1}{3}} \prod_{\rho=1\,; 
\\ j_\rho\neq 0}^k\,\prod_{j=1}^{j_{\rho}}\Big((c+1) \frac{t^{1/3}}{\rho^{x_d}}\Big)^{n_\rho^{(j)}}.
\end{equation}
From this we can deduce
$$\|\mathfrak{b}\|_{H^0}  \leq t^{\frac{r-1}{3}}\cdot \prod_{\sigma\in\mathcal{S}_{\Gamma_{\mathfrak{b}}}}w_\sigma$$
using the following observation: if we denote by $\mathcal{S}_{\mathcal{Z}_\rho^{(j)}}$ the set consisting  of  at most $2n_\rho^{(j)}-2$ steps between rectangles of  $\text{supp}\mathcal{Z}_\rho^{(j)}$ and the additional at most $2$ steps from rectangles of lower size and back, then we have
$$ \Big((c+1)\frac{t^{\frac{1}{3}}}{\rho^{x_d}}\Big)^{n_\rho^{(j)}}\leq \prod_{\sigma\in\mathcal{S}_{\mathcal{Z}_\rho^{(j)}} } w_\sigma\,,$$
since $w_\sigma$, $\sigma\in\mathcal{S}_{\mathcal{Z}_\rho^{(j)}} $, coincides with $\Big((c+1)\frac{t^{\frac{1}{3}}}{\rho^{x_d}}\Big)^{\frac{1}{2}}<1$ and $|\mathcal{S}_{\mathcal{Z}_\rho^{(j)}}|\leq 2n_\rho^{(j)} $, by construction.
\qed

\end{appendix}

\end{document}